\pdfoutput=1

\documentclass[11pt,twoside,a4paper,cmspaper,final,collab]{cms-tdr}
\def\svnVersion{473627P}\def\cmsCernNoTag{CERN-EP-2018-123}\def\cmsCernDate{\today}\def\cmsMessage{Published in the Journal of High Energy Physics as \href{http://dx.doi.org/10.1007/JHEP08(2018)130}{\doi{10.1007/JHEP08(2018)130.}}}

\begin{document}\cmsNoteHeader{EXO-16-056}

\hyphenation{had-ron-i-za-tion}
\hyphenation{cal-or-i-me-ter}
\hyphenation{de-vices}
\RCS$HeadURL: svn+ssh://svn.cern.ch/reps/tdr2/papers/EXO-16-056/trunk/EXO-16-056.tex $
\RCS$Id: EXO-16-056.tex 470254 2018-07-30 15:52:28Z alverson $
\newlength\cmsFigWidth
\ifthenelse{\boolean{cms@external}}{\setlength\cmsFigWidth{0.85\columnwidth}}{\setlength\cmsFigWidth{0.4\textwidth}}
\ifthenelse{\boolean{cms@external}}{\providecommand{\cmsLeft}{top\xspace}}{\providecommand{\cmsLeft}{left\xspace}}
\ifthenelse{\boolean{cms@external}}{\providecommand{\cmsRight}{bottom\xspace}}{\providecommand{\cmsRight}{right\xspace}}
\providecommand{\CL}{CL\xspace}
\ifthenelse{\boolean{cms@external}}{\providecommand{\CLp}{C.L}}{\providecommand{\CLp}{CL}}
\ifthenelse{\boolean{cms@external}}{\providecommand{\NA}{\ensuremath{\cdots}\xspace}}{\providecommand{\NA}{\text{---}\xspace}}
\providecommand{\cmsTable}[1]{\resizebox{\textwidth}{!}{#1}}

\newcommand{\mjj}{\ensuremath{m_{\mathrm{jj}}}\xspace}
\newcommand{\detajj}{\ensuremath{\abs{\Delta\eta}}\xspace}
\newcommand{\Qstar}{\ensuremath{\PQq^*}\xspace}
\newcommand{\RunLumi}{36\fbinv}
\newcommand{\CaloLumi}{27\fbinv}
\newcommand{\lumiUncert}{2.5\%\xspace}
\newcommand{\jecUncert}{2\%\xspace}
\newcommand{\RECOminMjjCut }{1.25\TeV}
\newcommand{\CALOminMjjCut }{0.49\TeV}
\newcommand{\minMassLow}{0.6\TeV}
\newcommand{\mDM}{\ensuremath{m_{\text{DM}}}\xspace}
\newcommand{\gDM}{\ensuremath{g_{\text{DM}}}\xspace}
\newcommand{\gq}{\ensuremath{g_\PQq}\xspace}
\newcommand{\mMed}{\ensuremath{M_{\text{Med}}}\xspace}

\cmsNoteHeader{EXO-16-056}
\title{Search for narrow and broad dijet resonances in proton-proton collisions at $\sqrt{s}=13\TeV$ and constraints on dark matter mediators and other new particles}

\date{\today}

\abstract{

Searches for resonances decaying into pairs of jets are performed using proton-proton
collision data collected at $\sqrt{s} = 13$\TeV corresponding to an integrated luminosity of up to 36\fbinv.
A low-mass search, for resonances with masses between 0.6 and 1.6\TeV, is performed based on events with dijets reconstructed at the trigger level from
calorimeter information.
A high-mass search, for resonances with masses above 1.6\TeV, is performed using dijets reconstructed offline with a particle-flow algorithm.
The dijet mass spectrum is well described by a smooth parameterization and no evidence for the production of new
particles is observed. Upper limits at $95\%$ confidence level are reported on the production cross section for narrow resonances with masses above 0.6\TeV. In
the context of specific models, the limits exclude string resonances with masses below 7.7\TeV, scalar diquarks below 7.2\TeV, axigluons and
colorons below 6.1\TeV, excited quarks below 6.0\TeV, color-octet scalars below 3.4\TeV, $\PWpr$ bosons below 3.3\TeV, $\PZpr$
bosons below 2.7\TeV, Randall--Sundrum gravitons below 1.8\TeV and in the range 1.9 to 2.5\TeV, and dark matter mediators below 2.6\TeV. The limits on both
vector and axial-vector mediators, in a simplified model of interactions between quarks and dark matter particles, are presented as functions of dark matter
particle mass and coupling to quarks. Searches are also presented for broad resonances, including for the first time spin-1 resonances with intrinsic widths as large
as 30\% of the resonance mass. The broad resonance search improves and extends the exclusions of a dark matter mediator to larger values of its mass and coupling to quarks.}

\hypersetup{%
pdfauthor={CMS Collaboration},%
pdftitle={Searches for narrow and broad dijet resonances in pp collisions at sqrt(s)=13 TeV and constraints on dark matter mediators and other new particles},%
pdfsubject={CMS},%
pdfkeywords={CMS, physics, search, exotica, dijet, resonance}}

\maketitle
\section{Introduction}

Models of physics that extend the standard model (SM) often require new particles that couple to quarks (\PQq) and/or gluons (\Pg) and
decay to dijets. The natural width of resonances in the dijet mass (\mjj) spectrum increases with the coupling,
and may vary from narrow to broad compared to the experimental resolution. For example, in a model in which dark matter (DM)
particles couple to quarks through a DM mediator, the mediator can decay to either a pair of DM particles or a pair of jets
and therefore can be observed as a dijet resonance~\cite{Chala:2015ama,Abercrombie:2015wmb} that is
either narrow or broad, depending on the strength
of the coupling. When the resonance is broad, its observed line-shape depends significantly on the resonance spin. Here we report a search for
narrow dijet resonances and a complementary search for broad resonances that considers multiple values of the resonance spin and widths as large as
30\% of the resonance mass. Both approaches are sensitive to resonances with intrinsic widths that are small compared to the experimental
resolution, but the broad resonance search is also sensitive to resonances with larger intrinsic widths.
We explore the implications for multiple specific models of dijet resonances and for a range of quark coupling strength for a DM mediator.

\subsection{Searches}

This paper presents the results of searches for dijet resonances that were performed with proton-proton ($\Pp\Pp$)
collision data collected at $\sqrt{s}=13$\TeV. The data correspond to an integrated luminosity of up to \RunLumi and were collected in 2016
with the CMS detector at the CERN LHC. Similar searches for narrow resonances have been published previously by
the ATLAS and CMS Collaborations at $\sqrt{s}=13$\TeV~\cite{Aaboud:2018fzt,Aaboud:2017yvp,Sirunyan:2016iap,Khachatryan:2015dcf,ATLAS:2015nsi},
8\TeV~\cite{Khachatryan:2016ecr,Khachatryan:2015sja,Aad:2014aqa,Chatrchyan:2013qhXX}, and
7\TeV~\cite{CMS:2012yf,Aad201237,ATLAS:2012pu,Chatrchyan2011123,Aad:2011aj,Khachatryan:2010jd,ATLAS2010} using
strategies reviewed in Ref.~\cite{Harris:2011bh}. A search for broad resonances considering natural widths as large as
30\% of the resonance mass, directly applicable to spin-2 resonances only, has been published once before by CMS at
$\sqrt{s}=8$\TeV~\cite{Khachatryan:2015sja}. Here we explicitly consider spin-1 and spin-2 resonances that are both broad.

The narrow resonance search is conducted in two regions of the dijet mass. The first is a low-mass search for resonances with
masses between 0.6 and 1.6\TeV. This search uses a dijet event sample
corresponding to an integrated luminosity of \CaloLumi, less than the full data
sample, as discussed in Section~\ref{sec:trigger}.
The events are reconstructed, selected, and recorded in a compact form by the high-level trigger (HLT)~\cite{Khachatryan:2016bia} in
a technique referred to as ``data scouting"~\cite{Mukherjee:2017wcl},
which is conceptually similar to the strategy that is reported in Ref.~\cite{Aaij:2016rxn}.
Data scouting was previously used for low-mass searches published
by CMS at $\sqrt{s}=13$\TeV~\cite{Sirunyan:2016iap} and at 8\TeV~\cite{Khachatryan:2016ecr}, and is similar to a trigger-level search at 13\TeV recently published by ATLAS~\cite{Aaboud:2018fzt}.
The second search is a high-mass search~\cite{Aaboud:2017yvp,Sirunyan:2016iap,Khachatryan:2015dcf,ATLAS:2015nsi,
Khachatryan:2015sja,Aad:2014aqa,Chatrchyan:2013qhXX,CMS:2012yf,Aad201237,ATLAS:2012pu,Chatrchyan2011123,Aad:2011aj,Khachatryan:2010jd,
ATLAS2010} for resonances with masses above 1.6\TeV, based on dijet events that are reconstructed offline in the full data sample
corresponding to an integrated luminosity of \RunLumi. The search for broad resonances uses the same selected events as does the
high-mass search for narrow resonances.

\subsection{Models}

We present model independent results for $s$-channel dijet resonances and apply the results
to the following narrow dijet resonances predicted by eleven benchmark models:
\begin{enumerate}
\item[1.] String resonances~\cite{Anchordoqui:2008di,Cullen:2000ef}, which are the Regge excitations of the
quarks and gluons in string theory. There are multiple mass-degenerate states with various spin and color multiplicities.
The $\Pq\Pg$ states dominate the cross section for all masses considered.
\item[2.] Scalar diquarks, which decay to $\cPq\cPq$ and \cPaq\hspace{1 pt}\cPaq, predicted by a grand unified theory based on the
E$_6$ gauge symmetry group~\cite{ref_diquark}. The coupling constant is conventionally assumed to be of electromagnetic strength.
\item[3.] Mass-degenerate excited quarks (\Qstar), which decay to \cPq\cPg, predicted in quark compositeness models~\cite{ref_qstar,Baur:1989kv}; the compositeness scale is set to be equal to the mass of the excited quark. We
consider production and decay of the first generation of excited quarks and antiquarks ($\PQu^*$, $\PQd^*$, $\PAQu^*$, and $\PAQd^*$) via quark-gluon fusion ($\cPq\cPg \rightarrow \Qstar \rightarrow \cPq\cPg$). We do not include production or decay via contact interactions ($\cPq\cPq \rightarrow \cPq\Qstar$)~\cite{Baur:1989kv}.
\item[4--5.] Axigluons and colorons, axial-vector and vector particles,
which are predicted in the chiral color~\cite{ref_axi} and the flavor-universal coloron~\cite{ref_coloron} models, respectively. These are massive color-octet
particles, which decay to \cPq\cPaq.
The coloron coupling parameter is set at its minimum value $\cot\theta=1$~\cite{ref_coloron}, which gives identical production cross section values for colorons and axigluons.
\item[6.] Color-octet scalars~\cite{Han:2010rf}, which decay to \cPg\cPg, appear in dynamical electroweak symmetry breaking
models such as technicolor.
The value of the squared anomalous coupling of color-octet scalars to gluons is chosen to be $k_\mathrm{s}^2=1/2$~\cite{Chivukula:2014pma}.
\item[7--8.] New gauge bosons (\PWpr and \PZpr), which decay to \cPq\cPaq, predicted by models that include new gauge
symmetries~\cite{ref_gauge}; the \PWpr and \PZpr\ bosons are assumed to have standard-model-like couplings.
\item[9.] Randall--Sundrum (RS) gravitons (G), which decay to \cPq\cPaq\ and \cPg\cPg, predicted in the RS model of extra
dimensions~\cite{ref_rsg}. The value of the dimensionless coupling $k/\overline{M}_\text{Pl}$ is chosen to be
0.1, where $k$ is the curvature scale and $\overline{M}_\text{Pl}$ is the reduced Planck mass.
\item[10.] Dark matter mediators, which decay to \cPq\cPaq\ and pairs of DM particles, are the mediators of an interaction between quarks and dark matter~\cite{Chala:2015ama,Abercrombie:2015wmb,Boveia:2016mrp,Abdallah:2015ter}.
For the DM mediator we follow the recommendations of Ref.~\cite{Boveia:2016mrp} on the model choice and coupling values, using a
simplified model~\cite{Abdallah:2015ter} of a spin-1 mediator decaying only to $\PQq\PAQq$ and pairs of DM particles,
with an unknown mass \mDM, and with a universal quark coupling $\gq = 0.25$ and a DM coupling $\gDM=1.0$.
\item[11.] Leptophobic $\PZpr$ resonances~\cite{Dobrescu:2013coa}, which decay to \cPq\cPaq\ only, with a universal quark coupling $\gq^{\prime}$ related to the
coupling of Ref.~\cite{Dobrescu:2013coa} by $\gq^{\prime}=g_B/6$.
\end{enumerate}

\section{Measurement}
\label{sec:reco}

\subsection{Detector}
A detailed description of the CMS detector and its coordinate system, including definitions of the azimuthal angle
$\phi$ (in radians) and
pseudorapidity variable $\eta$, is given in Ref.~\cite{refCMS}.
The central feature of the CMS apparatus is a superconducting
solenoid of 6\unit{m} internal diameter providing an axial field of 3.8\unit{T}.
Within the solenoid volume
are located the silicon pixel and strip tracker ($\abs{\eta}<2.4$) and the barrel and
endcap calorimeters ($\abs{\eta}<3.0$), consisting of a lead tungstate crystal electromagnetic
calorimeter, and a brass and scintillator hadron calorimeter.
An iron and quartz-fiber hadron forward calorimeter is located in the region ($3.0<\abs{\eta}<5.0$),
outside the solenoid volume.

\subsection{Reconstruction}
A particle-flow (PF) event algorithm is used to reconstruct and identify each individual particle with an optimized
combination of information from the various elements of the CMS detector~\cite{CMS-PRF-14-001}.
Particles are classified as muons, electrons, photons, and either charged or neutral hadrons.

Jets are reconstructed
either from particles identified by the PF algorithm, yielding ``PF-jets", or from energy deposits in the calorimeters,
yielding ``Calo-jets". The PF-jets, reconstructed offline, are used for the high-mass
search, while Calo-jets, reconstructed at the HLT, are used for the low-mass search.
To reconstruct either type of jet, we use the anti-\kt algorithm~\cite{Cacciari:2005hq,Cacciari:2008gp} with a distance
parameter of 0.4, as implemented in the \textsc{FastJet} package~\cite{Cacciari:2011ma}.
For the high-mass search, at least one reconstructed vertex is required.
The reconstructed vertex with the largest value of summed physics-object $\pt^2$ is taken to be the
primary $\Pp\Pp$ interaction vertex. Here the physics objects are the jets made of tracks, clustered using the jet finding
algorithm~\cite{Cacciari:2008gp,Cacciari:2011ma} with the tracks assigned to the vertex as inputs,
and the associated missing transverse momentum, taken as the negative vector sum of the \pt of those jets.
For PF-jets, charged PF candidates not originating from the primary vertex
are removed prior to the jet finding.
For both PF-jets and Calo-jets, an event-by-event correction based on the jet
area~\cite{jetarea_fastjet_pu,Khachatryan:2016kdb}
is applied to the jet energy to remove the estimated contribution from additional collisions in
the same or adjacent bunch crossings (pileup).

\subsection{Trigger and minimum dijet mass}
\label{sec:trigger}
Events are selected using a two-tier trigger system~\cite{Khachatryan:2016bia}. Events satisfying
loose jet requirements at the first level (L1) trigger are examined by the HLT. We use single-jet triggers
that require a jet in the event to satisfy a predefined \pt threshold. We also use triggers that require \HT to exceed
a predefined threshold, where \HT is the scalar sum of the \pt of all jets in the event with $\abs{\eta}<3.0$. Both PF-jets and Calo-jets are available at the HLT.

\begin{figure}[htbp]
  \centering
    \includegraphics[width=0.48\textwidth]{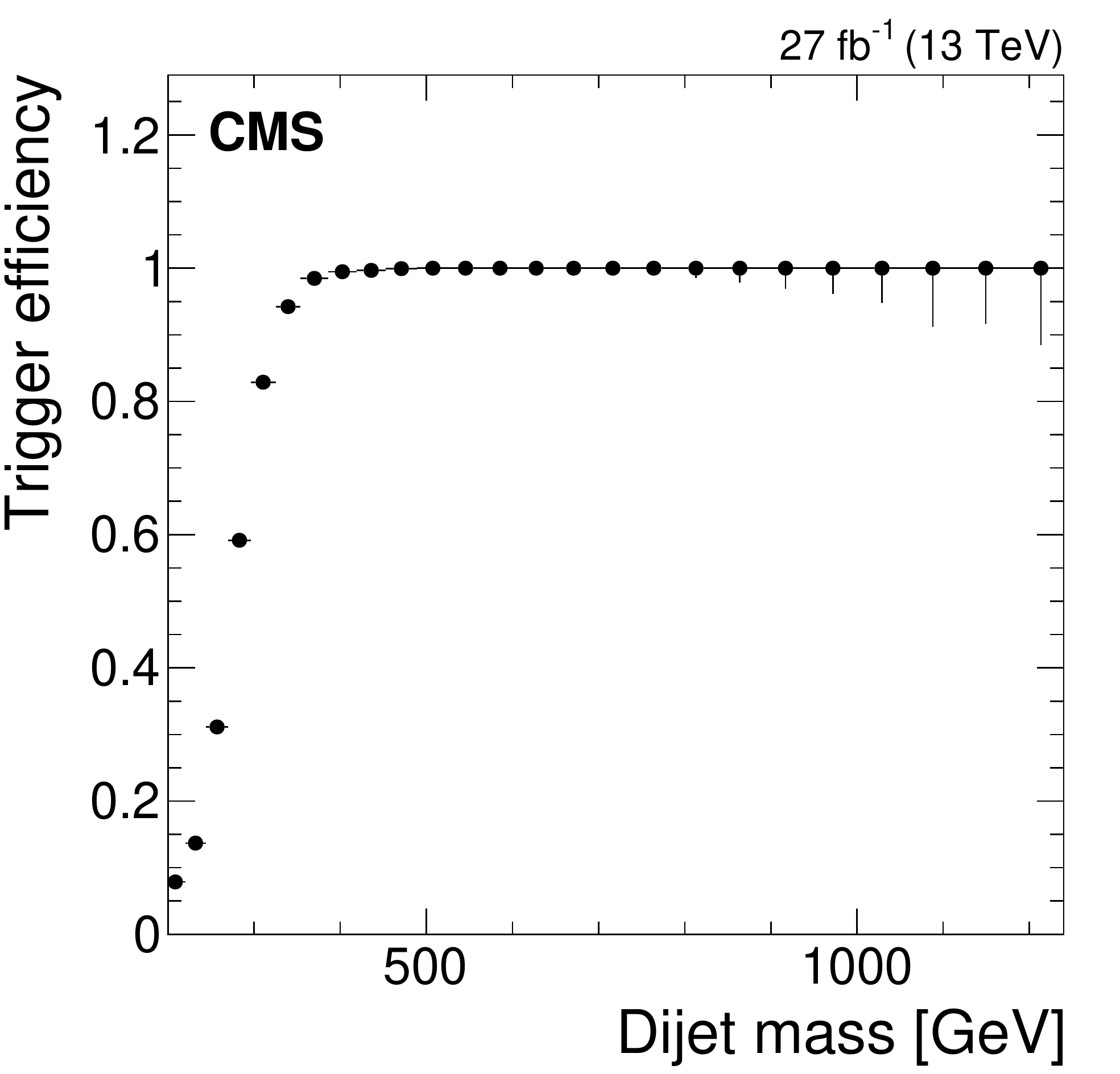}
    \includegraphics[width=0.48\textwidth]{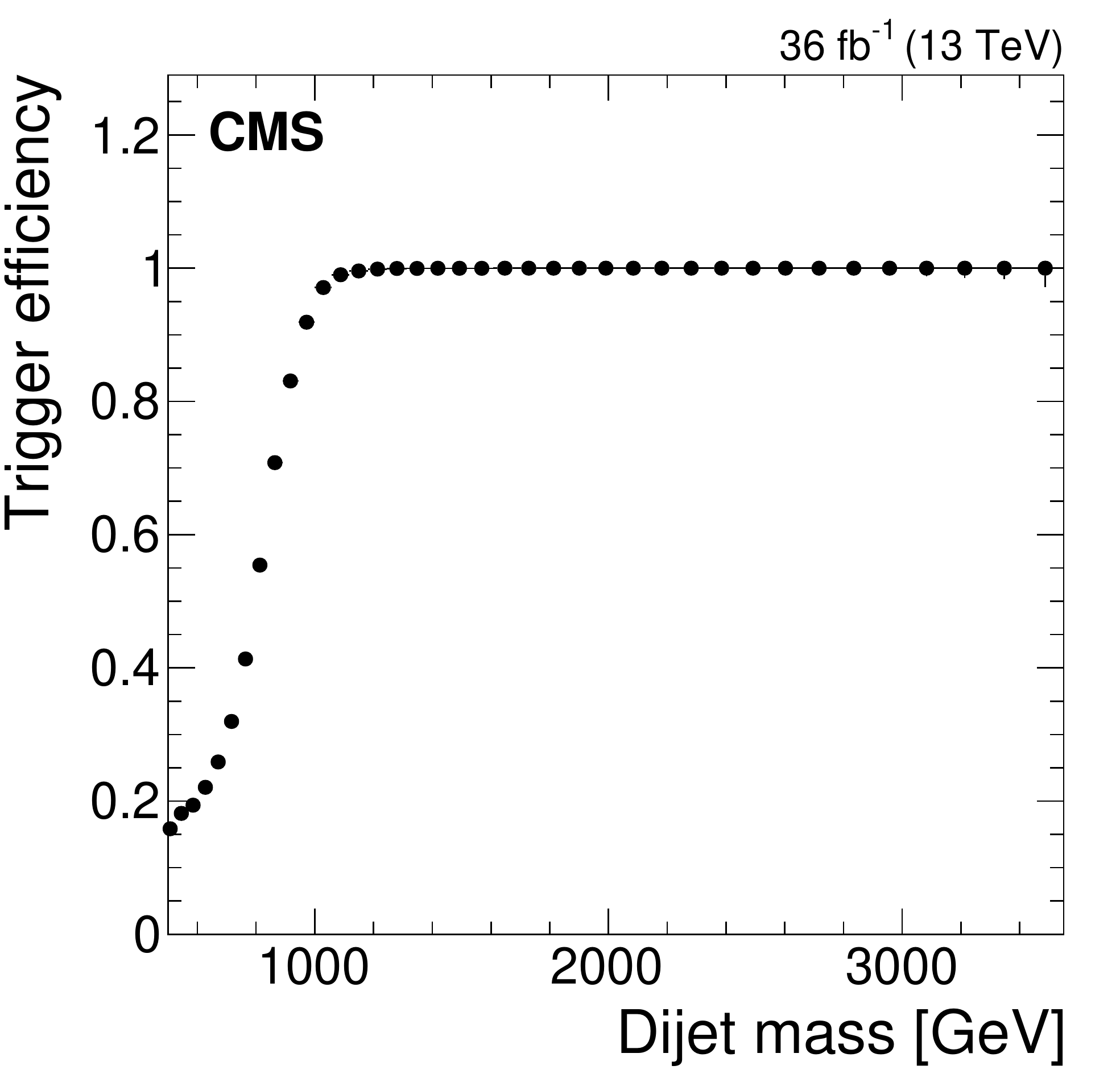}
   \caption{
The efficiency of the trigger for the low-mass search (\cmsLeft) and the high-mass search (\cmsRight) as a function
of dijet mass for wide jets, defined in Section~\ref{sec:wideJets}, after all jet calibrations and event selections discussed in Section~\ref{sec:reco}.
The horizontal lines on the data points show the variable bin sizes.}
    \label{figTriggerEff}
\end{figure}

For the high-mass search, the full event information is reconstructed if the event satisfies the HLT trigger.
In the early part of the data taking period, the HLT trigger required $\HT>800$\GeV,
with \HT calculated using PF-jets with $\pt>30$\GeV.
For the remainder of the run, an HLT requiring $\HT>900$\GeV with this same jet \pt threshold was used.
The latter \HT trigger suffered from an inefficiency. The efficiency loss occurred within the \HT trigger at L1, towards the end of the data taking period used in this analysis.
To recover the lost efficiency we used single-jet triggers at the HLT that did not
rely on the \HT trigger at L1 but instead used an efficient single-jet trigger at L1.
There were three such triggers at the HLT: the first
requiring a PF-jet with $\pt>500$\GeV, a second requiring a Calo-jet with $\pt>500$\GeV, and a
third requiring a PF-jet with an increased distance parameter of 0.8 and $\pt>450$ GeV. The trigger used for the high-mass search was
the logical OR of these five triggers. We select events with $\mjj>\RECOminMjjCut$, where the dijet mass is fully reconstructed offline
using wide jets, defined later. For this selection, the combined L1 trigger and HLT was
found to be fully efficient for the full \RunLumi sample, as shown in Fig.~\ref{figTriggerEff}. Here the absolute
trigger efficiency is measured using a sample acquired with an orthogonal trigger requiring muons with $\pt>45$ GeV at the HLT.

The data scouting technique is used for the low-mass search. When an event passes a data scouting trigger, the Calo-jets
reconstructed at the HLT are saved along with the event energy density and the missing transverse momentum
reconstructed from the calorimeter. The energy density is defined for each event as the median calorimeter energy per unit area calculated
in a grid of $\eta-\phi$ cells~\cite{Khachatryan:2016kdb} covering the calorimeter acceptance.
The shorter time required for the reconstruction of the calorimetric quantities and the reduced size of the data recorded for these events allow a reduced
\HT threshold compared to the high-mass search. For the low-mass search, Calo-jets with $\pt>40$\GeV are used to compute \HT. The trigger threshold
is $\HT>250$\GeV, and we select events with $\mjj>\CALOminMjjCut$ for which the trigger
is fully efficient, as shown in Fig.~\ref{figTriggerEff}. Here the trigger efficiency is measured using a prescaled sample acquired with a
data scouting trigger which required only that the event passed the jet trigger at L1 with $\HT>175$\GeV.
This L1 trigger is also fully efficient for $\mjj>\CALOminMjjCut$, measured using another prescaled sample acquired with an even looser trigger with effectively no requirements
(zero-bias) at L1 and requiring at least one Calo-jet with $\pt>40$\GeV at the HLT.
Unlike the high-mass search, there were no
single-jet triggers at the HLT in data scouting that would allow for the recovery of the inefficiency in the L1 trigger in 9\fbinv of data at the end of the run,
so only the first \CaloLumi of integrated luminosity was used for the low-mass search.

The trigger efficiencies for the low-mass and high-mass regions are shown as functions of dijet mass in Fig.~\ref{figTriggerEff}.  The binning choices
are the same as those adopted for the dijet mass spectra: bins of width approximately equal to the dijet mass resolution determined from simulation. All dijet mass bin edges and widths throughout this paper are the same as those used by previous
dijet resonances searches performed by the CMS collaboration~\cite{Sirunyan:2016iap,Khachatryan:2015dcf,Khachatryan:2016ecr,
Khachatryan:2015sja,Chatrchyan:2013qhXX,CMS:2012yf,Chatrchyan2011123,Khachatryan:2010jd}.
Fig.~\ref{figTriggerEff} illustrates that the searches are fully efficient for the chosen
dijet mass thresholds.
For the purpose of our search, full efficiency requires the measured trigger inefficiency in a bin to be less
than the fractional statistical uncertainty in the number of events in the same bin in the dijet mass spectrum.  For example, the measured
trigger efficiency in the bin between 1246 and 1313\GeV in Fig.~\ref{figTriggerEff}~(right) is $99.95\pm0.02$\%, giving a trigger inefficiency of 0.05\%
in that bin, which is less than the statistical uncertainty of 0.08\% arising from the 1.6 million events in that same bin of the dijet mass spectrum.
This criterion for choosing the dijet mass thresholds, $\mjj>\RECOminMjjCut$ for the high mass search and  $\mjj>\CALOminMjjCut$ for the low mass search,
ensures that the search results are not biased by the trigger inefficiency.

\subsection{Offline calibration and jet identification}
The jet momenta and energies are corrected using calibration constants
obtained from simulation, test beam results, and pp collision
data at $\sqrt{s}=13$\TeV. The methods described in
Ref.~\cite{Khachatryan:2016kdb} are applied using all \textit{in-situ}
calibrations obtained from the current data,
and fit with analytic functions so the calibrations are forced to be smooth functions of \pt.
All jets, the PF-jets in the high-mass search and Calo-jets in the
low-mass search, are required to have $\pt>30$\GeV and $\abs{\eta}<2.5$.  The two jets with largest
\pt are defined as the leading jets.
Jet identification (ID) criteria are applied to remove spurious jets associated with calorimeter noise as well as those
associated with muon and electron candidates that are either mis-reconstructed or isolated~\cite{CMS:2017wyc}.
For all PF-jets, the jet ID requires that the neutral hadron and
photon energies are less than 90\% of the total jet energy.
For PF-jets that satisfy $\abs{\eta}<2.4$, within the fiducial tracker coverage, the jet ID additionally requires that the jet has non-zero charged hadron
energy, and muon and electron energies less than 80 and 90\% of the total jet energy, respectively.
The jet ID for Calo-jets requires that the jet be detected by both the electromagnetic and hadron
calorimeters with the fraction of jet energy deposited within
the electromagnetic calorimeter between 5 and 95\% of the
total jet energy. An event is rejected if either of the two leading jets fails the jet ID criteria. These requirements are sufficient to reduce background events from detector noise and other sources to a negligible level.

\subsection{Wide jet reconstruction and event selection}
\label{sec:wideJets}
Spatially close jets are combined into ``wide jets'' and
used to determine the dijet mass, as in the previous CMS
searches~\cite{Sirunyan:2016iap,Khachatryan:2015dcf,Khachatryan:2016ecr,Chatrchyan2011123,CMS:2012yf,Chatrchyan:2013qhXX,Khachatryan:2015sja}.  The wide-jet algorithm, designed
for dijet resonance event reconstruction, reduces the analysis sensitivity to gluon radiation from the
final-state partons.  The two leading jets are used as seeds and the four-vectors of all other jets, if within $\Delta R=\sqrt{\smash[b]{(\Delta\eta)^2 +
  (\Delta\phi)^2}}<1.1$, are added to the
nearest leading jet to obtain two wide jets, which then form
the dijet system. The dijet mass is the magnitude of the momentum-energy 4-vector of the dijet system, which is the invariant mass of the two wide jets.
The wide jet algorithm thereby collects hard gluon radiation, satisfying the jet requirement $\pt>30$\GeV and
found nearby the leading two final state partons, in order to improve the dijet mass resolution.
This is preferable to only increasing the distance parameter within the anti-\kt algorithm to 1.1, which would include in the leading
jets the unwanted soft energy from pile-up and initial state radiation.
The wide jet algorithm is similar to first increasing the distance parameter and then
applying jet trimming~\cite{Krohn:2009th} to remove unwanted soft energy.

The angular distribution of background from $t$-channel dijet events is similar to that for Rutherford scattering, approximately proportional to
$1/[1-\tanh(\detajj/2)]^2$, which peaks at large values of $\detajj$.
This background is suppressed by requiring the pseudorapidity separation of the two wide
jets to satisfy $\detajj<1.3$.
This requirement also makes the trigger efficiency in Fig.~\ref{figTriggerEff} turn on quickly, reaching a plateau at 100\% for
relatively low values of dijet mass. This is because the jet \pt threshold of the trigger at a fixed dijet mass is more easily satisfied at
low $\detajj$, as seen by the approximate relation $\mjj\approx 2\pt\cosh(\detajj/2)$.

The above requirements maximize the search sensitivity for
isotropic decays of dijet resonances in the presence of dijet background from quantum chromodynamics (QCD).

\subsection{Calibration of wide jets in the low-mass search}
\begin{figure}[hbt]
\begin{center}
\includegraphics[width=0.5\textwidth]{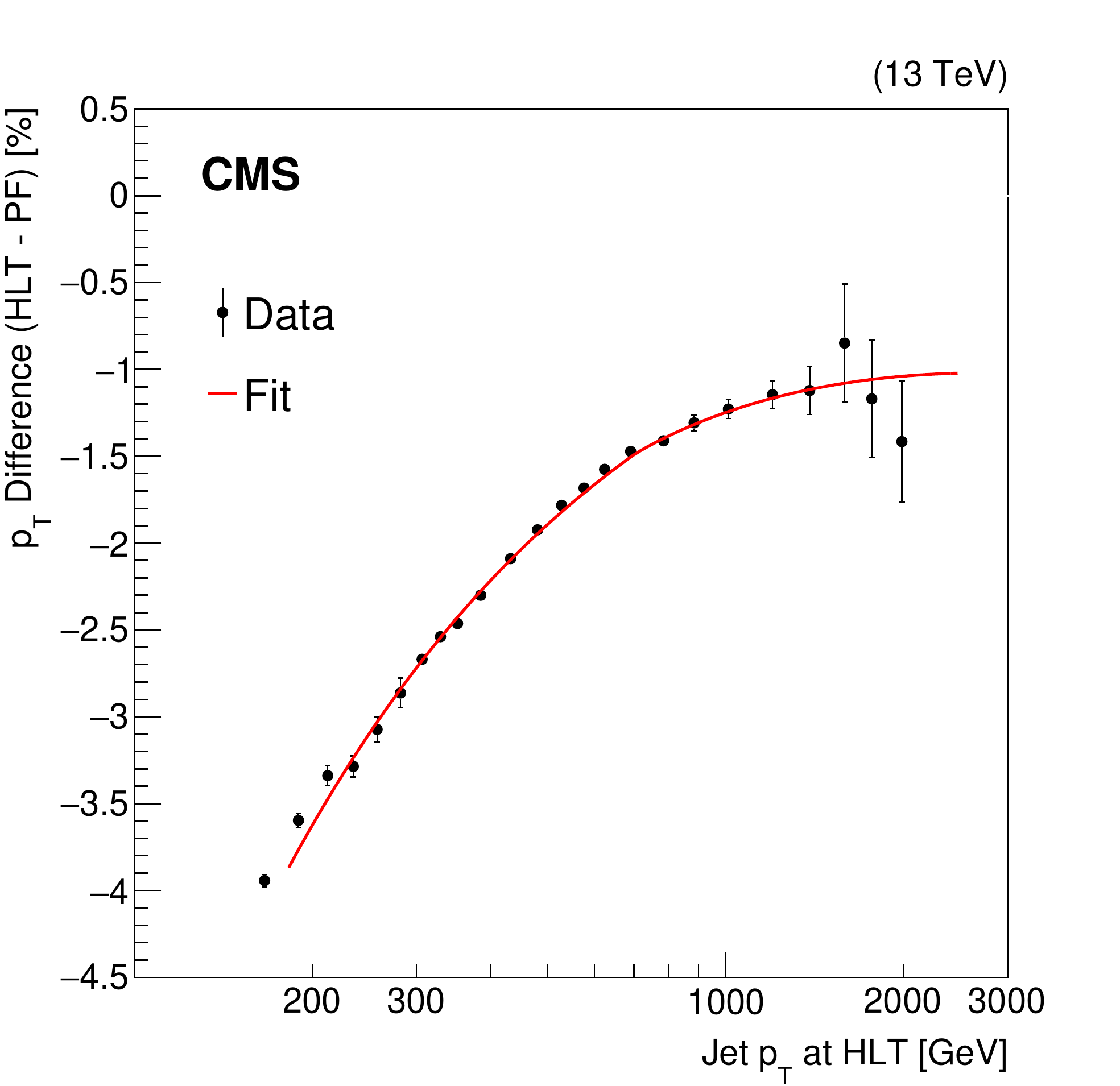}
\caption{The calibration of jets in the low-mass analysis. The percent difference in data (points), between the \pt of the wide jets reconstructed from
Calo-jets at the HLT and the wide jets reconstructed from PF-jets, is fit to a smooth parameterization (curve), as a function of the HLT \pt.}
\label{fig:FinalJEC}
\end{center}
\end{figure}

The jet energy scale of the low-mass search has been calibrated to be the same as the jet energy scale of the high-mass search.
For the low-mass search, after wide jet reconstruction and event selection,
we calibrate the wide jets reconstructed from Calo-jets at the HLT to have the same
average response as the wide jets reconstructed from PF-jets.  We use a smaller \textit{monitoring} data set,
which includes both Calo-jets at the HLT and the fully reconstructed PF-jets, to measure the \pt difference
between the two types of wide jets, as shown in Fig.~\ref{fig:FinalJEC}.
A dijet balance ``tag-and-probe" method similar to that
discussed in Ref.~\cite{Khachatryan:2016kdb} is used.
One of the two jets in the dijet system is designated as the tag jet, and the other is designated as the probe jet,
and the \pt difference between Calo-Jets at the HLT and fully reconstructed PF-jets is measured for the probe jet as a
function of the \pt of the tag PF-jet.
We avoid jet selection bias of the probe Calo-Jet \pt, which would result from
resolution effects on the steeply falling \pt spectrum, by measuring the \pt difference as a function of
the \pt of the tag PF-jet instead of the \pt of the probe Calo-jet at the HLT.
This calibration is then translated into a function of
the average \pt of the probe Calo-jets measured within each bin of \pt of the tag PF-jets. Figure~\ref{fig:FinalJEC} shows this measurement of
the \pt difference, as a function of jet \pt, from the monitoring data set. The measured points are fit with a parameterization
and the resulting smooth curve is used to calibrate the wide jets in the low-mass search.

\subsection{Dijet data and QCD background predictions}
\begin{figure}[htbp]
  \centering
    \includegraphics[width=0.48\textwidth]{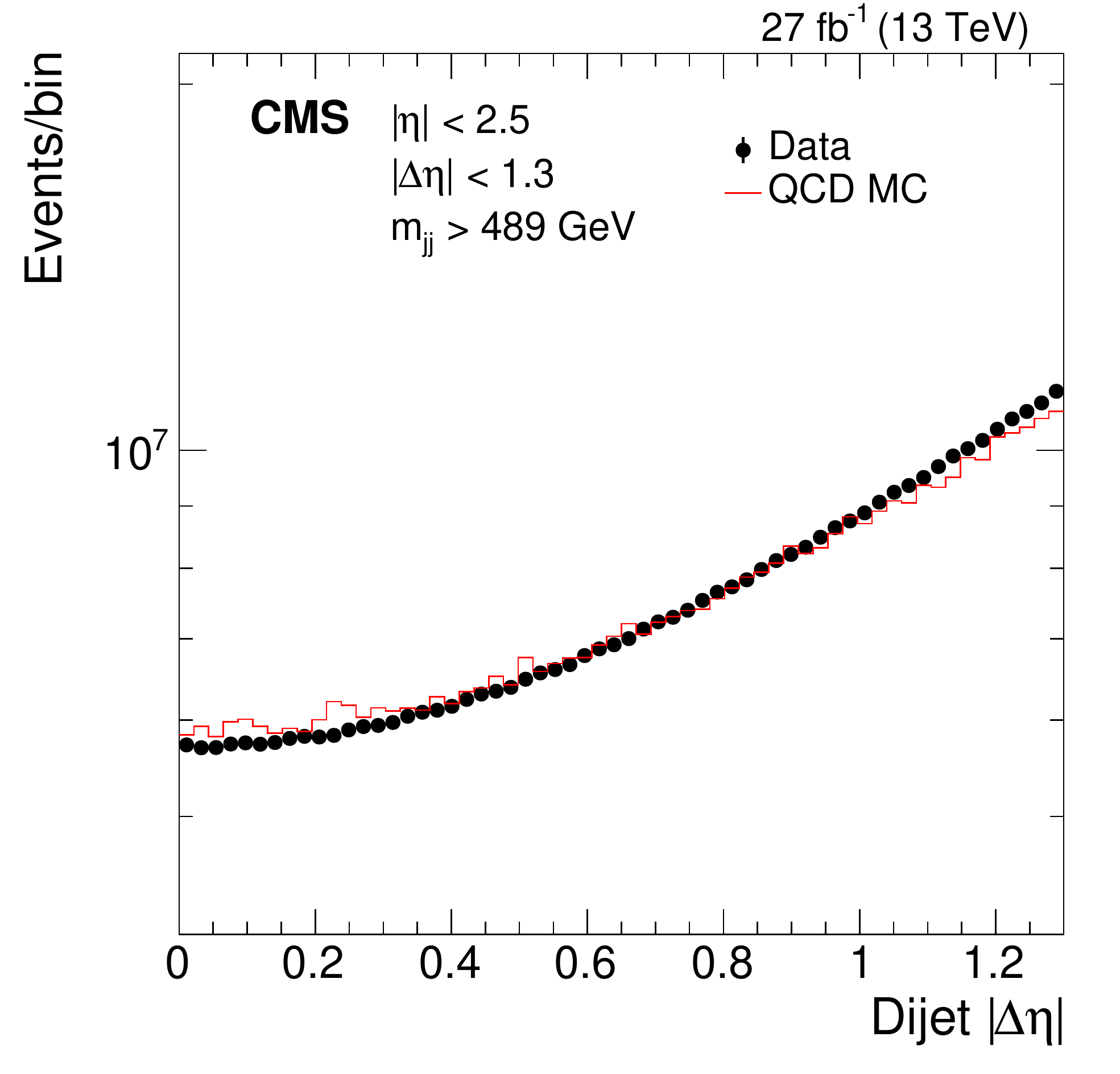}
    \includegraphics[width=0.48\textwidth]{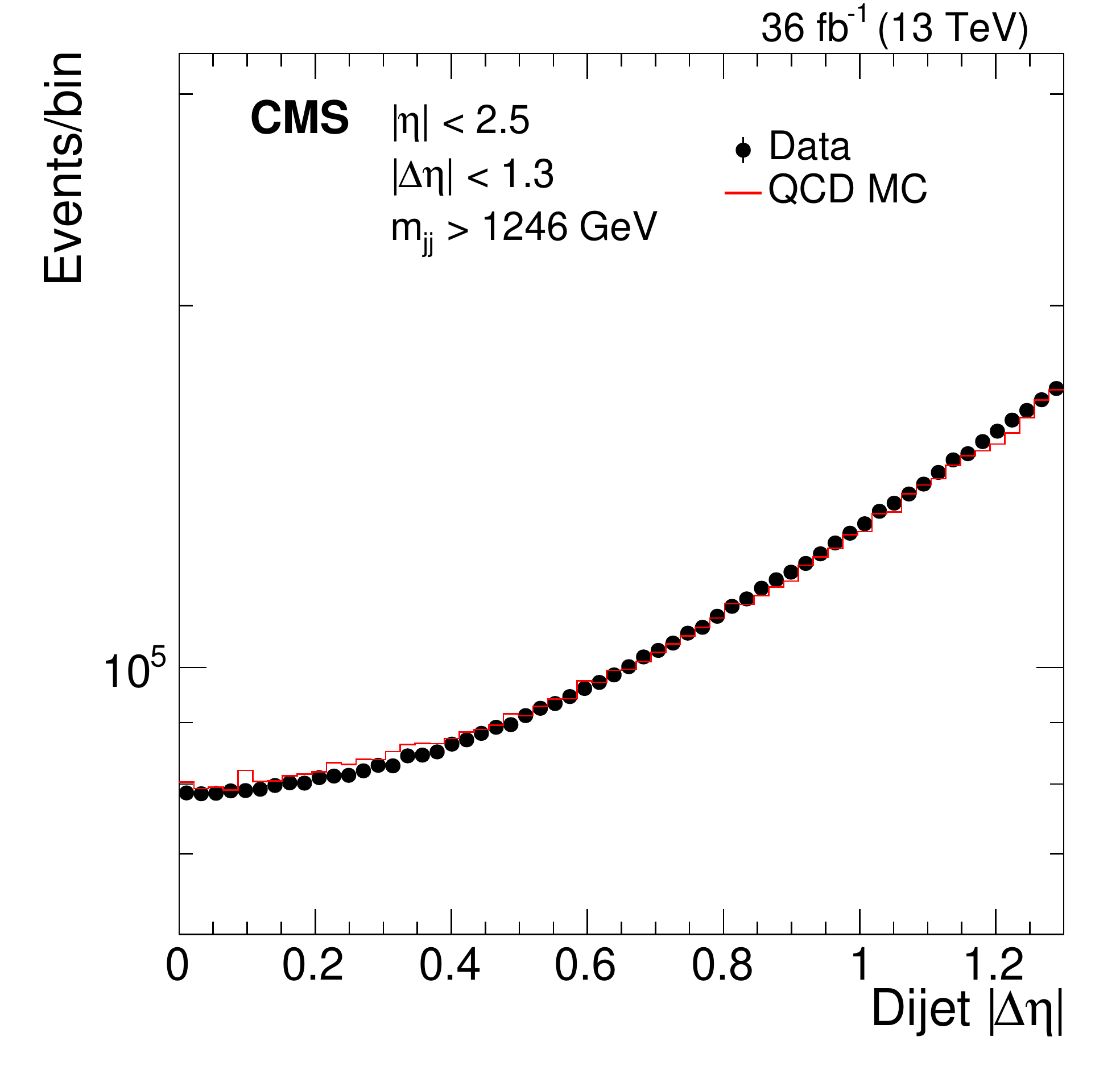}
   \caption{
The pseudorapidity separation between the two wide jets
from the low-mass search (\cmsLeft) and the high-mass search (\cmsRight). Data
(points) are compared to QCD predictions from the \PYTHIA8 MC including detector simulation (histogram) normalized to the data.}
    \label{figDeltaEta}
\end{figure}

\begin{figure}[htbp]
  \centering
    \includegraphics[width=0.48\textwidth]{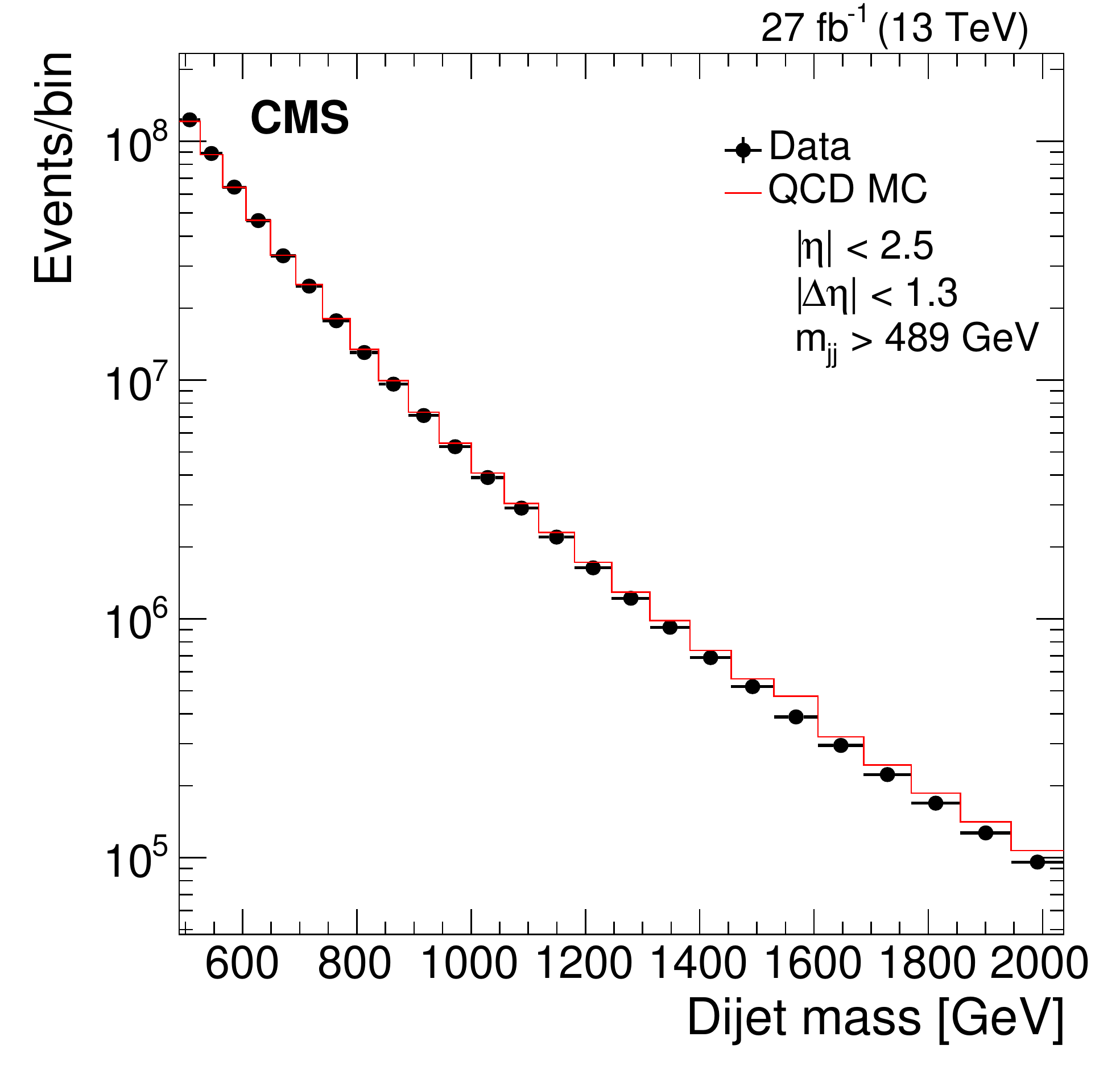}
    \includegraphics[width=0.48\textwidth]{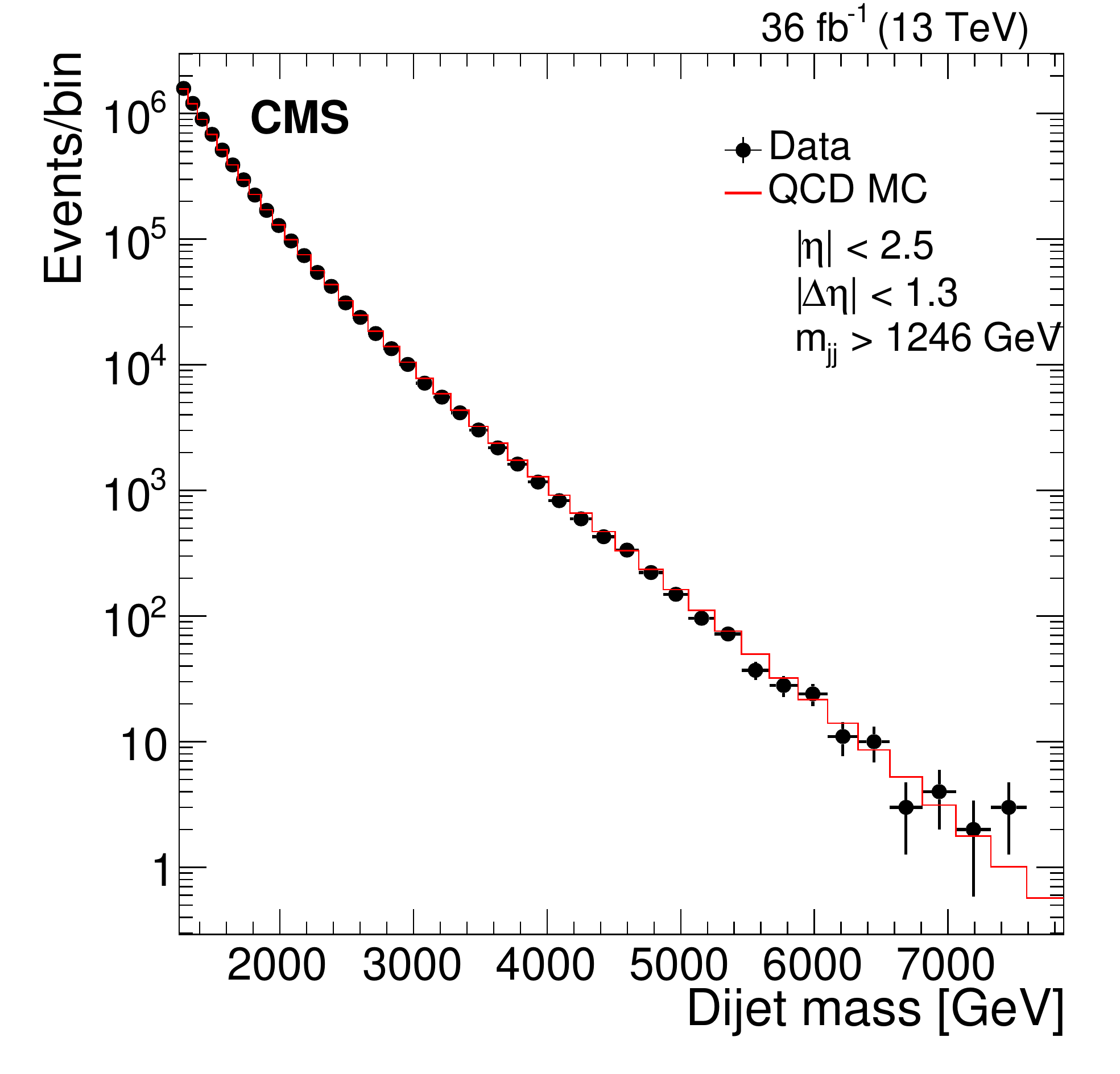}
   \caption{
The dijet mass of the two wide jets from the low-mass search (\cmsLeft) and the high-mass search (\cmsRight). Data
(points) are compared to QCD predictions from the \PYTHIA8 MC including detector simulation (histogram) normalized to the data.
The horizontal lines on the data points show the variable bin sizes.}
    \label{figDijetMass}
\end{figure}

\begin{figure}[htbp]
  \centering
    \includegraphics[width=0.5\textwidth]{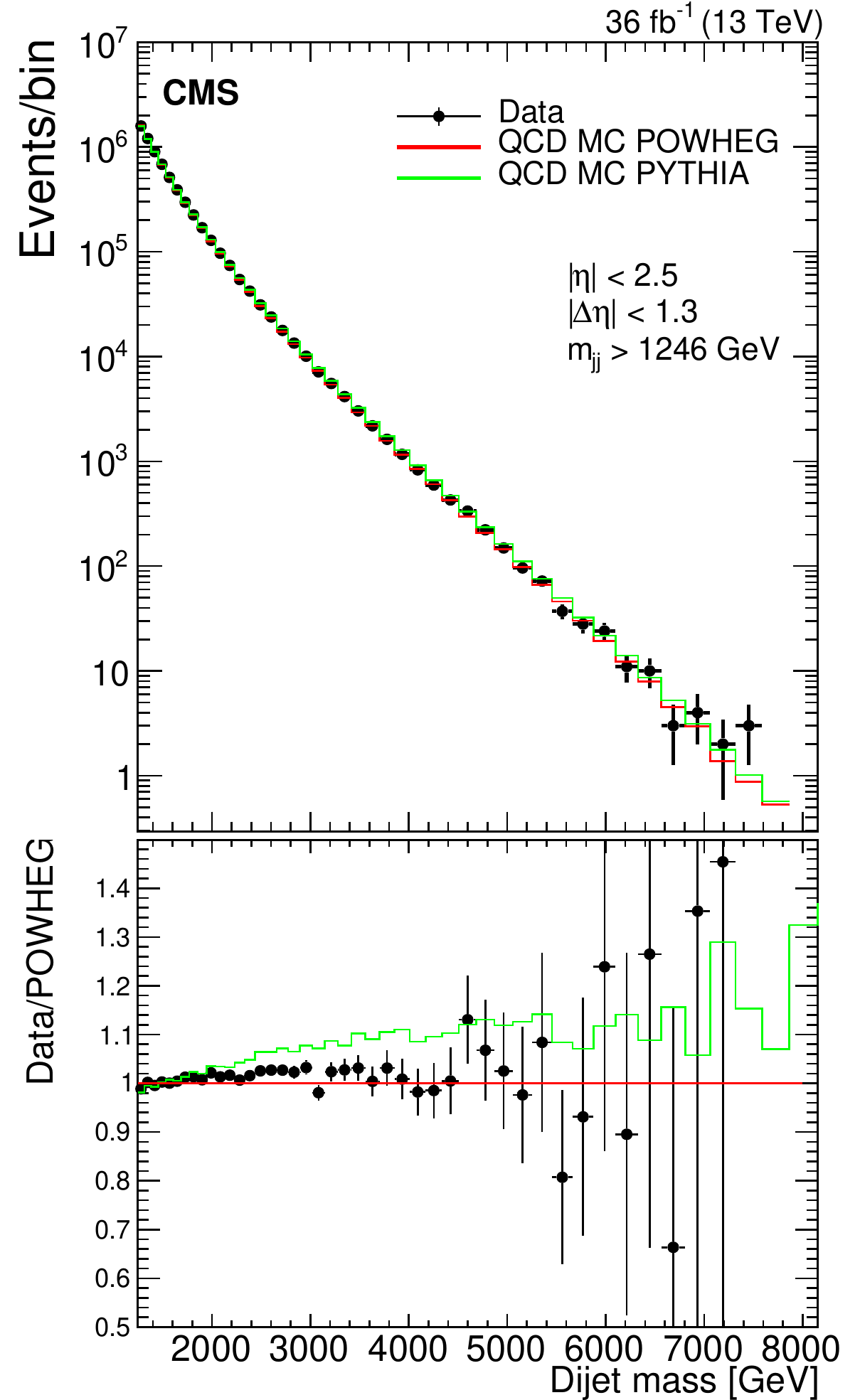}
   \caption{
The dijet mass distribution of the two wide jets from the high-mass search. (Upper) Data (points) are compared to
predictions from the \POWHEG MC  in red (darker) and the \PYTHIA8 MC in green
(lighter), including detector simulation, each normalized
to the data.  (Lower) The ratio of data to the \POWHEG prediction, compared to
unity and compared to the ratio of the \PYTHIA8 MC to the \POWHEG prediction.
The horizontal lines on the data points show the variable bin sizes.}
    \label{figDijetMassPowheg}
\end{figure}

As the dominant background for this analysis is expected to be the QCD production of two or more jets,
we begin by performing comparisons of the data to QCD background predictions for the dijet events.
The predictions are based upon a sample of 56 million Monte Carlo events produced with the \PYTHIA8.205~\cite{Sjostrand:2007gs}
program with the CUETP8M1 tune~\cite{Khachatryan:2015pea,Skands:2014pea} and including
a \GEANTfour-based \cite{refGEANT}
simulation of the CMS detector. The QCD background predictions are normalized to the data by multiplying
them by a factor of 0.87 for the high-mass search and by a factor of
0.96 for the low-mass search, so that for each search the prediction for the total number of events agrees with the number observed.
\begin{figure}[htbp]
  \centering

    \includegraphics[width=0.48\textwidth]{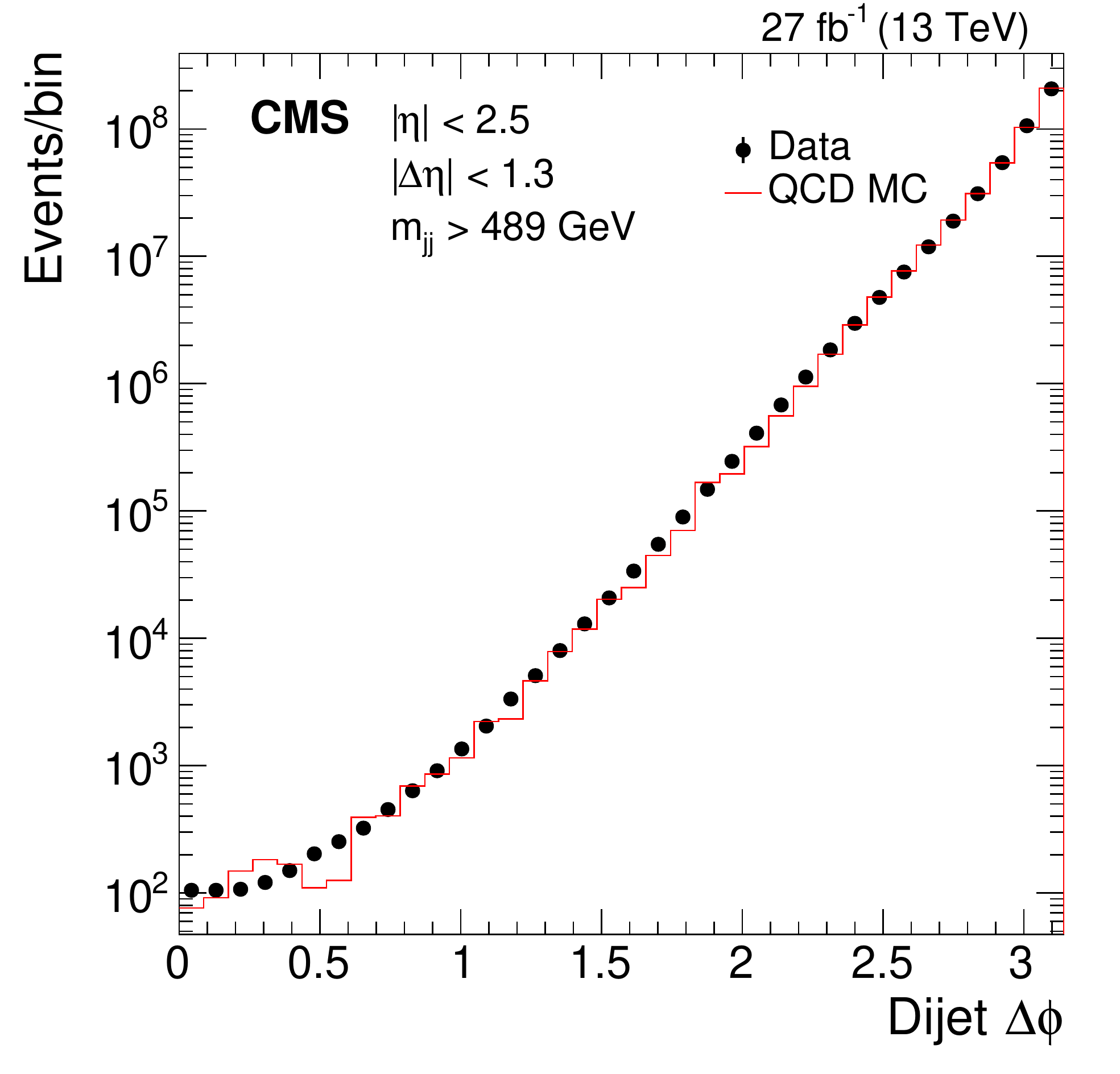}
    \includegraphics[width=0.48\textwidth]{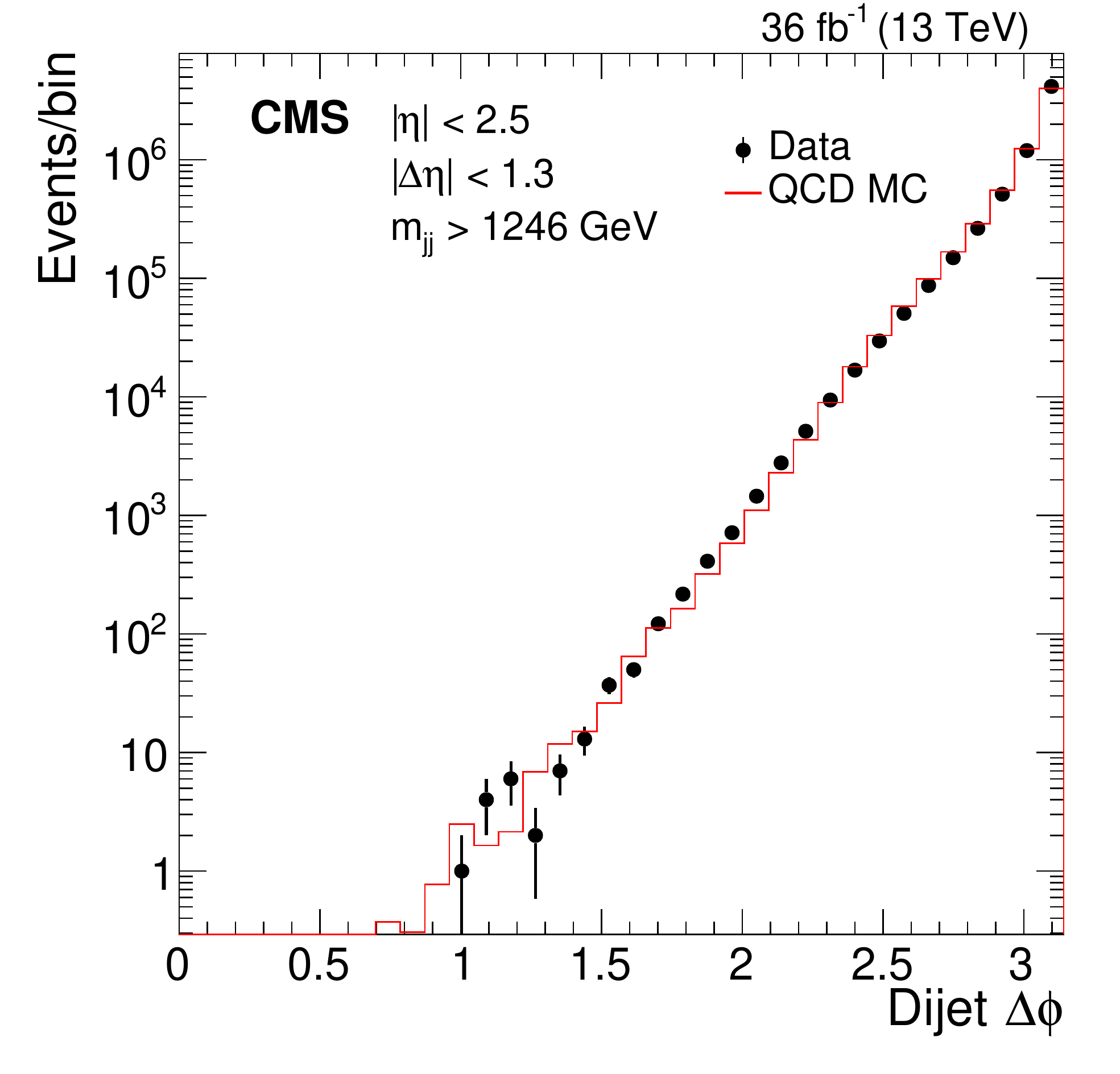}
   \caption{
The azimuthal angular separation between the two wide jets (in radians)
from the low-mass search (\cmsLeft) and the high-mass search (\cmsRight). Data
(points) are compared to QCD predictions from the \PYTHIA8 MC including detector simulation (histogram) normalized to the data.}
    \label{figDeltaPhi}
\end{figure}
In Fig.~\ref{figDeltaPhi}, we observe that the measured azimuthal
separation of the two wide jets, $\Delta\phi$, displays the "back-to-back" distribution expected from QCD dijet production.
The strong peak at $\Delta\phi=\pi$,
with very few events in the region $\Delta\phi\sim 0$, shows that the data sample is dominated by genuine parton-parton
scattering, with negligible backgrounds from detector noise or other nonphysical sources that would produce events more isotropic
in $\Delta\phi$.
In Fig.~\ref{figDeltaEta}, we observe that dijet $\detajj$ has a distribution dominated by the $t$-channel parton exchange
as does the QCD production of two jets. Note that the production rate increases with increasing $\detajj$,
whereas $s$-channel signals from most models of dijet resonances would decrease with increasing $\detajj$.
In Fig.~\ref{figDijetMass}, we observe
that the number of dijets produced falls steeply and smoothly as a function of dijet mass.
The observed dijet mass distributions are very similar to the QCD prediction from \PYTHIA, which includes
a leading order QCD calculation and parton shower effects.
In Fig.~\ref{figDijetMassPowheg}, we also compare the dijet
mass data to a next-to-leading order (NLO) QCD prediction from \POWHEG 2.0~\cite{Alioli:2010xd} normalized to the data. For this
prediction, we used 10 million dijet events from an NLO calculation of two jet production~\cite{Alioli:2010xa} using NNPDF3.0 NLO parton distribution
functions~\cite{Ball:2014uwa}, interfaced with the aforementioned \PYTHIA8 parton shower and simulation of the CMS
detector. The \POWHEG prediction models the data better than the \PYTHIA prediction does. It is clear from these comparisons that
the dijet mass data behave approximately as expected from QCD predictions.
However, the intrinsic uncertainties associated with QCD calculations make them unreliable estimators of the backgrounds in dijet resonance searches.
Instead we will use the dijet data to estimate the background.
\clearpage

\section{Search for narrow dijet resonances}
\subsection{Dijet mass spectra and background parameterizations}

\begin{figure}[htbp]
  \centering
    \includegraphics[width=0.48\textwidth]{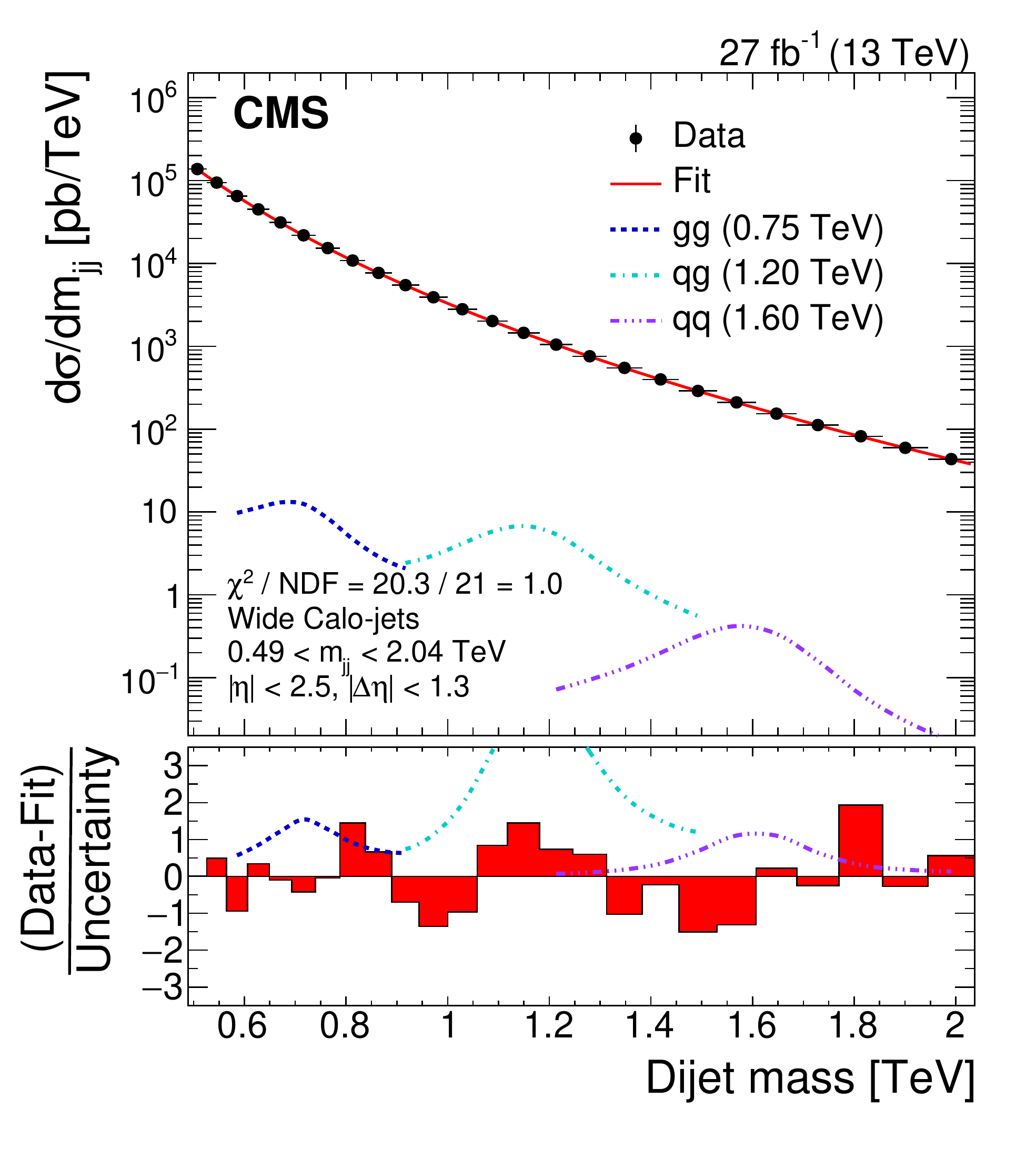}
    \includegraphics[width=0.48\textwidth]{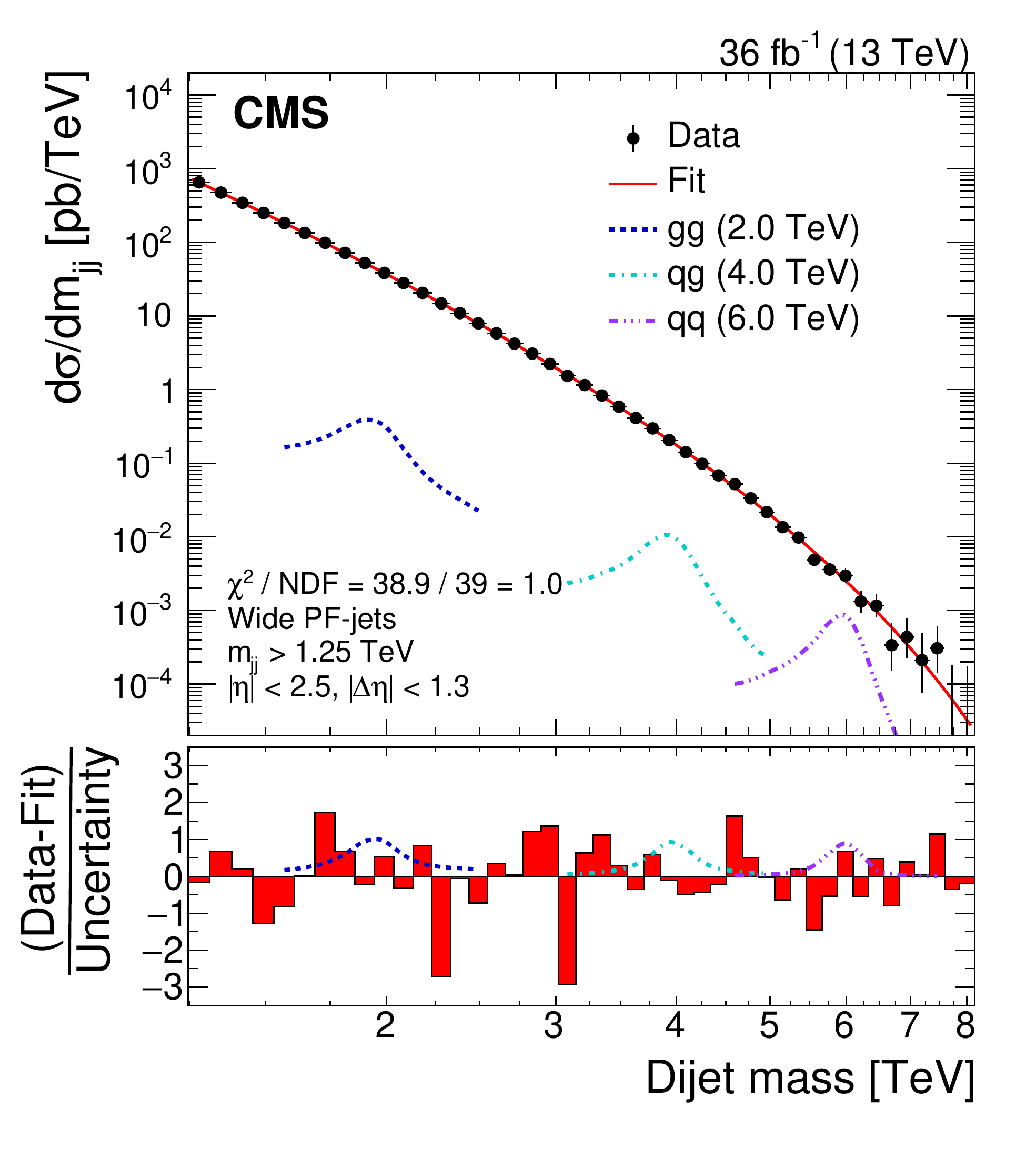}
   \caption{
    Dijet mass spectra (points) compared to a fitted
  parameterization of the background (solid curve) for the low-mass search (\cmsLeft) and
  the high-mass search (\cmsRight). The horizontal lines on the data points show the variable bin sizes.
  The lower panel in each plot shows the difference between the data and the
  fitted parametrization, divided by the statistical uncertainty of the data. Examples of predicted
  signals from narrow gluon-gluon, quark-gluon, and quark-quark resonances are shown with cross sections
  equal to the observed upper limits at 95\% \CL.}
    \label{figDataAndFit}
\end{figure}

Figure~\ref{figDataAndFit} shows
the dijet mass spectra, defined as the observed number of events in each bin divided by the
integrated luminosity and the bin width.
The dijet mass spectrum for the high-mass search is fit with the parameterization
\begin{equation}
\frac{{\rd}\sigma}{{\rd}\mjj} =
\frac{P_{0} (1 - x)^{P_{1}}}{x^{P_{2} + P_{3} \ln{(x)}}},
\label{eqBackgroundParam}
\end{equation}
where $x=\mjj/\sqrt{s}$; and $P_0$, $P_1$, $P_2$, and $P_3$ are four free fit parameters. The chi-squared per number of degrees of freedom of the fit is $\chi^2/\mathrm{NDF}=38.9/39$.
The functional form in Eq.~(\ref{eqBackgroundParam}) was also used in previous
searches \cite{Sirunyan:2016iap,Khachatryan:2016ecr,Khachatryan:2015dcf,Khachatryan:2010jd,Chatrchyan2011123,CMS:2012yf,Chatrchyan:2013qhXX,Khachatryan:2015sja,
ATLAS:2015nsi,ATLAS2010,Aad:2011aj,Aad201237,ATLAS:2012pu,Aad:2014aqa,refCDFrun2} to describe the data.
For the low-mass search
we used the following parameterization, which includes one additional parameter $P_4$, to fit the dijet mass spectrum:
\begin{equation}
\frac{{\rd}\sigma}{{\rd}\mjj} =
\frac{P_{0} (1 - x)^{P_{1}}}{x^{P_{2} + P_{3} \ln{(x)} + P_{4} \ln^2{(x)}}}.
\label{eqBackgroundParam5}
\end{equation}
Equation~(\ref{eqBackgroundParam5}) with five parameters gives $\chi^2/\mathrm{NDF}=20.3/20$ when fit to the low-mass data,
which is better than the  $\chi^2/\mathrm{NDF}=27.9/21$ obtained using the four parameter functional form in Eq.~(\ref{eqBackgroundParam}).
An F-test with a size $\alpha=0.05$~\cite{FisherTest} was used to confirm that no additional parameters
are needed to model these distributions, i.e. in the low-mass search including an additional term $P_5\ln^3{(x)}$
in Eq.(~\ref{eqBackgroundParam5}) gave $\chi^2/\mathrm{NDF}=20.1/19$, which corresponds to a smaller $p$-value than the fit with five parameters,
and this six parameter functional form was found to be unnecessary by the Fisher F-test.
The historical development of this family of parameterizations is discussed in Ref.~\cite{Harris:2011bh}. The functional forms of Eqs.~(\ref{eqBackgroundParam})
and (\ref{eqBackgroundParam5}) are motivated by QCD calculations, where the term in the numerator behaves like the parton distribution
functions at an average fractional momentum $x$ of the two partons, and the term in the denominator gives a mass dependence similar to the QCD matrix elements.
In Fig.~\ref{figDataAndFit}, we show the result of the binned maximum likelihood fits, performed independently for the low-mass and high-mass searches.
The dijet mass spectra are well modeled by the background fits.
The lower panels of Fig.~\ref{figDataAndFit} show the pulls of the fit, which
are the bin-by-bin differences between the data and the
background fit divided by the statistical uncertainty of the data. In the overlap region of the dijet mass between 1.2 and 2.0\TeV,
the pulls of the fit are not identical in the two searches because the fluctuations in reconstructed dijet mass for Calo-jets
and PF-jets are not fully correlated.

\subsection{Signal shapes, injection tests, and significance}
Examples of dijet mass distributions
for narrow resonances generated with the \PYTHIA8.205 program with the
CUETP8M1 tune and including a \GEANTfour-based simulation of the CMS detector
are shown in Fig.~\ref{figDataAndFit}. The quark-quark ($\PQq\PQq$) resonances are modeled by $\PQq\PAQq\to \PXXG \to \PQq\PAQq$,
the quark-gluon ($\PQq\Pg$) resonances are modeled by $\PQq\Pg\to \Qstar \to \PQq\Pg$,
and the gluon-gluon ($\Pg\Pg$) resonances are modeled by $\Pg\Pg\to \PXXG \to \Pg\Pg$.
The signal distributions shown in Fig.~\ref{figDataAndFit} are for $\PQq\PQq$, $\PQq\Pg$, and $\Pg\Pg$
resonances with signal cross sections corresponding to the limits at 95\% confidence level (\CL) obtained by this analysis, as described below.

\begin{figure}[htbp]
  \centering
    \includegraphics[width=0.48\textwidth]{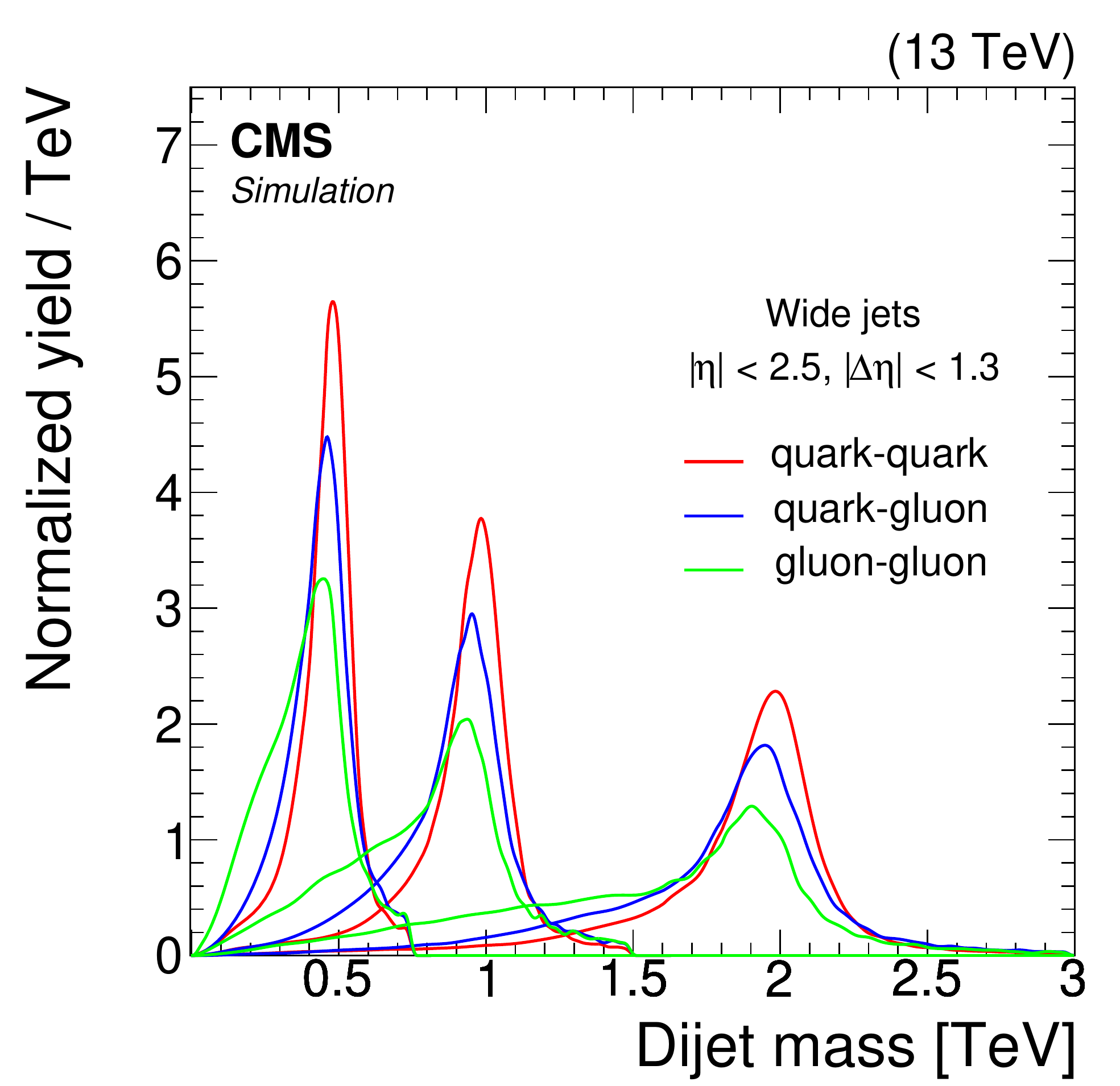}
    \includegraphics[width=0.48\textwidth]{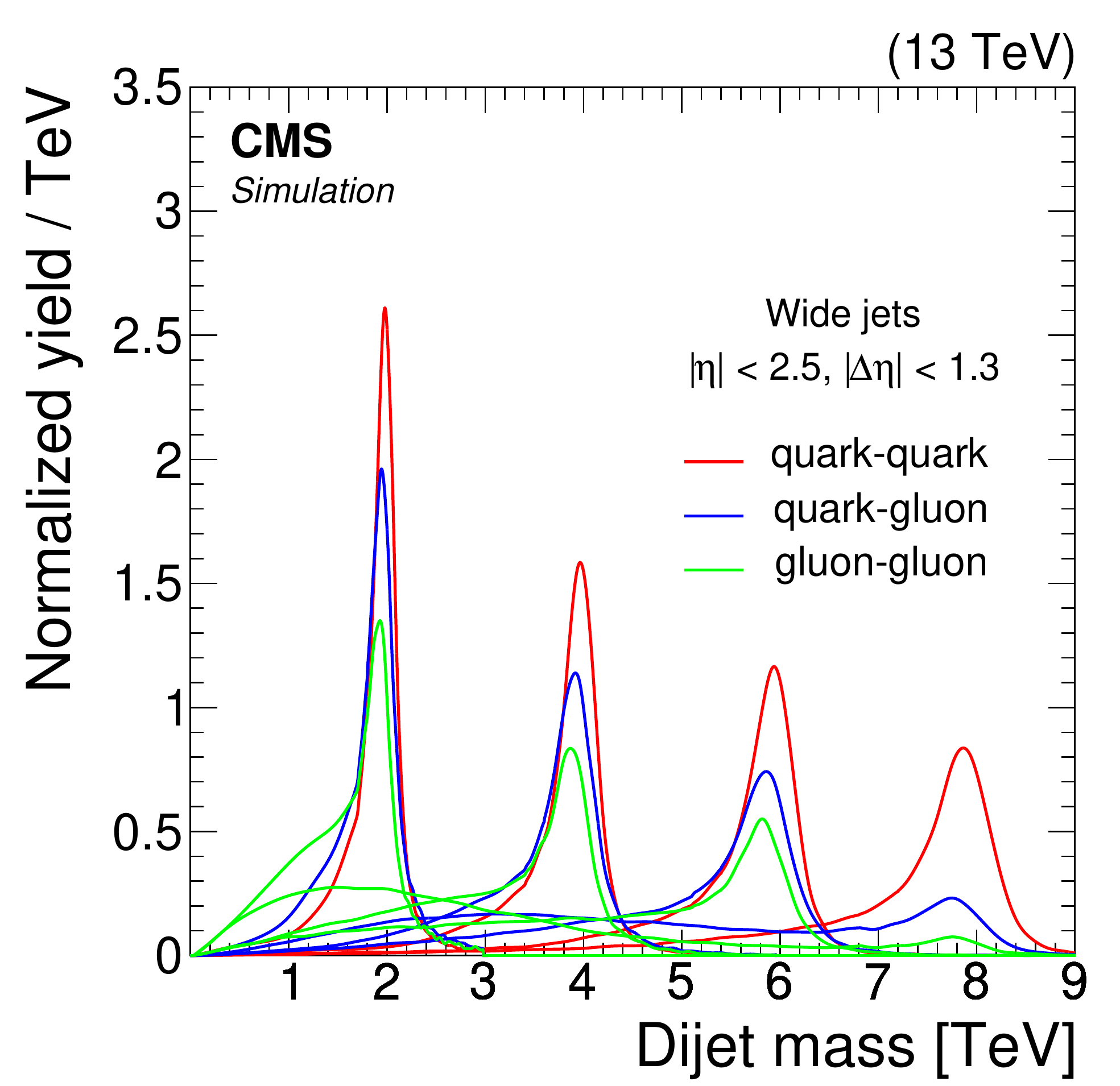}
   \caption{ Signal shapes of narrow resonances with masses of 0.5, 1, and 2\TeV in the low-mass search (\cmsLeft) and masses of 2, 4, 6, and
   8\TeV in the high-mass search (\cmsRight). These reconstructed dijet mass spectra show
wide jets from the {\PYTHIA8} MC event generator including simulation of the CMS detector. }
    \label{figSignalShapes}
\end{figure}

A more detailed view of the narrow-resonance signal shapes is provided in Fig.~\ref{figSignalShapes}.
The predicted mass distributions have Gaussian cores from jet energy resolution,
and tails towards lower mass values primarily from QCD radiation.  The observed width depends on the parton content of the
resonance ($\PQq\PQq$, $\PQq\Pg$, or $\Pg\Pg$).
The dijet mass resolution within the Gaussian core of gluon-gluon (quark-quark) resonances in Fig.~\ref{figSignalShapes} varies from 15\,(11)\% at a resonance mass of 0.5\TeV to 7.5\,(6.3)\% at 2\TeV
for wide jets reconstructed using Calo-Jets, and varies from 6.2\,(5.2)\% at 2\TeV to 4.8\,(4.0)\% at 8\TeV for wide jets reconstructed
using PF-Jets.
This total observed resolution for the parton-parton resonance includes theoretical contributions, arising from the parton shower and other sources, in addition to purely experimental contributions arising from uncertainties in measurements of the particles forming the jets.
The contribution of the low mass tail to the line shape also depends on the
parton content of the resonance.  Resonances
decaying to gluons, which emit more QCD radiation than
quarks, are broader and have a more pronounced tail. For the high-mass resonances, there is also a significant contribution that depends both on the
parton distribution functions and on the natural width of the Breit--Wigner distribution.  The low-mass component of
the Breit--Wigner distribution of the resonance is amplified by the rise of the parton distribution function at
low fractional momentum, as discussed in Section~7.3 of Ref.~\cite{Sjostrand:2006za}. These effects
cause a large tail at low mass values. Interference between the signal and the background processes is model
dependent and not considered in this analysis. In some cases interference can modify the effective signal shape
appreciably~\cite{Martin:2016bgw}. The signal shapes in the quark-quark channel come from
quark-antiquark (\Pq\Paq) resonances, which likely has a longer tail caused by parton distribution effects than that for diquark (\Pq\Pq) resonances,
tending to make the quoted limits in the quark-quark channel conservative when applied to diquark signals.

Signal injection tests were performed to investigate the potential bias introduced through the choice of
background parameterization. Two alternative parameterizations were found that model the dijet mass
data using different functional forms:
\begin{equation}
\frac{{\rd}\sigma}{{\rd}\mjj} = P_0\exp(P_1x^{P_2} + P_3(1-x)^{P_4})
\label{eqAltParam1}
\end{equation}
and
\begin{equation}
\frac{{\rd}\sigma}{{\rd}\mjj} = \frac{P_0}{x^{P_1}}\exp(-P_2x - P_3x^2 -P_4x^3).
\label{eqAltParam2}
\end{equation}
Pseudo-data were generated, assuming a signal and these alternative
parameterizations of the background, and then were fit with the
nominal parameterization given in Eq.~(\ref{eqBackgroundParam5}). The bias in the extracted signal was found to be negligible.

\begin{figure}[!htb] \centering
\includegraphics[width=0.48\textwidth]{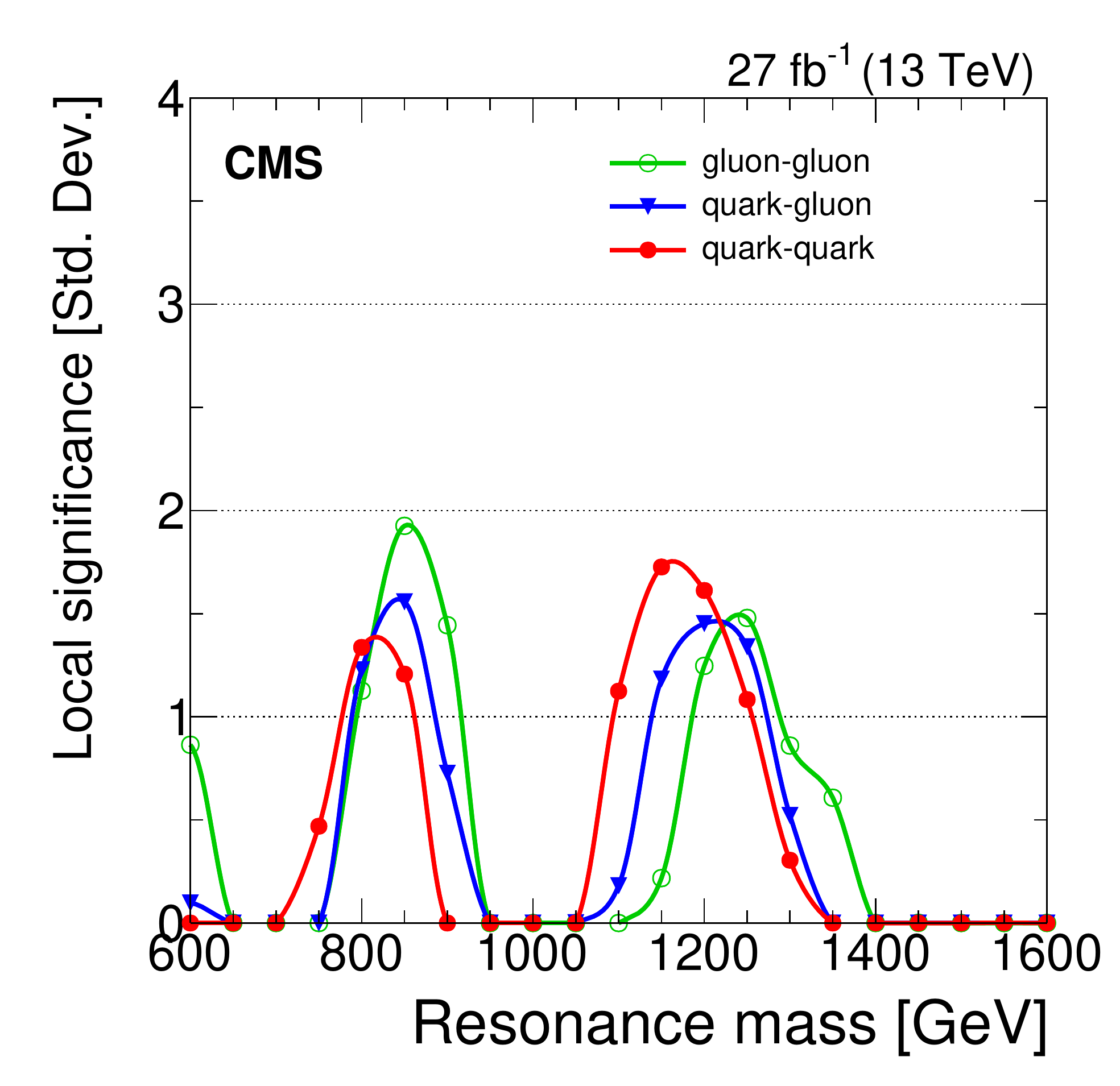}
\includegraphics[width=0.48\textwidth]{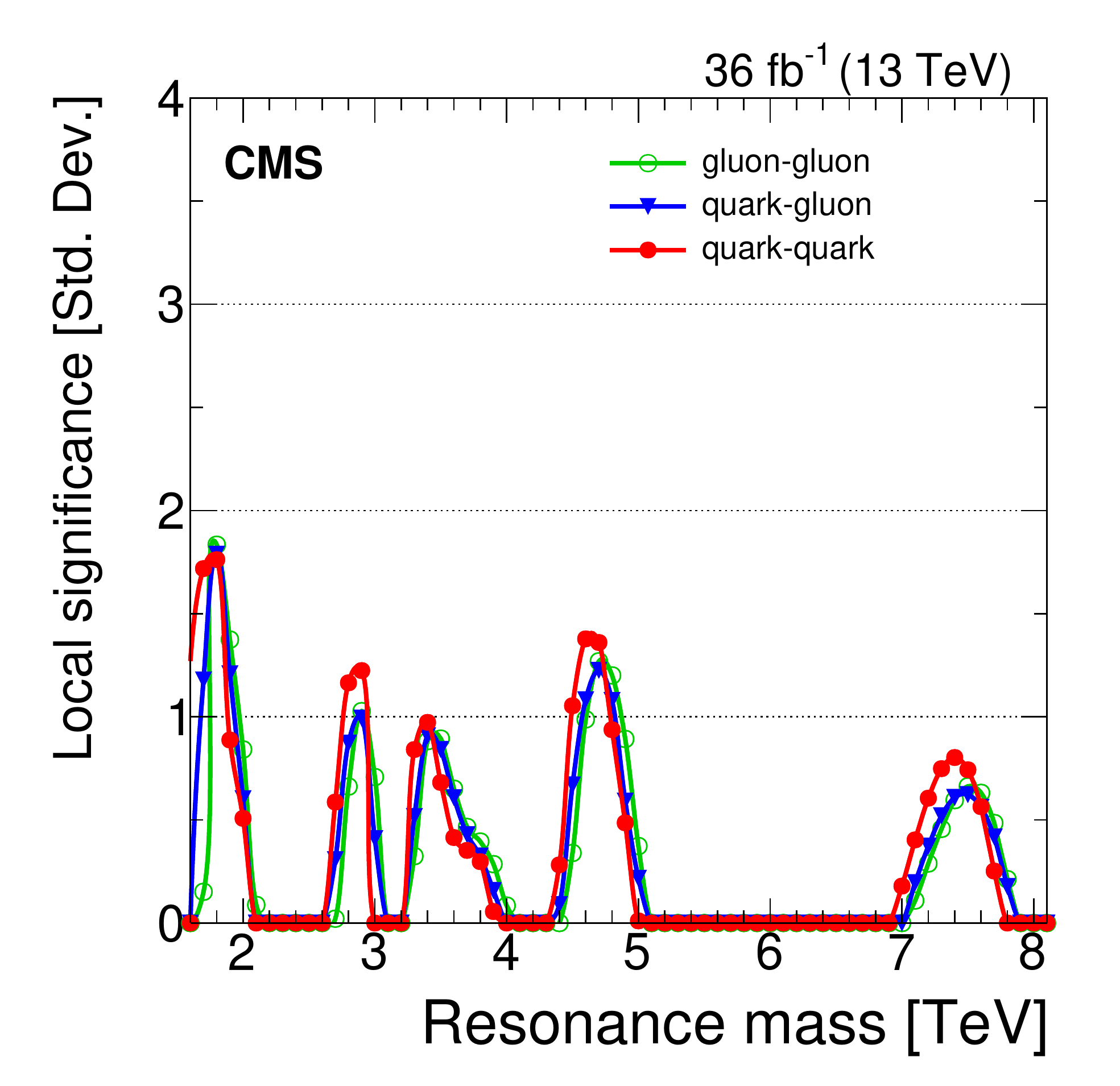}
\caption{Local significance for a narrow resonance from
the low-mass search (\cmsLeft) and the high-mass search (\cmsRight).}
\label{fig:pfsignif}
\end{figure}

There is no evidence for a narrow resonance in the data. The p-values of the background fits are $0.47$ for the high-mass search and $0.44$ for the low-mass search, indicating that the background hypothesis is an adequate description of the data. Using the statistical methodology discussed in Section~\ref{sec:statistics},
the local significance for $\PQq\PQq$, $\PQq\Pg$, and $\Pg\Pg$ resonance signals was
measured from 0.6 to 1.6\TeV in 50-\GeVns steps in the low-mass search, and from 1.6 to 8.1\TeV in 100-\GeVns steps in the high-mass search.
The significance values obtained for $\PQq\PQq$ resonances are shown in Fig.~\ref{fig:pfsignif}.
The most significant excess of the data relative to the background fit comes from the two consecutive bins between 0.79 and 0.89\TeV.
Fitting these data to $\PQq\PQq$, $\PQq\Pg$, and $\Pg\Pg$ resonances with a mass of 0.85\TeV
yields local significances of 1.2, 1.6, and 1.9 standard deviations, including systematic uncertainties, respectively.

\section{Limits on narrow resonances}

We use the dijet mass spectrum from wide jets, the background
parameterization, and the dijet resonance shapes to set
limits on the production cross section of new particles decaying to the parton pairs $\PQq\PQq$ (or $\PQq\PAQq$), $\Pq\Pg$, and $\Pg\Pg$. A separate limit is
determined for each final state because of the dependence of the
dijet resonance shape on the types of the two final-state partons.

\subsection{Systematic uncertainty and statistical methodology}
\label{sec:statistics}
The dominant sources of systematic uncertainty are the jet energy scale and resolution,
integrated luminosity, and the value of the parameters within the functional form
modeling the background shape in the dijet mass distribution. The uncertainty in the jet energy scale in both the
low-mass and the high-mass search is \jecUncert and is
determined from $\sqrt{s}=13$\TeV data using the methods described in Ref.~\cite{Khachatryan:2016kdb}.
This uncertainty is propagated to the limits by shifting the dijet mass shape for signal by $\pm$\jecUncert.
The uncertainty in the jet energy
resolution translates into an uncertainty of 10\% in the resolution of the dijet
mass~\cite{Khachatryan:2016kdb}, and is propagated to the limits by observing the effect of increasing and decreasing by 10\% the reconstructed
width of the dijet mass shape for signal.
The uncertainty in the
integrated luminosity is \lumiUncert~\cite{CMS-PAS-LUM-17-001}, and is propagated to the normalization of
the signal.
Changes in the values of the parameters describing the background introduce a change in the signal yield,
which is accounted for as a systematic uncertainty as discussed in the next paragraph.

\begin{figure}[!htb] \centering
\includegraphics[width=0.48\textwidth]{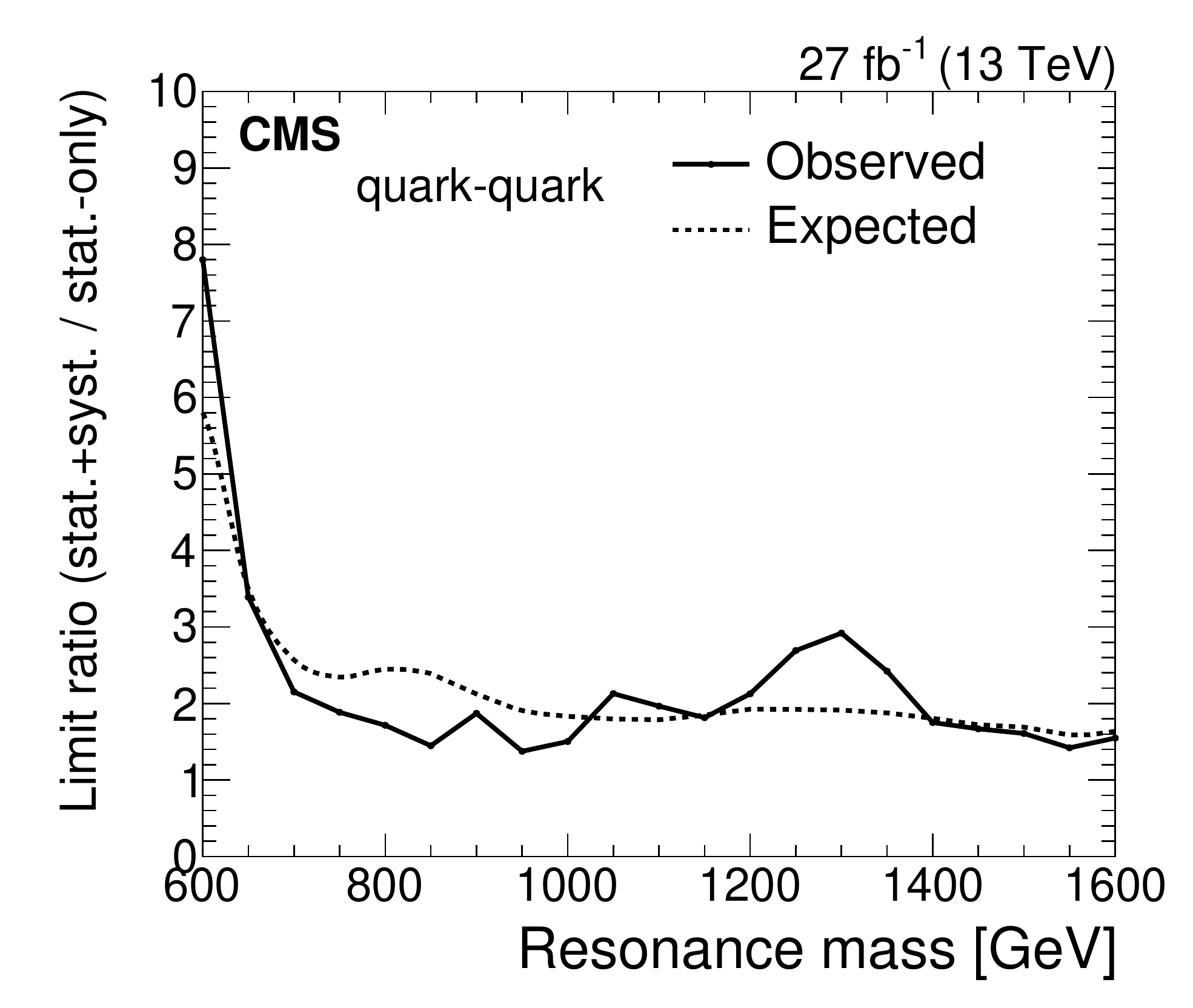}
\includegraphics[width=0.48\textwidth]{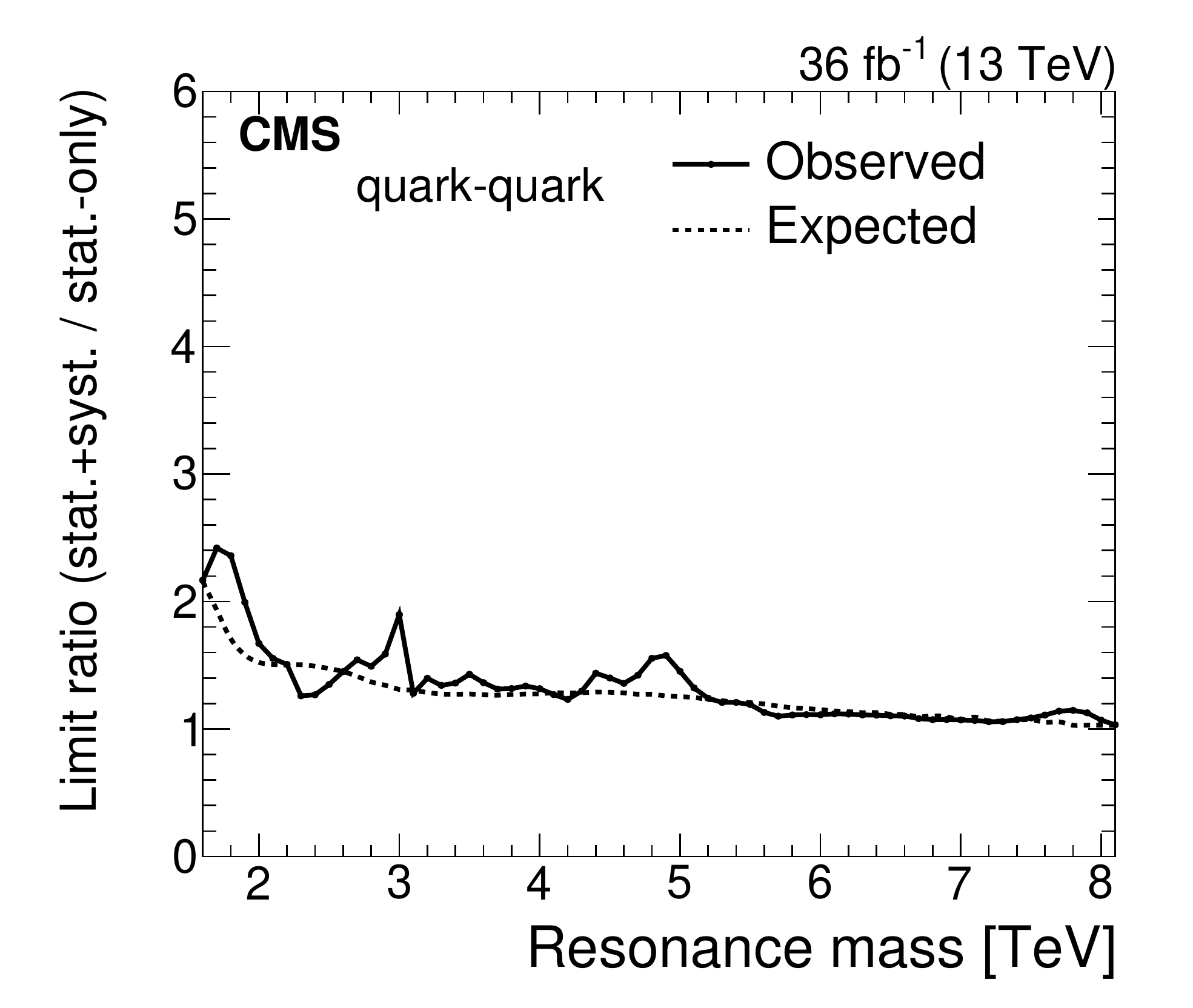}
\caption{The observed (points) and expected (dashed) ratio between the 95\% \CL  limit on the cross section, including systematic uncertainties, and the limit including statistical uncertainties only for
dijet resonances decaying to quark-quark in the low-mass search (\cmsLeft) and in the high-mass search (\cmsRight).}
\label{fig:systematics}
\end{figure}

The asymptotic approximation~\cite{Cowan:2010js} of the modified frequentist \CLs method~\cite{Junk1999,bib-cls} is
utilized to set upper limits on signal cross sections, following the prescription
described in Ref.~\cite{ATLAS:1379837}.  We use a multi-bin counting experiment likelihood, which is
a product of Poisson distributions corresponding to different bins.
We evaluate the likelihood independently at each value of the resonance pole mass from 0.6 to 1.6\TeV in 50-\GeVns
steps in the low-mass search, and from 1.6 to 8.1\TeV in 100-\GeVns steps in the high-mass search.
The contribution from each hypothetical resonance signal is
evaluated in every bin of dijet mass greater than the minimum dijet mass requirement in the search and less than 150\% of the resonance mass (e.g. the high mass tail of a 1\TeV resonance
is truncated, removing any contribution above a dijet mass of 1.5\TeV, but the low mass tail is not truncated).
The systematic uncertainties are implemented as nuisance parameters in the likelihood model, with Gaussian constraints for the jet
energy scale and resolution, and log-normal constraints for the integrated luminosity.
The systematic uncertainty in the background is automatically evaluated via profiling, effectively
refitting for the optimal values of the background parameters for each value of resonance cross section.
This allows the background parameters to float freely to their most likely value
for every signal cross section value within the likelihood function. Since the
observed data are effectively constraining the sum of signal and background, the
most likely value of the background decreases as the signal cross section
increases within the likelihood function. This statistical methodology therefore
gives a smaller background for larger signals within the likelihood function
than methodologies that hold the background parameters fixed within the
likelihood.  This leads to larger probabilities for larger signals and hence higher upper limits on the signal cross section.
The extent to which the
background uncertainty affects the limit depends significantly on the signal shape and the resonance mass, with the largest
effect occurring for
the $\Pg\Pg$ resonances, because they are broader, and the smallest effect
occurring for $\PQq\PQq$ resonances.
The effect increases as the resonance mass decreases, and is most severe at
the lowest resonance masses within each search, where the sideband used to
constrain the background, available at lower dijet mass, is smaller.
The effect of the systematic uncertainties on the limit for $\PQq\PQq$ resonances is shown
in Fig.~\ref{fig:systematics}. For almost all resonance mass values, the background systematic uncertainty
produces the majority of the effect on the limit shown here.

\subsection{Limits on the resonance production cross section}

\begin{figure*}[hbtp!]
  \centering
    \includegraphics[width=0.48\textwidth]{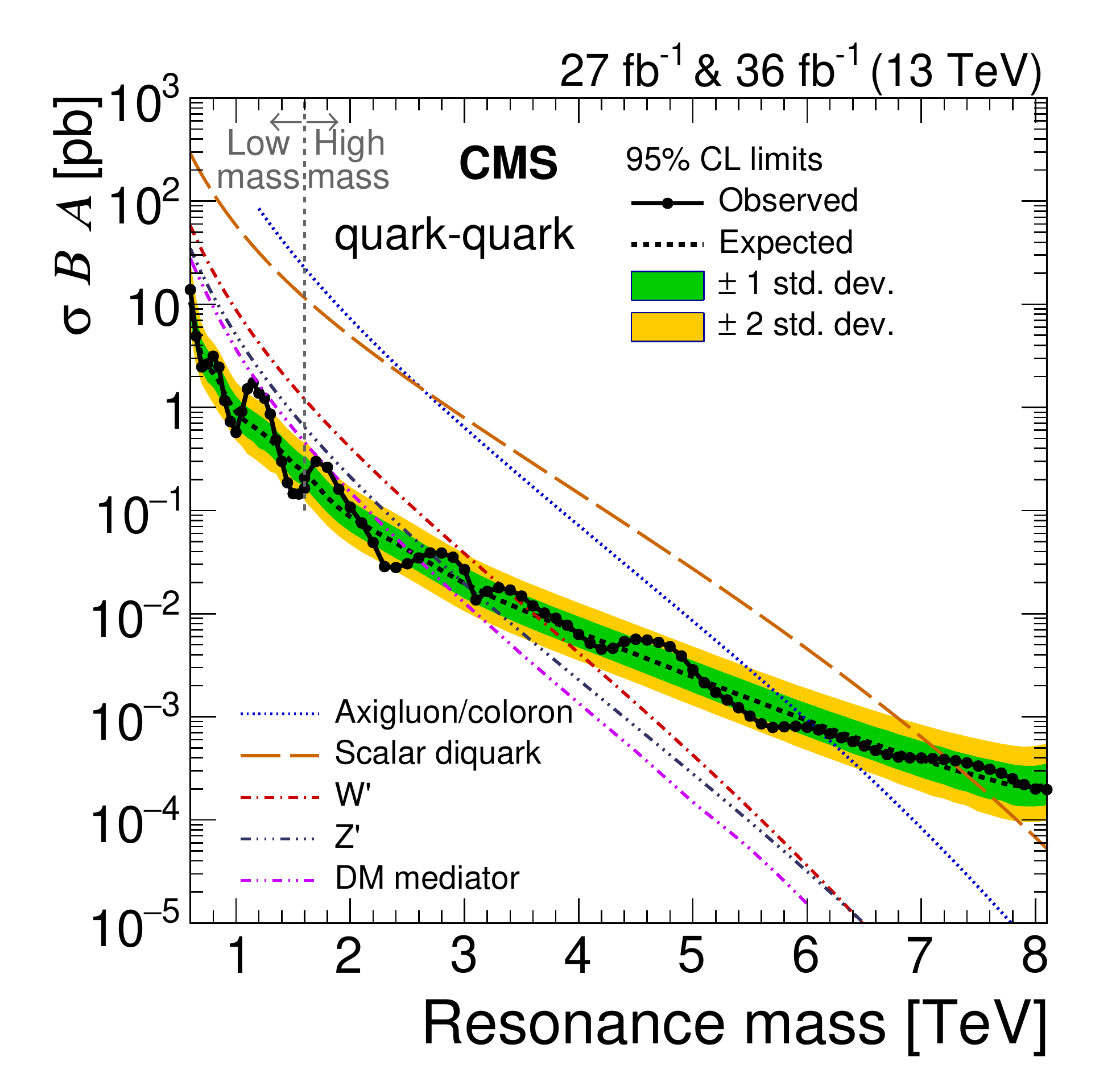}
    \includegraphics[width=0.48\textwidth]{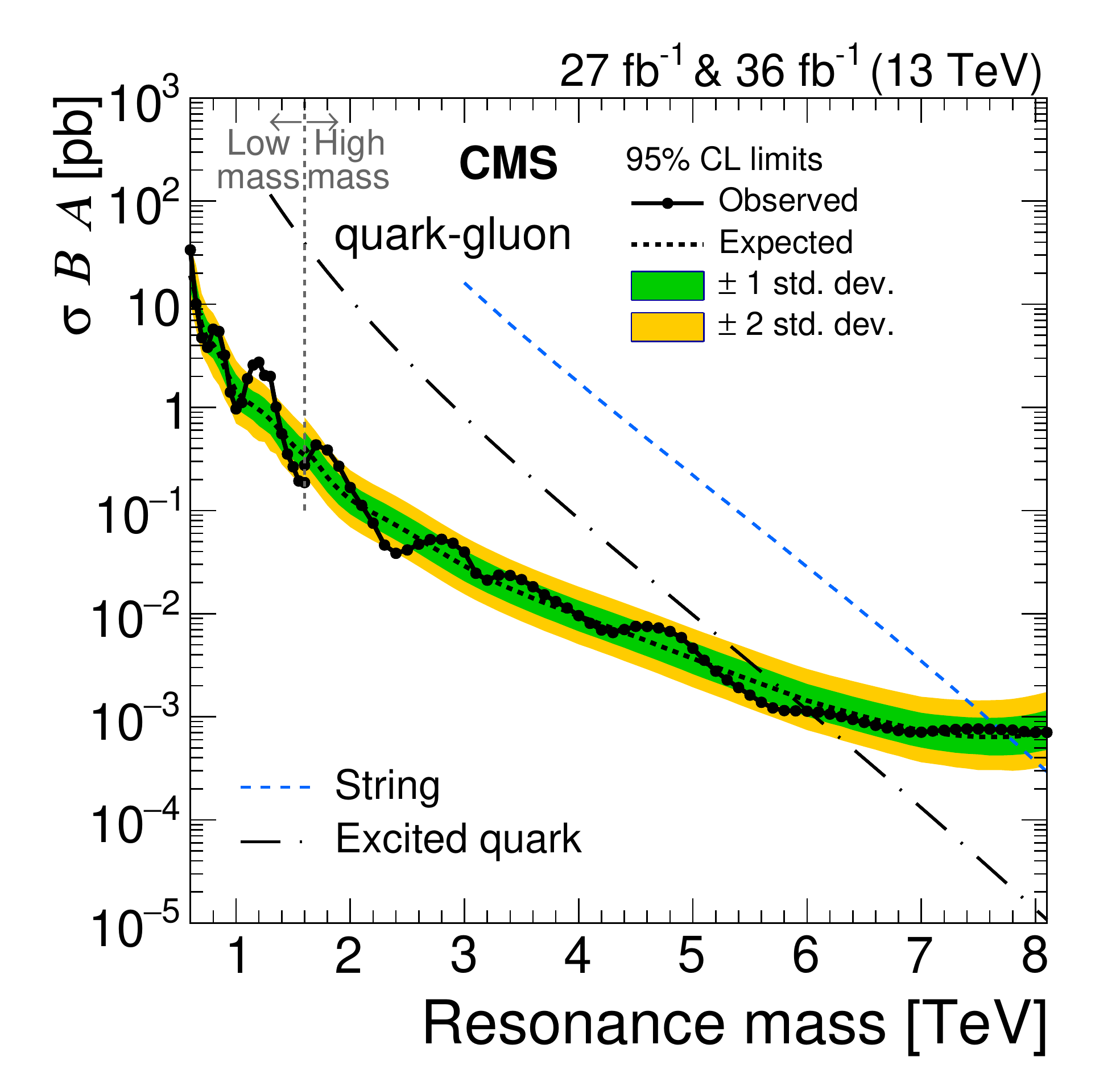}
    \includegraphics[width=0.48\textwidth]{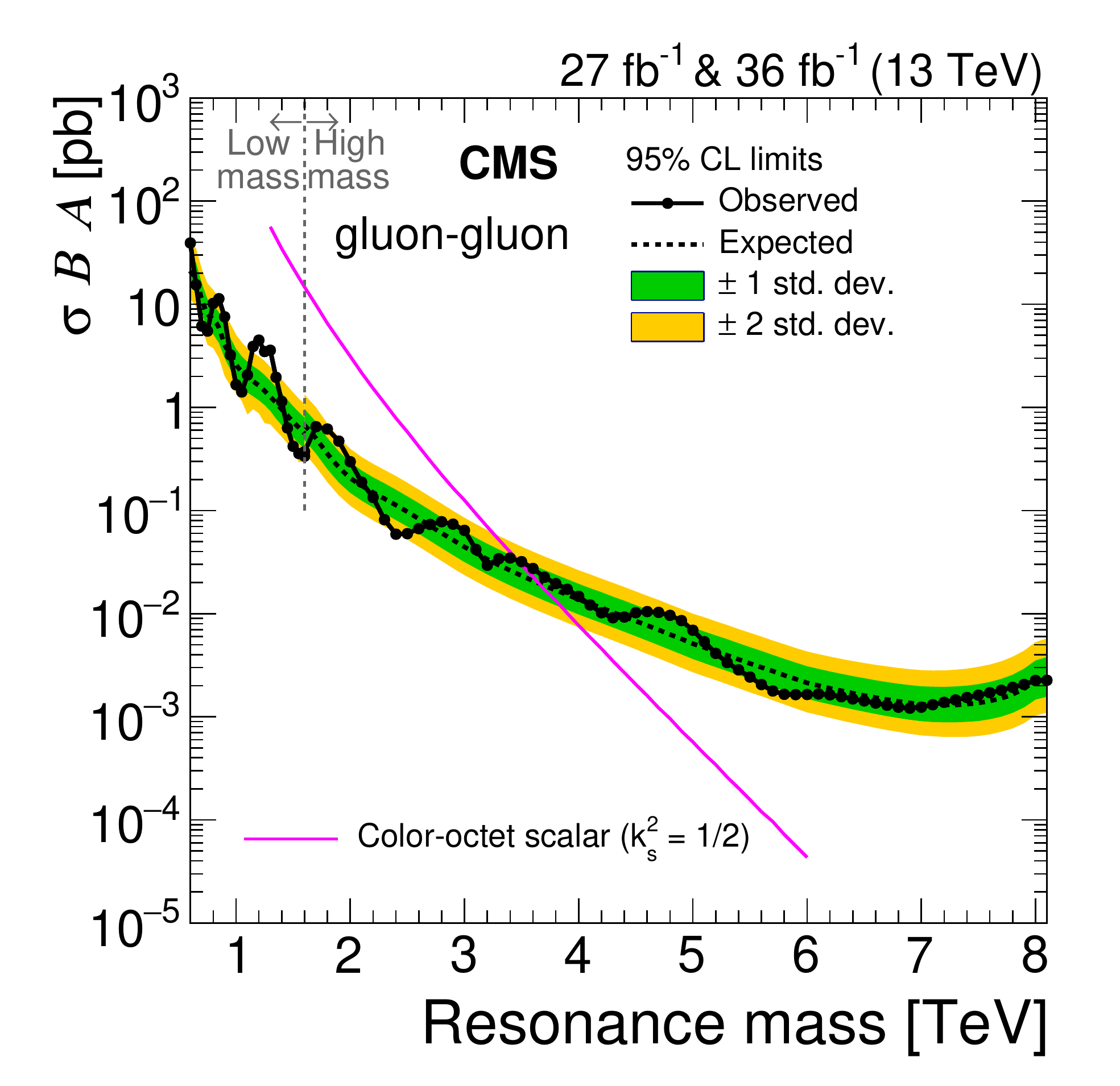}
    \includegraphics[width=0.48\textwidth]{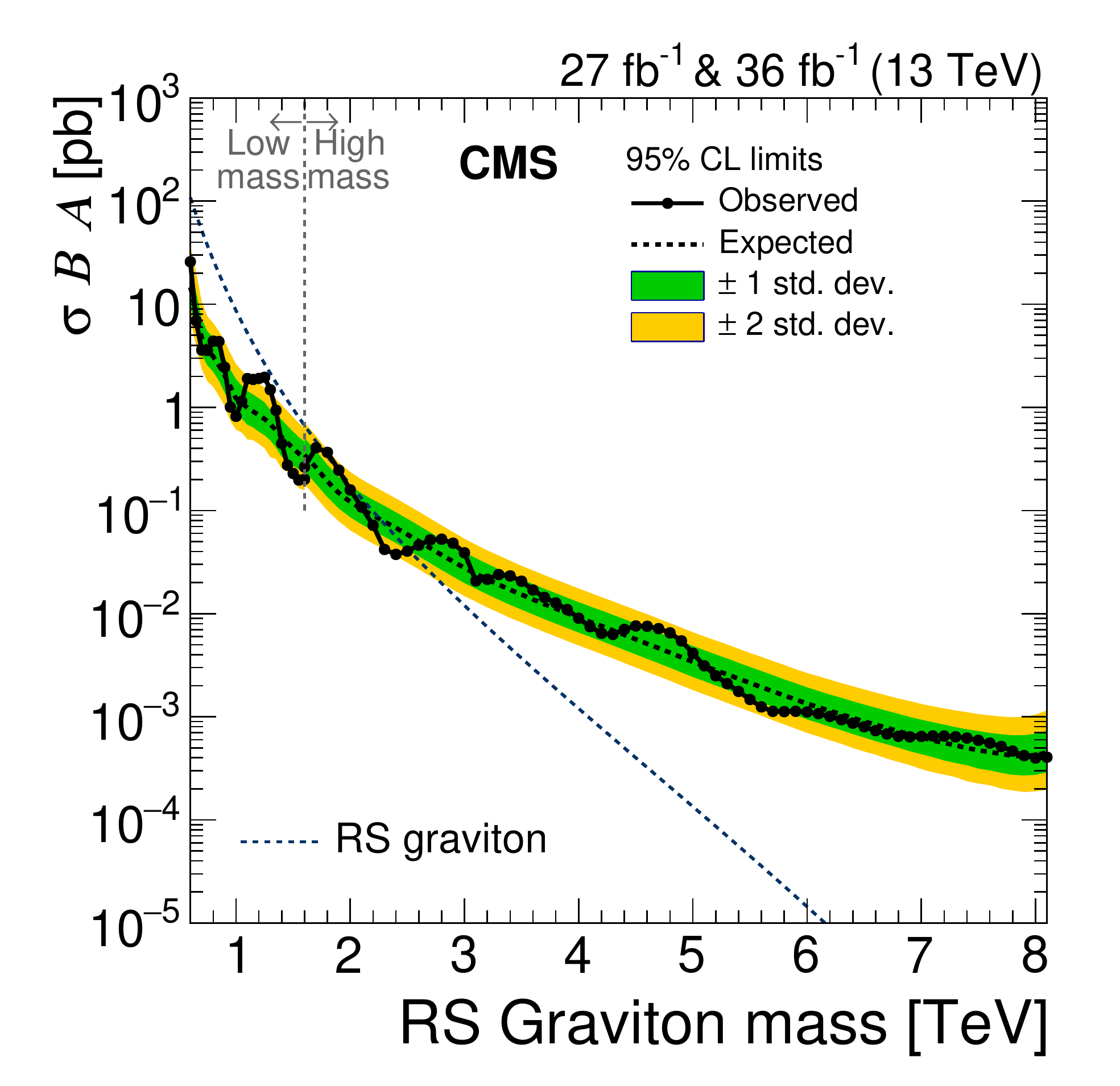}
    \caption{The observed 95\% \CL upper limits on the product of the cross section, branching fraction, and acceptance for dijet resonances
    decaying to quark-quark (upper left), quark-gluon (upper right), gluon-gluon (lower left), and for RS gravitons (lower right).
    The corresponding expected limits (dashed) and their variations
    at the 1 and 2 standard deviation levels (shaded bands) are also shown.
    Limits are compared to predicted cross sections for string resonances~\cite{Anchordoqui:2008di,Cullen:2000ef},
    excited quarks~\cite{ref_qstar,Baur:1989kv}, axigluons~\cite{ref_axi}, colorons~\cite{ref_coloron},  scalar diquarks~\cite{ref_diquark},
    color-octet scalars~\cite{Han:2010rf}, new gauge bosons $\PWpr$ and $\PZpr$ with SM-like couplings~\cite{ref_gauge},
    dark matter mediators for $m_{\mathrm{DM}}=1$\GeV~\cite{Boveia:2016mrp,Abdallah:2015ter}, and RS gravitons~\cite{ref_rsg}.
    }
    \label{figLimitAll}
\end{figure*}

\begin{figure*}[hbtp!]
  \centering
    \includegraphics[width=0.48\textwidth]{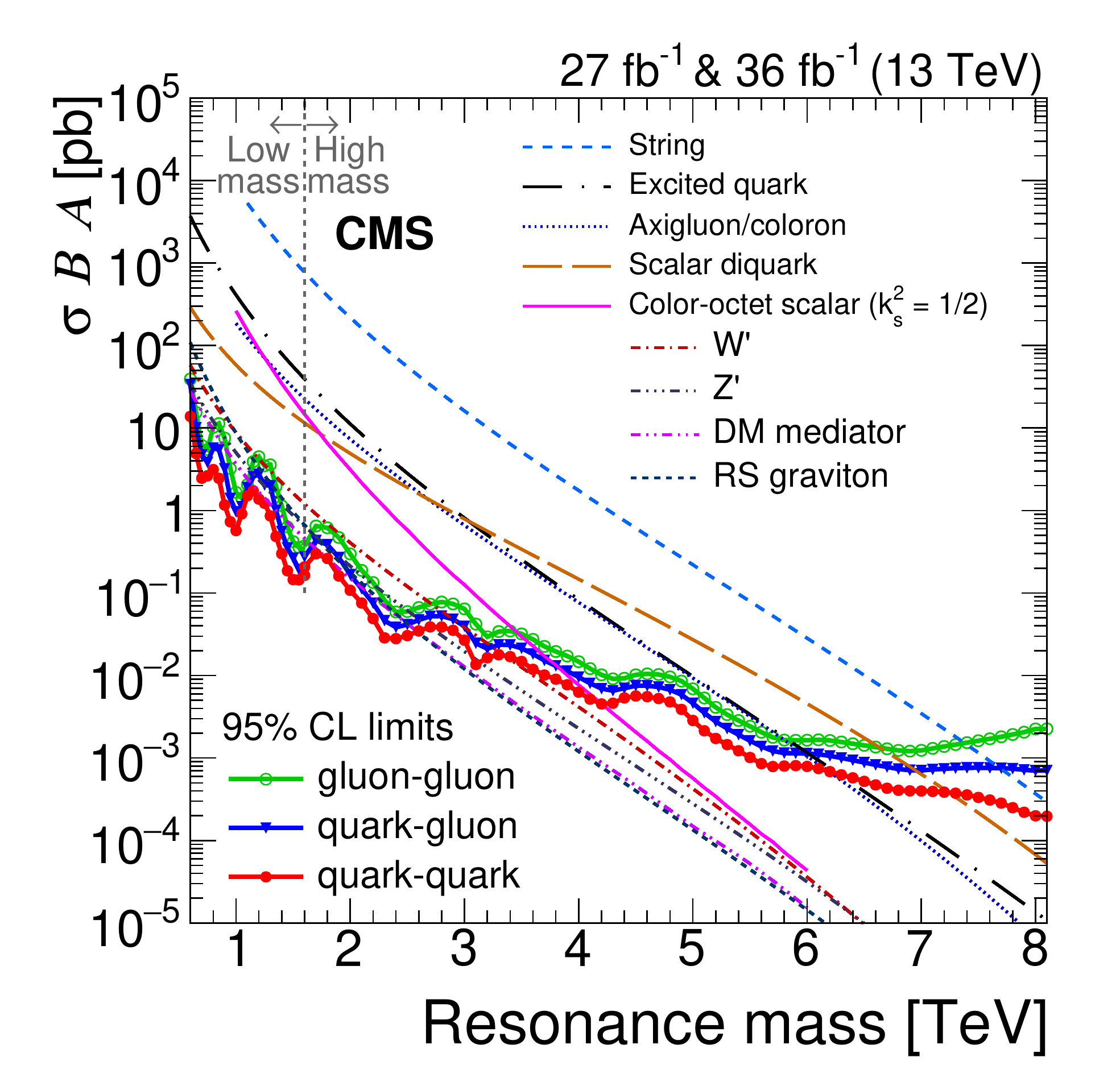}
    \caption{The observed 95\% \CL upper limits on the product of the cross section, branching fraction, and acceptance for
    quark-quark, quark-gluon, and gluon-gluon dijet resonances.
    Limits are compared to predicted cross sections for string resonances~\cite{Anchordoqui:2008di,Cullen:2000ef},
    excited quarks~\cite{ref_qstar,Baur:1989kv}, axigluons~\cite{ref_axi}, colorons~\cite{ref_coloron},  scalar diquarks~\cite{ref_diquark},
    color-octet scalars~\cite{Han:2010rf}, new gauge bosons $\PWpr$ and $\PZpr$ with SM-like couplings~\cite{ref_gauge},
    dark matter mediators for $m_{\mathrm{DM}}=1$\GeV~\cite{Boveia:2016mrp,Abdallah:2015ter}, and RS gravitons~\cite{ref_rsg}.
    }
    \label{figLimitSummary}
\end{figure*}

Tables~\ref{tab:limits} and ~\ref{tab:pflimits}, and Figs.~\ref{figLimitAll} and \ref{figLimitSummary}, show the model-independent observed
upper limits at 95\% \CL on the product of
the cross section ($\sigma$), the branching fraction to dijets ($B$), and the acceptance ($A$) for narrow resonances, with the
kinematic requirements $\detajj<1.3$ for the dijet system and $\abs{\eta}<2.5$ for each of the jets. The acceptance of the minimum dijet mass requirement in each search has been evaluated separately for
$\PQq\PQq$, $\PQq\Pg$, and $\Pg\Pg$ resonances, and has been taken into account by correcting the limits and therefore does not appear in the acceptance $A$.
The resonance mass boundary of 1.6\TeV between the high- and low-mass searches was chosen to maintain a reasonable acceptance for the minimum dijet
mass requirement imposed by the high-mass search.  For a 1.6\TeV dijet
resonance, the acceptance of the 1.25\TeV dijet mass requirement
is 57\% for a gluon-gluon resonance, 76\% for a quark-gluon resonance, and 85\% for a quark-quark resonance.
At this resonance mass, the expected limits we find on $\sigma B A$ for a
quark-quark resonance are the same in the high and low mass search.
Figure \ref{figLimitAll} also shows the expected limits on $\sigma B A$ and their bands of uncertainty.
The difference in the limits for $\PQq\PQq$, $\PQq\Pg$, and $\Pg\Pg$ resonances at the same resonance mass
originates from the difference in their line shapes. For the RS graviton model, which decays to both $\PQq\PAQq$ and $\Pg\Pg$ final states,
the upper limits on the cross section are derived using a weighted average of the $\PQq\PQq$
and $\Pg\Pg$ resonance shapes, where the weights correspond to the relative branching fractions for
the two final states.

\begin{table*}[hbtp]
  \topcaption{Limits from the low-mass search. The observed and expected upper limits at 95\% \CL on $\sigma B A$ for
    gluon-gluon, quark-gluon, and quark-quark resonances, and an RS graviton are given as
    a function
    of the resonance mass.
           \label{tab:limits}}
  \centering
\cmsTable{
  \begin{tabular}{cllllllll}
      \hline
      \multirow{3}{*}{Mass~[\TeVns{}]} & \multicolumn{8}{c}{95\% \CL upper limit [pb]} \\
        &\multicolumn{2}{c}{$\cPg\cPg$}
        &\multicolumn{2}{c}{$\cPq\cPg$}
        &\multicolumn{2}{c}{$\cPq\cPq$}
        &\multicolumn{2}{c}{RS graviton} \\
       & Observed & Expected & Observed & Expected  & Observed & Expected & Observed & Expected
    \\\hline
    0.60 & 3.93$\times 10^{+1}$ & 2.10$\times 10^{+1}$ & 3.37$\times 10^{+1}$ & 1.90$\times 10^{+1}$ & 1.38$\times 10^{+1}$ & 1.05$\times 10^{+1}$ & 2.59$\times 10^{+1}$ & 1.46$\times 10^{+1}$\\
    0.65 & 1.55$\times 10^{+1}$ & 1.77$\times 10^{+1}$ & 1.01$\times 10^{+1}$ & 1.14$\times 10^{+1}$ & 4.92$\times 10^0$ & 5.15$\times 10^0$ & 6.92$\times 10^0$ & 8.28$\times 10^0$\\
    0.70 & 6.14$\times 10^0$ & 1.12$\times 10^{+1}$ & 4.73$\times 10^0$ & 6.32$\times 10^0$ & 2.47$\times 10^0$ & 3.16$\times 10^0$ & 3.59$\times 10^0$ & 4.77$\times 10^0$\\
    0.75 & 5.50$\times 10^0$ & 8.13$\times 10^0$ & 3.82$\times 10^0$ & 4.68$\times 10^0$ & 2.64$\times 10^0$ & 2.49$\times 10^0$ & 3.57$\times 10^0$ & 3.67$\times 10^0$\\
    0.80 & 1.02$\times 10^{+1}$ & 7.15$\times 10^0$ & 5.73$\times 10^0$ & 4.06$\times 10^0$ & 3.14$\times 10^0$ & 2.14$\times 10^0$ & 4.39$\times 10^0$ & 3.11$\times 10^0$\\
    0.85 & 1.13$\times 10^{+1}$ & 5.93$\times 10^0$ & 5.45$\times 10^0$ & 3.33$\times 10^0$ & 2.46$\times 10^0$ & 1.79$\times 10^0$ & 4.35$\times 10^0$ & 2.55$\times 10^0$\\
    0.90 & 7.56$\times 10^0$ & 4.04$\times 10^0$ & 3.21$\times 10^0$ & 2.42$\times 10^0$ & 1.17$\times 10^0$ & 1.36$\times 10^0$ & 2.45$\times 10^0$ & 2.04$\times 10^0$\\
    0.95 & 3.23$\times 10^0$ & 3.32$\times 10^0$ & 1.40$\times 10^0$ & 1.86$\times 10^0$ & 7.30$\times 10^{-1}$ & 1.10$\times 10^0$ & 1.01$\times 10^0$ & 1.59$\times 10^0$\\
    1.00 & 1.66$\times 10^0$ & 2.60$\times 10^0$ & 9.67$\times 10^{-1}$ & 1.45$\times 10^0$ & 5.72$\times 10^{-1}$ & 9.06$\times 10^{-1}$ & 8.19$\times 10^{-1}$ & 1.23$\times 10^0$\\
    1.05 & 1.41$\times 10^0$ & 2.22$\times 10^0$ & 1.11$\times 10^0$ & 1.26$\times 10^0$ & 9.11$\times 10^{-1}$ & 7.90$\times 10^{-1}$ & 1.14$\times 10^0$ & 1.11$\times 10^0$\\
    1.10 & 2.06$\times 10^0$ & 1.96$\times 10^0$ & 1.90$\times 10^0$ & 1.13$\times 10^0$ & 1.51$\times 10^0$ & 7.11$\times 10^{-1}$ & 1.90$\times 10^0$ & 9.86$\times 10^{-1}$\\
    1.15 & 3.90$\times 10^0$ & 1.79$\times 10^0$ & 2.58$\times 10^0$ & 1.04$\times 10^0$ & 1.74$\times 10^0$ & 6.55$\times 10^{-1}$ & 1.87$\times 10^0$ & 9.12$\times 10^{-1}$\\
    1.20 & 4.49$\times 10^0$ & 1.63$\times 10^0$ & 2.74$\times 10^0$ & 9.49$\times 10^{-1}$ & 1.38$\times 10^0$ & 6.00$\times 10^{-1}$ & 1.91$\times 10^0$ & 8.39$\times 10^{-1}$\\
    1.25 & 3.48$\times 10^0$ & 1.45$\times 10^0$ & 2.04$\times 10^0$ & 8.64$\times 10^{-1}$ & 1.23$\times 10^0$ & 5.40$\times 10^{-1}$ & 1.96$\times 10^0$ & 7.72$\times 10^{-1}$\\
    1.30 & 3.58$\times 10^0$ & 1.26$\times 10^0$ & 2.00$\times 10^0$ & 7.60$\times 10^{-1}$ & 8.61$\times 10^{-1}$ & 4.85$\times 10^{-1}$ & 1.48$\times 10^0$ & 6.87$\times 10^{-1}$\\
    1.35 & 1.96$\times 10^0$ & 1.11$\times 10^0$ & 1.01$\times 10^0$ & 6.62$\times 10^{-1}$ & 4.85$\times 10^{-1}$ & 4.24$\times 10^{-1}$ & 9.35$\times 10^{-1}$ & 6.01$\times 10^{-1}$\\
    1.40 & 1.14$\times 10^0$ & 9.55$\times 10^{-1}$ & 5.56$\times 10^{-1}$ & 5.71$\times 10^{-1}$ & 3.00$\times 10^{-1}$ & 3.69$\times 10^{-1}$ & 4.47$\times 10^{-1}$ & 5.16$\times 10^{-1}$\\
    1.45 & 6.32$\times 10^{-1}$ & 8.33$\times 10^{-1}$ & 3.52$\times 10^{-1}$ & 4.97$\times 10^{-1}$ & 1.86$\times 10^{-1}$ & 3.27$\times 10^{-1}$ & 2.75$\times 10^{-1}$ & 4.55$\times 10^{-1}$\\
    1.50 & 4.20$\times 10^{-1}$ & 7.23$\times 10^{-1}$ & 2.66$\times 10^{-1}$ & 4.30$\times 10^{-1}$ & 1.45$\times 10^{-1}$ & 2.84$\times 10^{-1}$ & 2.29$\times 10^{-1}$ & 4.00$\times 10^{-1}$\\
    1.55 & 3.57$\times 10^{-1}$ & 6.38$\times 10^{-1}$ & 1.93$\times 10^{-1}$ & 3.81$\times 10^{-1}$ & 1.44$\times 10^{-1}$ & 2.59$\times 10^{-1}$ & 1.97$\times 10^{-1}$ & 3.57$\times 10^{-1}$\\
    1.60 & 3.37$\times 10^{-1}$ & 5.58$\times 10^{-1}$ & 1.87$\times 10^{-1}$ & 3.45$\times 10^{-1}$ & 1.64$\times 10^{-1}$ & 2.35$\times 10^{-1}$ & 2.01$\times 10^{-1}$ & 3.20$\times 10^{-1}$\\
    \hline
\end{tabular}}
\end{table*}

\begin{table*}[hbtp]
  \topcaption{Limits from the high-mass search. The observed and expected upper limits at 95\% \CL on $\sigma B A$ for
    gluon-gluon, quark-gluon, and quark-quark resonances, and an RS graviton are shown as
    functions
    of the resonance mass.
            \label{tab:pflimits}}
  \centering
\resizebox{0.80\textwidth}{!}{
\begin{tabular}{ccccccccc}
      \hline
      \multirow{3}{*}{Mass~[\TeVns{}]} & \multicolumn{8}{c}{95\% \CL upper limit [pb]} \\
        &\multicolumn{2}{c}{$\cPg\cPg$}
        &\multicolumn{2}{c}{$\cPq\cPg$}
        &\multicolumn{2}{c}{$\cPq\cPq$}
        &\multicolumn{2}{c}{RS graviton}
	\\
       & Observed & Expected & Observed & Expected  & Observed & Expected & Observed & Expected

    \\\hline

1.6 & 3.72$\times 10^{-1}$ & 6.72$\times 10^{-1}$ & 2.74$\times 10^{-1}$ & 4.08$\times 10^{-1}$ & 2.07$\times 10^{-1}$ & 2.38$\times 10^{-1}$ & 2.65$\times 10^{-1}$ & 3.46$\times 10^{-1}$\\
1.7 & 6.50$\times 10^{-1}$ & 5.02$\times 10^{-1}$ & 4.33$\times 10^{-1}$ & 2.96$\times 10^{-1}$ & 2.99$\times 10^{-1}$ & 1.79$\times 10^{-1}$ & 4.06$\times 10^{-1}$ & 2.61$\times 10^{-1}$\\
1.8 & 6.17$\times 10^{-1}$ & 3.55$\times 10^{-1}$ & 3.86$\times 10^{-1}$ & 2.10$\times 10^{-1}$ & 2.62$\times 10^{-1}$ & 1.34$\times 10^{-1}$ & 3.66$\times 10^{-1}$ & 1.92$\times 10^{-1}$\\
1.9 & 4.71$\times 10^{-1}$ & 2.63$\times 10^{-1}$ & 2.69$\times 10^{-1}$ & 1.60$\times 10^{-1}$ & 1.61$\times 10^{-1}$ & 1.06$\times 10^{-1}$ & 2.46$\times 10^{-1}$ & 1.48$\times 10^{-1}$\\
2.0 & 2.97$\times 10^{-1}$ & 2.07$\times 10^{-1}$ & 1.67$\times 10^{-1}$ & 1.29$\times 10^{-1}$ & 1.08$\times 10^{-1}$ & 8.71$\times 10^{-2}$& 1.59$\times 10^{-1}$ & 1.22$\times 10^{-1}$\\
2.1 & 1.88$\times 10^{-1}$ & 1.74$\times 10^{-1}$ & 1.12$\times 10^{-1}$ & 1.10$\times 10^{-1}$ & 7.56$\times 10^{-2}$& 7.44$\times 10^{-2}$& 1.08$\times 10^{-1}$ & 1.03$\times 10^{-1}$\\
2.2 & 1.34$\times 10^{-1}$ & 1.50$\times 10^{-1}$ & 7.53$\times 10^{-2}$& 9.49$\times 10^{-2}$& 4.90$\times 10^{-2}$& 6.43$\times 10^{-2}$& 7.18$\times 10^{-2}$& 8.95$\times 10^{-2}$\\
2.3 & 8.15$\times 10^{-2}$& 1.30$\times 10^{-1}$ & 4.62$\times 10^{-2}$& 8.32$\times 10^{-2}$& 2.86$\times 10^{-2}$& 5.57$\times 10^{-2}$& 4.19$\times 10^{-2}$& 7.78$\times 10^{-2}$\\
2.4 & 5.89$\times 10^{-2}$& 1.13$\times 10^{-1}$ & 3.84$\times 10^{-2}$& 7.21$\times 10^{-2}$& 2.80$\times 10^{-2}$& 4.82$\times 10^{-2}$& 3.75$\times 10^{-2}$& 6.78$\times 10^{-2}$\\
2.5 & 5.96$\times 10^{-2}$& 9.73$\times 10^{-2}$& 4.15$\times 10^{-2}$& 6.23$\times 10^{-2}$& 3.05$\times 10^{-2}$& 4.16$\times 10^{-2}$& 4.04$\times 10^{-2}$& 5.86$\times 10^{-2}$\\
2.6 & 6.67$\times 10^{-2}$& 8.32$\times 10^{-2}$& 4.71$\times 10^{-2}$& 5.33$\times 10^{-2}$& 3.47$\times 10^{-2}$& 3.58$\times 10^{-2}$& 4.61$\times 10^{-2}$& 5.05$\times 10^{-2}$\\
2.7 & 7.32$\times 10^{-2}$& 7.09$\times 10^{-2}$& 5.22$\times 10^{-2}$& 4.55$\times 10^{-2}$& 3.88$\times 10^{-2}$& 3.08$\times 10^{-2}$& 5.19$\times 10^{-2}$& 4.33$\times 10^{-2}$\\
2.8 & 7.79$\times 10^{-2}$& 6.04$\times 10^{-2}$& 5.26$\times 10^{-2}$& 3.91$\times 10^{-2}$& 3.87$\times 10^{-2}$& 2.63$\times 10^{-2}$& 5.27$\times 10^{-2}$& 3.70$\times 10^{-2}$\\
2.9 & 7.37$\times 10^{-2}$& 5.18$\times 10^{-2}$& 4.82$\times 10^{-2}$& 3.35$\times 10^{-2}$& 3.53$\times 10^{-2}$& 2.28$\times 10^{-2}$& 4.82$\times 10^{-2}$& 3.20$\times 10^{-2}$\\
3.0 & 6.42$\times 10^{-2}$& 4.43$\times 10^{-2}$& 3.96$\times 10^{-2}$& 2.90$\times 10^{-2}$& 2.68$\times 10^{-2}$& 1.96$\times 10^{-2}$& 3.89$\times 10^{-2}$& 2.77$\times 10^{-2}$\\
3.1 & 4.20$\times 10^{-2}$& 3.86$\times 10^{-2}$& 2.46$\times 10^{-2}$& 2.53$\times 10^{-2}$& 1.36$\times 10^{-2}$& 1.74$\times 10^{-2}$& 2.08$\times 10^{-2}$& 2.43$\times 10^{-2}$\\
3.2 & 2.95$\times 10^{-2}$& 3.37$\times 10^{-2}$& 2.11$\times 10^{-2}$& 2.24$\times 10^{-2}$& 1.64$\times 10^{-2}$& 1.54$\times 10^{-2}$& 2.15$\times 10^{-2}$& 2.16$\times 10^{-2}$\\
3.3 & 3.41$\times 10^{-2}$& 2.96$\times 10^{-2}$& 2.36$\times 10^{-2}$& 1.96$\times 10^{-2}$& 1.78$\times 10^{-2}$& 1.36$\times 10^{-2}$& 2.39$\times 10^{-2}$& 1.91$\times 10^{-2}$\\
3.4 & 3.47$\times 10^{-2}$& 2.63$\times 10^{-2}$& 2.34$\times 10^{-2}$& 1.75$\times 10^{-2}$& 1.69$\times 10^{-2}$& 1.22$\times 10^{-2}$& 2.32$\times 10^{-2}$& 1.70$\times 10^{-2}$\\
3.5 & 3.19$\times 10^{-2}$& 2.33$\times 10^{-2}$& 2.14$\times 10^{-2}$& 1.58$\times 10^{-2}$& 1.48$\times 10^{-2}$& 1.10$\times 10^{-2}$& 2.06$\times 10^{-2}$& 1.53$\times 10^{-2}$\\
3.6 & 2.74$\times 10^{-2}$& 2.08$\times 10^{-2}$& 1.82$\times 10^{-2}$& 1.41$\times 10^{-2}$& 1.19$\times 10^{-2}$& 9.81$\times 10^{-3}$ & 1.70$\times 10^{-2}$& 1.37$\times 10^{-2}$\\
3.7 & 2.25$\times 10^{-2}$& 1.87$\times 10^{-2}$& 1.52$\times 10^{-2}$& 1.27$\times 10^{-2}$& 1.01$\times 10^{-2}$& 8.86$\times 10^{-3}$ & 1.44$\times 10^{-2}$& 1.24$\times 10^{-2}$\\
3.8 & 1.96$\times 10^{-2}$& 1.68$\times 10^{-2}$& 1.31$\times 10^{-2}$& 1.16$\times 10^{-2}$& 9.02$\times 10^{-3}$ & 8.03$\times 10^{-3}$ & 1.27$\times 10^{-2}$& 1.12$\times 10^{-2}$\\
3.9 & 1.72$\times 10^{-2}$& 1.53$\times 10^{-2}$& 1.13$\times 10^{-2}$& 1.05$\times 10^{-2}$& 7.72$\times 10^{-3}$ & 7.25$\times 10^{-3}$ & 1.09$\times 10^{-2}$& 1.01$\times 10^{-2}$\\
4.0 & 1.47$\times 10^{-2}$& 1.37$\times 10^{-2}$& 9.57$\times 10^{-3}$ & 9.45$\times 10^{-3}$ & 6.29$\times 10^{-3}$ & 6.57$\times 10^{-3}$ & 9.04$\times 10^{-3}$ & 9.16$\times 10^{-3}$\\
4.1 & 1.21$\times 10^{-2}$& 1.25$\times 10^{-2}$& 8.06$\times 10^{-3}$ & 8.67$\times 10^{-3}$ & 5.17$\times 10^{-3}$ & 5.98$\times 10^{-3}$ & 7.46$\times 10^{-3}$ & 8.33$\times 10^{-3}$\\
4.2 & 1.02$\times 10^{-2}$& 1.14$\times 10^{-2}$& 6.93$\times 10^{-3}$ & 7.89$\times 10^{-3}$ & 4.52$\times 10^{-3}$ & 5.40$\times 10^{-3}$ & 6.45$\times 10^{-3}$ & 7.59$\times 10^{-3}$\\
4.3 & 9.12$\times 10^{-3}$ & 1.03$\times 10^{-2}$& 6.55$\times 10^{-3}$ & 7.20$\times 10^{-3}$ & 4.61$\times 10^{-3}$ & 4.91$\times 10^{-3}$ & 6.29$\times 10^{-3}$ & 6.86$\times 10^{-3}$\\
4.4 & 9.27$\times 10^{-3}$ & 9.35$\times 10^{-3}$ & 7.01$\times 10^{-3}$ & 6.57$\times 10^{-3}$ & 5.35$\times 10^{-3}$ & 4.46$\times 10^{-3}$ & 7.02$\times 10^{-3}$ & 6.23$\times 10^{-3}$\\
4.5 & 1.02$\times 10^{-2}$& 8.47$\times 10^{-3}$ & 7.52$\times 10^{-3}$ & 5.98$\times 10^{-3}$ & 5.65$\times 10^{-3}$ & 4.04$\times 10^{-3}$ & 7.60$\times 10^{-3}$ & 5.64$\times 10^{-3}$\\
4.6 & 1.05$\times 10^{-2}$& 7.69$\times 10^{-3}$ & 7.51$\times 10^{-3}$ & 5.44$\times 10^{-3}$ & 5.55$\times 10^{-3}$ & 3.65$\times 10^{-3}$ & 7.54$\times 10^{-3}$ & 5.10$\times 10^{-3}$\\
4.7 & 1.03$\times 10^{-2}$& 6.96$\times 10^{-3}$ & 7.27$\times 10^{-3}$ & 4.91$\times 10^{-3}$ & 5.26$\times 10^{-3}$ & 3.31$\times 10^{-3}$ & 7.16$\times 10^{-3}$ & 4.63$\times 10^{-3}$\\
4.8 & 9.62$\times 10^{-3}$ & 6.27$\times 10^{-3}$ & 6.72$\times 10^{-3}$ & 4.46$\times 10^{-3}$ & 4.79$\times 10^{-3}$ & 2.99$\times 10^{-3}$ & 6.51$\times 10^{-3}$ & 4.19$\times 10^{-3}$\\
4.9 & 8.56$\times 10^{-3}$ & 5.69$\times 10^{-3}$ & 5.86$\times 10^{-3}$ & 4.04$\times 10^{-3}$ & 3.88$\times 10^{-3}$ & 2.70$\times 10^{-3}$ & 5.44$\times 10^{-3}$ & 3.77$\times 10^{-3}$\\
5.0 & 6.90$\times 10^{-3}$ & 5.10$\times 10^{-3}$ & 4.62$\times 10^{-3}$ & 3.67$\times 10^{-3}$ & 2.85$\times 10^{-3}$ & 2.44$\times 10^{-3}$ & 4.12$\times 10^{-3}$ & 3.41$\times 10^{-3}$\\
5.1 & 5.34$\times 10^{-3}$ & 4.70$\times 10^{-3}$ & 3.53$\times 10^{-3}$ & 3.33$\times 10^{-3}$ & 2.14$\times 10^{-3}$ & 2.22$\times 10^{-3}$ & 3.12$\times 10^{-3}$ & 3.11$\times 10^{-3}$\\
5.2 & 4.11$\times 10^{-3}$ & 4.28$\times 10^{-3}$ & 2.77$\times 10^{-3}$ & 3.04$\times 10^{-3}$ & 1.73$\times 10^{-3}$ & 2.01$\times 10^{-3}$ & 2.50$\times 10^{-3}$ & 2.82$\times 10^{-3}$\\
5.3 & 3.35$\times 10^{-3}$ & 3.94$\times 10^{-3}$ & 2.28$\times 10^{-3}$ & 2.77$\times 10^{-3}$ & 1.45$\times 10^{-3}$ & 1.81$\times 10^{-3}$ & 2.09$\times 10^{-3}$ & 2.58$\times 10^{-3}$\\
5.4 & 2.85$\times 10^{-3}$ & 3.60$\times 10^{-3}$ & 1.92$\times 10^{-3}$ & 2.50$\times 10^{-3}$ & 1.22$\times 10^{-3}$ & 1.64$\times 10^{-3}$ & 1.76$\times 10^{-3}$ & 2.34$\times 10^{-3}$\\
5.5 & 2.43$\times 10^{-3}$ & 3.28$\times 10^{-3}$ & 1.62$\times 10^{-3}$ & 2.29$\times 10^{-3}$ & 1.01$\times 10^{-3}$ & 1.50$\times 10^{-3}$ & 1.47$\times 10^{-3}$ & 2.13$\times 10^{-3}$\\
5.6 & 2.05$\times 10^{-3}$ & 3.02$\times 10^{-3}$ & 1.38$\times 10^{-3}$ & 2.08$\times 10^{-3}$ & 8.54$\times 10^{-4}$ & 1.36$\times 10^{-3}$ & 1.25$\times 10^{-3}$ & 1.93$\times 10^{-3}$\\
5.7 & 1.78$\times 10^{-3}$ & 2.77$\times 10^{-3}$ & 1.22$\times 10^{-3}$ & 1.90$\times 10^{-3}$ & 7.88$\times 10^{-4}$ & 1.23$\times 10^{-3}$ & 1.13$\times 10^{-3}$ & 1.76$\times 10^{-3}$\\
5.8 & 1.65$\times 10^{-3}$ & 2.53$\times 10^{-3}$ & 1.15$\times 10^{-3}$ & 1.73$\times 10^{-3}$ & 8.00$\times 10^{-4}$ & 1.11$\times 10^{-3}$ & 1.12$\times 10^{-3}$ & 1.61$\times 10^{-3}$\\
5.9 & 1.64$\times 10^{-3}$ & 2.33$\times 10^{-3}$ & 1.14$\times 10^{-3}$ & 1.58$\times 10^{-3}$ & 8.09$\times 10^{-4}$ & 1.02$\times 10^{-3}$ & 1.13$\times 10^{-3}$ & 1.47$\times 10^{-3}$\\
6.0 & 1.64$\times 10^{-3}$ & 2.13$\times 10^{-3}$ & 1.13$\times 10^{-3}$ & 1.43$\times 10^{-3}$ & 7.91$\times 10^{-4}$ & 9.25$\times 10^{-4}$ & 1.11$\times 10^{-3}$ & 1.34$\times 10^{-3}$\\
6.1 & 1.66$\times 10^{-3}$ & 2.01$\times 10^{-3}$ & 1.11$\times 10^{-3}$ & 1.34$\times 10^{-3}$ & 7.45$\times 10^{-4}$ & 8.39$\times 10^{-4}$ & 1.07$\times 10^{-3}$ & 1.24$\times 10^{-3}$\\
6.2 & 1.63$\times 10^{-3}$ & 1.89$\times 10^{-3}$ & 1.06$\times 10^{-3}$ & 1.24$\times 10^{-3}$ & 6.84$\times 10^{-4}$ & 7.66$\times 10^{-4}$ & 1.01$\times 10^{-3}$ & 1.14$\times 10^{-3}$\\
6.3 & 1.56$\times 10^{-3}$ & 1.79$\times 10^{-3}$ & 1.00$\times 10^{-3}$ & 1.16$\times 10^{-3}$ & 6.26$\times 10^{-4}$ & 6.99$\times 10^{-4}$ & 9.36$\times 10^{-4}$ & 1.05$\times 10^{-3}$\\
6.4 & 1.49$\times 10^{-3}$ & 1.69$\times 10^{-3}$ & 9.41$\times 10^{-4}$ & 1.08$\times 10^{-3}$ & 5.75$\times 10^{-4}$ & 6.44$\times 10^{-4}$ & 8.66$\times 10^{-4}$ & 9.74$\times 10^{-4}$\\
6.5 & 1.42$\times 10^{-3}$ & 1.61$\times 10^{-3}$ & 8.82$\times 10^{-4}$ & 1.00$\times 10^{-3}$ & 5.21$\times 10^{-4}$ & 5.89$\times 10^{-4}$ & 8.00$\times 10^{-4}$ & 9.00$\times 10^{-4}$\\
6.6 & 1.36$\times 10^{-3}$ & 1.53$\times 10^{-3}$ & 8.26$\times 10^{-4}$ & 9.37$\times 10^{-4}$ & 4.72$\times 10^{-4}$ & 5.40$\times 10^{-4}$ & 7.33$\times 10^{-4}$ & 8.39$\times 10^{-4}$\\
6.7 & 1.29$\times 10^{-3}$ & 1.47$\times 10^{-3}$ & 7.79$\times 10^{-4}$ & 8.82$\times 10^{-4}$ & 4.30$\times 10^{-4}$ & 4.91$\times 10^{-4}$ & 6.81$\times 10^{-4}$ & 7.78$\times 10^{-4}$\\
6.8 & 1.24$\times 10^{-3}$ & 1.41$\times 10^{-3}$ & 7.35$\times 10^{-4}$ & 8.27$\times 10^{-4}$ & 4.06$\times 10^{-4}$ & 4.55$\times 10^{-4}$ & 6.46$\times 10^{-4}$ & 7.23$\times 10^{-4}$\\
6.9 & 1.21$\times 10^{-3}$ & 1.36$\times 10^{-3}$ & 7.11$\times 10^{-4}$ & 7.78$\times 10^{-4}$ & 4.00$\times 10^{-4}$ & 4.18$\times 10^{-4}$ & 6.38$\times 10^{-4}$ & 6.74$\times 10^{-4}$\\
7.0 & 1.24$\times 10^{-3}$ & 1.32$\times 10^{-3}$ & 7.08$\times 10^{-4}$ & 7.29$\times 10^{-4}$ & 3.98$\times 10^{-4}$ & 3.81$\times 10^{-4}$ & 6.44$\times 10^{-4}$ & 6.32$\times 10^{-4}$\\
7.1 & 1.31$\times 10^{-3}$ & 1.30$\times 10^{-3}$ & 7.27$\times 10^{-4}$ & 7.05$\times 10^{-4}$ & 3.94$\times 10^{-4}$ & 3.57$\times 10^{-4}$ & 6.52$\times 10^{-4}$ & 5.89$\times 10^{-4}$\\
7.2 & 1.38$\times 10^{-3}$ & 1.30$\times 10^{-3}$ & 7.40$\times 10^{-4}$ & 6.81$\times 10^{-4}$ & 3.86$\times 10^{-4}$ & 3.27$\times 10^{-4}$ & 6.50$\times 10^{-4}$ & 5.58$\times 10^{-4}$\\
7.3 & 1.46$\times 10^{-3}$ & 1.30$\times 10^{-3}$ & 7.53$\times 10^{-4}$ & 6.62$\times 10^{-4}$ & 3.74$\times 10^{-4}$ & 3.02$\times 10^{-4}$ & 6.39$\times 10^{-4}$ & 5.28$\times 10^{-4}$\\
7.4 & 1.54$\times 10^{-3}$ & 1.32$\times 10^{-3}$ & 7.61$\times 10^{-4}$ & 6.50$\times 10^{-4}$ & 3.57$\times 10^{-4}$ & 2.84$\times 10^{-4}$ & 6.22$\times 10^{-4}$ & 4.97$\times 10^{-4}$\\
7.5 & 1.62$\times 10^{-3}$ & 1.36$\times 10^{-3}$ & 7.62$\times 10^{-4}$ & 6.38$\times 10^{-4}$ & 3.33$\times 10^{-4}$ & 2.66$\times 10^{-4}$ & 5.91$\times 10^{-4}$ & 4.73$\times 10^{-4}$\\
7.6 & 1.71$\times 10^{-3}$ & 1.42$\times 10^{-3}$ & 7.59$\times 10^{-4}$ & 6.38$\times 10^{-4}$ & 3.10$\times 10^{-4}$ & 2.47$\times 10^{-4}$ & 5.55$\times 10^{-4}$ & 4.55$\times 10^{-4}$\\
7.7 & 1.81$\times 10^{-3}$ & 1.51$\times 10^{-3}$ & 7.53$\times 10^{-4}$ & 6.38$\times 10^{-4}$ & 2.84$\times 10^{-4}$ & 2.29$\times 10^{-4}$ & 5.15$\times 10^{-4}$ & 4.36$\times 10^{-4}$\\
7.8 & 1.93$\times 10^{-3}$ & 1.65$\times 10^{-3}$ & 7.43$\times 10^{-4}$ & 6.44$\times 10^{-4}$ & 2.50$\times 10^{-4}$ & 2.17$\times 10^{-4}$ & 4.65$\times 10^{-4}$ & 4.18$\times 10^{-4}$\\
7.9 & 2.06$\times 10^{-3}$ & 1.87$\times 10^{-3}$ & 7.19$\times 10^{-4}$ & 6.56$\times 10^{-4}$ & 2.20$\times 10^{-4}$ & 2.11$\times 10^{-4}$ & 4.20$\times 10^{-4}$ & 4.18$\times 10^{-4}$\\
8.0 & 2.25$\times 10^{-3}$ & 2.24$\times 10^{-3}$ & 7.03$\times 10^{-4}$ & 6.93$\times 10^{-4}$ & 1.99$\times 10^{-4}$ & 2.11$\times 10^{-4}$ & 3.98$\times 10^{-4}$ & 4.24$\times 10^{-4}$\\
8.1 & 2.26$\times 10^{-3}$ & 2.41$\times 10^{-3}$ & 7.05$\times 10^{-4}$ & 7.35$\times 10^{-4}$ & 1.97$\times 10^{-4}$ & 2.17$\times 10^{-4}$ & 4.05$\times 10^{-4}$ & 4.55$\times 10^{-4}$\\

      \hline
    \end{tabular}}
\end{table*}

\clearpage

\subsection{Limits on the resonance mass for benchmark models}
All upper limits presented can be compared to the parton-level predictions of $\sigma B A$, without detector simulation,
to determine mass limits on new particles.
The model predictions shown in Figs.~\ref{figLimitAll} and \ref{figLimitSummary} are calculated in the narrow-width
approximation~\cite{Harris:2011bh} using the CTEQ6L1~\cite{refCTEQ} parton distribution functions at leading order.
A next-to-leading order correction factor of $K=1+ 8\pi\alpS/9\approx 1.3$ is applied to the leading order predictions
for the $\PWpr$ model and $K=1+(4\alpS/6\pi)(1+4\pi^2/3)\approx 1.3$ for the $\PZpr$ model
(see pages 248 and 233 of Ref.~\cite{Barger:1987nn}), where $\alpS$ is the
strong coupling constant evaluated at a renormalization scale equal to the resonance mass.
Similarly, for the axigluon/coloron models a correction factor is applied which varies between $K=1.08$ at a resonance mass of
0.6\TeV and $K=1.33$ at 8.1\TeV~\cite{Chivukula:2013xla}.
The branching fraction includes the direct decays of the resonance into the five light quarks and
gluons only, excluding top quarks from the decay, although top quarks are included in the calculation of
the resonance width.
The signal acceptance evaluated at the parton level for the resonance decay to two partons can be written as $A=A_{\Delta}A_{\eta}$,
where $A_{\Delta}$ is the acceptance of requiring $\detajj<1.3$ alone, and $A_{\eta}$ is the acceptance of also requiring
$\abs{\eta}<2.5$. The acceptance $A_{\Delta}$ is model dependent. In the case of isotropic decays, the dijet angular distribution as a function of $\tanh{(\detajj/2)}$
is approximately constant, and $A_{\Delta}\approx\tanh(1.3/2)=0.57$, independent
of resonance mass. The acceptance
$A_{\eta}$ is maximal for resonance masses above 1\TeV---greater than 0.99 for all models considered.
The acceptance $A_{\eta}$ decreases as the resonance mass decreases below 1\TeV, and for a resonance mass of 0.6\TeV it is 0.92 for
excited quarks, 0.98 for RS gravitons, and between those two values for the other models.
For a given model, new particles are excluded at 95\% \CL in mass regions where the theoretical prediction
lies at or above the observed upper limit for the appropriate final state of Figs.~\ref{figLimitAll} and \ref{figLimitSummary}.
Mass limits on all benchmark models are summarized in Table~\ref{tab:MassLimit}.
\clearpage

\begin{table}[hbtp]
  \topcaption{ Observed and expected mass limits at 95\% \CL from this analysis compared to previously published limits on narrow
  resonances from CMS with 12.9\fbinv~\cite{Sirunyan:2016iap}.
  The listed models are excluded between \minMassLow and the indicated mass limit by this analysis. In addition to the observed mass limits
  listed below, this analysis also excludes the RS graviton model within the
  mass interval between 1.9 and 2.5\TeV and the $\PZpr$ model within roughly a 50\GeV window around 3.1\TeV.}
  \centering
  \begin{tabular}{lccc}
  \hline
    Model        & Final                              & \multicolumn{2}{c}{\ \ \ Observed (expected) mass limit [\TeVns{}]} \\
               & State                              & \hspace{0.5in} \RunLumi                    & Ref.~\cite{Sirunyan:2016iap}, 12.9\fbinv    \\   \hline
String resonance & $\PQq\Pg$                        & \hspace{0.5in} 7.7\ (7.7)                  & 7.4\ (7.4)    \\
Scalar diquark  & $\PQq\PQq$                        & \hspace{0.5in} 7.2\ (7.4)                  & 6.9\ (6.8)    \\
Axigluon/coloron  & $\PQq\PAQq$                     & \hspace{0.5in} 6.1\ (6.0)                  & 5.5\ (5.6)    \\
Excited quark  &  $\PQq\Pg$                         & \hspace{0.5in} 6.0\ (5.8)                  & 5.4\ (5.4)    \\
Color-octet scalar ($k_s^2=1/2$) & $\Pg\Pg$         & \hspace{0.5in} 3.4\ (3.6)                  & 3.0\ (3.3)    \\
$\PWpr$ SM-like& $\PQq\PAQq$                        & \hspace{0.5in} 3.3\ (3.6)                  & 2.7\ (3.1)    \\
$\PZpr$ SM-like& $\PQq\PAQq$                        & \hspace{0.5in} 2.7\ (2.9)                  & 2.1\ (2.3)    \\
RS graviton  ($k/\overline{M}_\text{Pl}=0.1$)
             & $\PQq\PAQq$, $\Pg\Pg$                & \hspace{0.5in} 1.8\ (2.3)                  & 1.9\ (1.8)    \\
DM mediator  ($m_{\text{DM}}=1$~GeV) & $\PQq\PAQq$  & \hspace{0.5in} 2.6\ (2.5)                  & 2.0\ (2.0)    \\  \hline
 \end{tabular}
\label{tab:MassLimit}
\end{table}

\subsection{Limits on the coupling to quarks of a leptophobic \texorpdfstring{$\PZpr$}{Z'}}
\begin{figure}[hbt]
  \centering
    \includegraphics[width=0.48\textwidth]{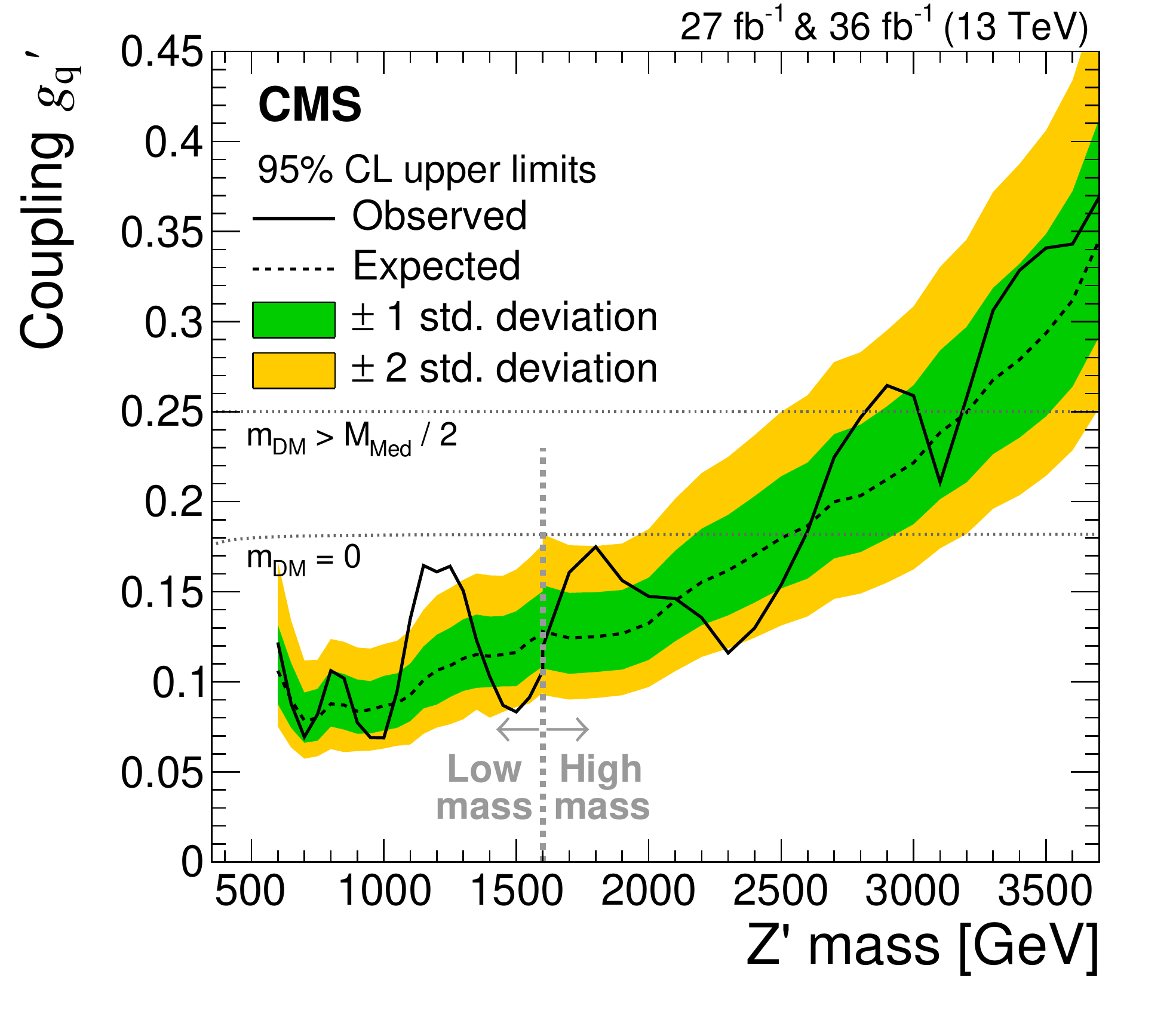}
\caption{  The 95\% \CL upper limits on the universal
quark coupling $\gq^{\prime}$ as a function of resonance mass for a leptophobic $\PZpr$ resonance that only couples to quarks.
The observed limits (solid), expected
limits (dashed) and their variation at the 1 and 2 standard deviation levels (shaded bands) are shown.
The dotted horizontal lines show the coupling strength for which the cross section for dijet production in this model is the same
as for a DM mediator (see text).}
    \label{figCoupling}
\end{figure}

Mass limits on new particles are sensitive to the assumptions about their coupling.
Furthermore, at a fixed resonance mass, as the search sensitivity increases we can exclude models with smaller couplings.
Figure~\ref{figCoupling} shows upper limits on the coupling as a function of
mass for a leptophobic $\PZpr$ resonance which has a
natural width
\begin{equation}
\Gamma = \frac{3(\gq^{\prime})^2 M}{2\pi}
\label{eqWidthZp}
\end{equation}
where $M$ is the resonance mass.
Limits are only shown in Fig.~\ref{figCoupling} for coupling values
$\gq^{\prime}<0.45$, corresponding to a width less than 10\% of the resonance mass, for which our narrow
resonance limits are approximately valid.
Up to this width value, for resonance masses less than roughly 4\TeV, the
Breit-Wigner natural line shape of the quark-quark resonance does not
significantly change the observed line shape, and the dijet resonance can be
considered effectively narrow. To constrain larger values of the
coupling we will consider broad resonances in Section~\ref{sec:Wide}.

\section{Limits on a dark matter mediator}
\begin{figure}[hbtp]
  \centering
    \includegraphics[width=0.75\textwidth]{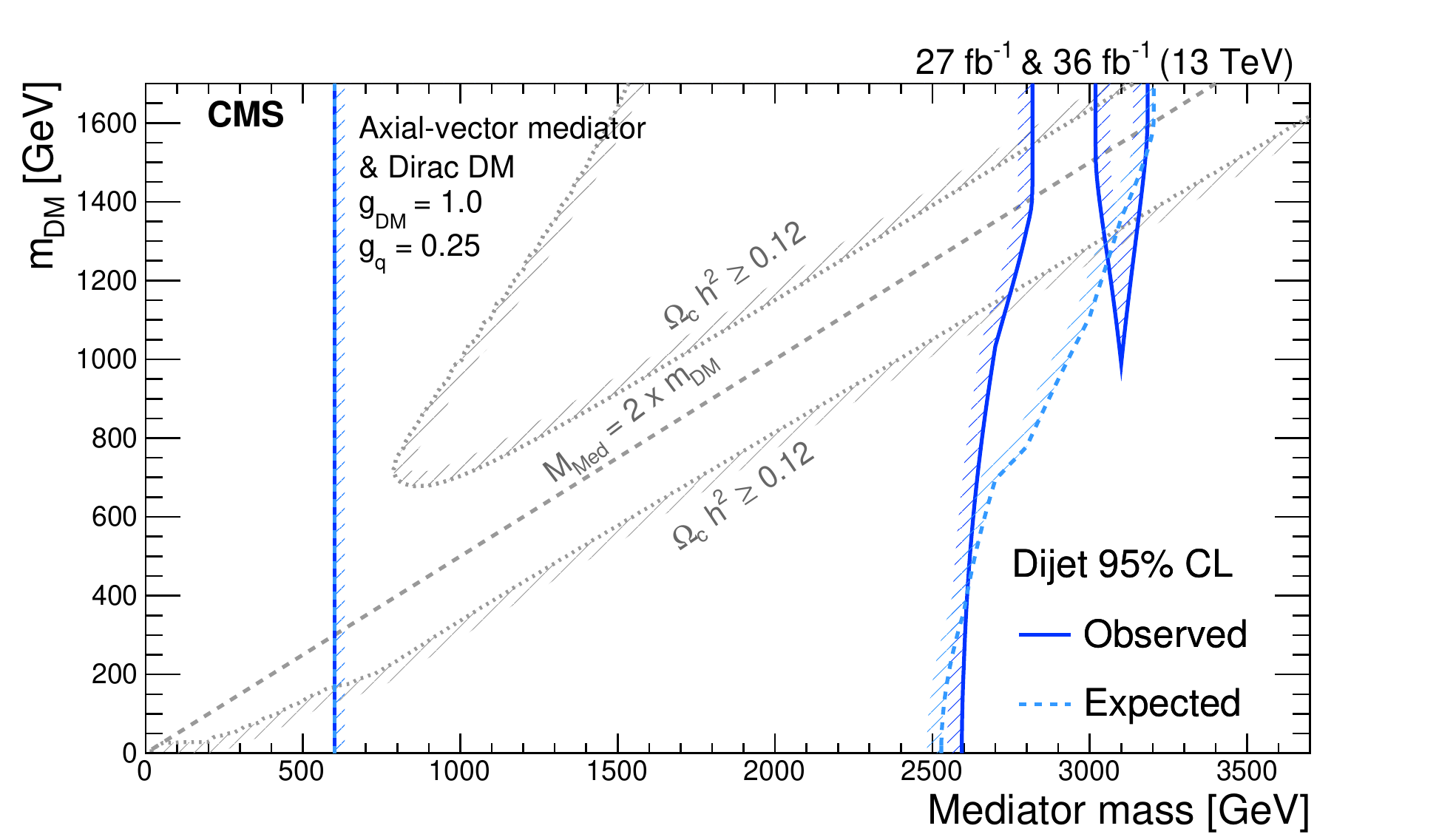}
    \includegraphics[width=0.75\textwidth]{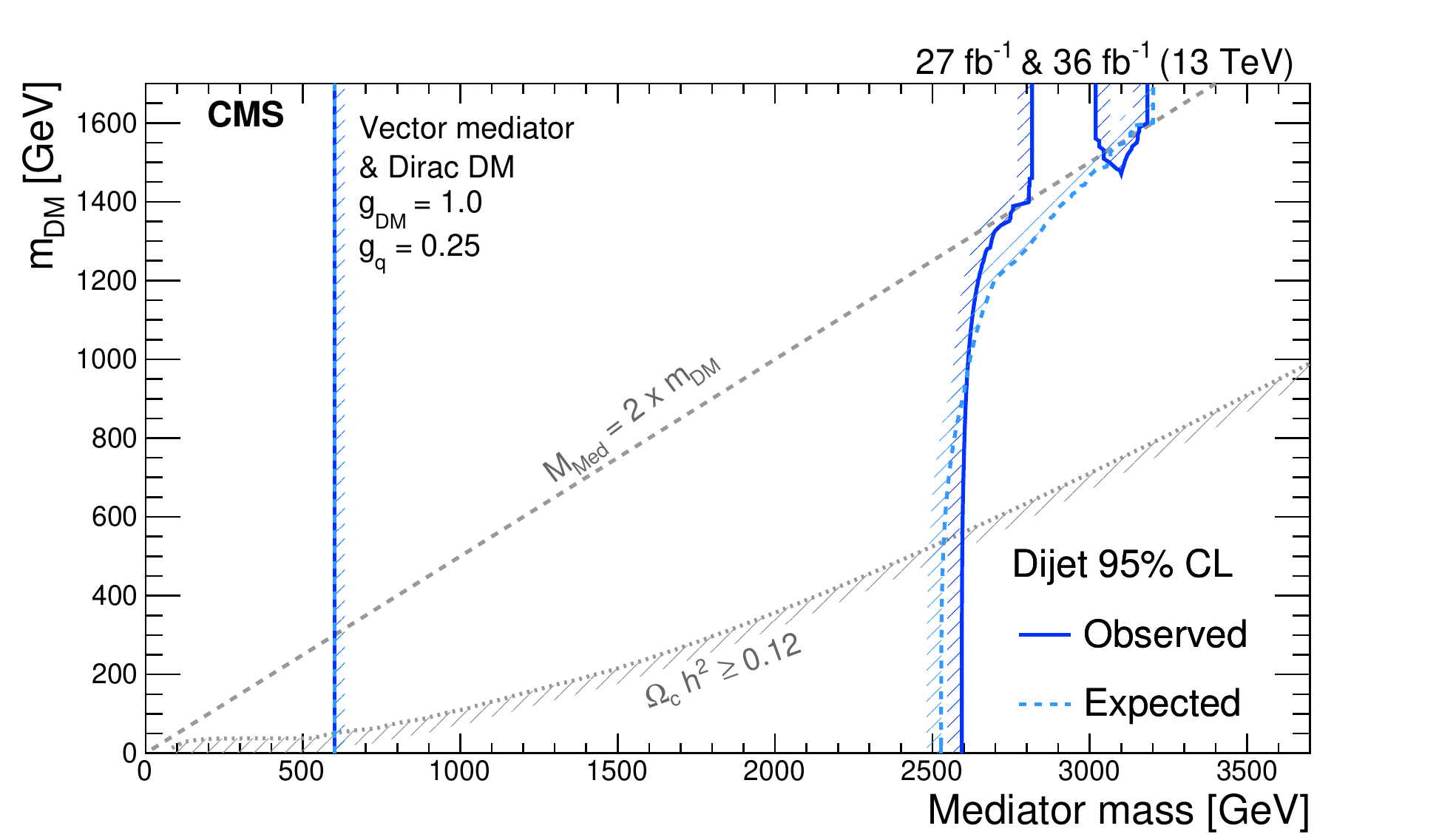}
\caption{  The 95\% \CL observed (solid) and expected (dashed) excluded regions in the plane of dark matter mass vs.
mediator mass, for an axial-vector mediator (upper) and a vector mediator (lower), compared to the excluded regions where
the abundance of DM exceeds the cosmological relic density (light gray). Following the recommendation of the
LHC DM working group~\cite{Boveia:2016mrp, Abdallah:2015ter}, the exclusions are computed for Dirac DM and for a universal quark coupling $\gq = 0.25$ and for a DM
coupling of $\gDM=1.0$. It should also be noted that the excluded region
strongly depends on the chosen coupling and model scenario. Therefore, the excluded regions and relic density contours shown in this plot are
not applicable to other choices of coupling values or models.}
    \label{figDM}
\end{figure}

We use our limits to constrain simplified models of DM, with leptophobic vector and axial-vector
mediators that couple only to quarks and DM particles~\cite{Boveia:2016mrp,Abdallah:2015ter}.
Figure~\ref{figDM} shows the excluded values of mediator mass as a function of \mDM,
for both types of mediators. For \mDM = 1\GeV the observed excluded range of the
mediator mass (\mMed) is between 0.6 and 2.6\TeV, as
also shown in Fig.~\ref{figLimitAll} and listed in Table~\ref{tab:MassLimit}. The limits on a dark matter mediator are
indistinguishable for $\mDM = 0$ and 1\GeV.
In Fig.~\ref{figDM} the expected upper value of excluded \mMed increases with \mDM because
the branching fraction to $\PQq\PAQq$ increases with \mDM.
In Fig.~\ref{figDM} our exclusions are compared to constraints from the cosmological
relic density of DM determined from astrophysical measurements~\cite{Spergel:2006hy,Ade:2013zuv} and from \textsc{MadDM} version
2.0.6~\cite{Backovic:2013dpa,Backovic:2015cra} as described in Ref.~\cite{Pree:2016hwc}.

\subsection{Relationship of the DM mediator model to the leptophobic \texorpdfstring{$\PZpr$}{Z'} model}
If $\mDM>\mMed/2$, the mediator cannot decay to DM particles ``on-shell", and the dijet cross section from the mediator
models~\cite{Boveia:2016mrp} becomes identical to that
in the leptophobic $\PZpr$ model ~\cite{Dobrescu:2013coa} used in Fig.~\ref{figCoupling} with a coupling $\gq^{\prime}=\gq=0.25$. Therefore, for these values of \mDM
the limits on the mediator mass in Fig.~\ref{figDM} are identical to the limits on the $\PZpr$ mass at $\gq^{\prime}=0.25$ in Fig.~\ref{figCoupling}.
Similarly, if $\mDM=0$, the limits on the mediator mass in Fig.~\ref{figDM} are identical to the limits on the $\PZpr$ mass at
$\gq^{\prime}=\gq/\sqrt{\smash[b]{1+16/(3N_f)}}\approx 0.182$ in Fig.~\ref{figCoupling}. Here $N_f$ is the effective number of quark flavors contributing
to the width of the resonance, $N_f=5+\sqrt{\smash[b]{1-4m_\cPqt^2/\mMed^2}}$, where $m_\cPqt$ is the top quark mass.

\subsection{Limits on the coupling to quarks of a narrow DM mediator}
\begin{figure}[htb]
  \begin{center}

  \includegraphics[width=0.75\textwidth]{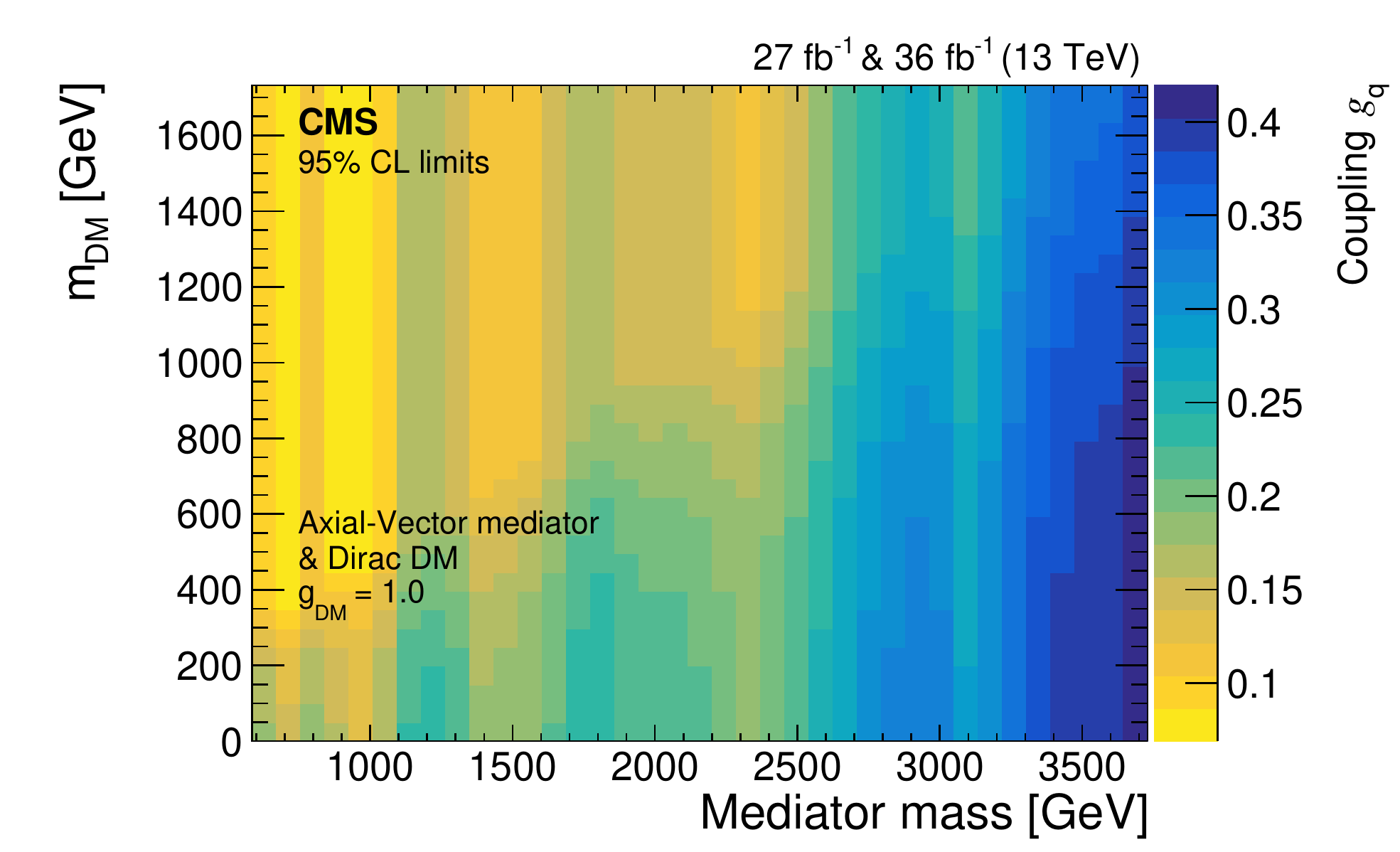}
  \includegraphics[width=0.75\textwidth]{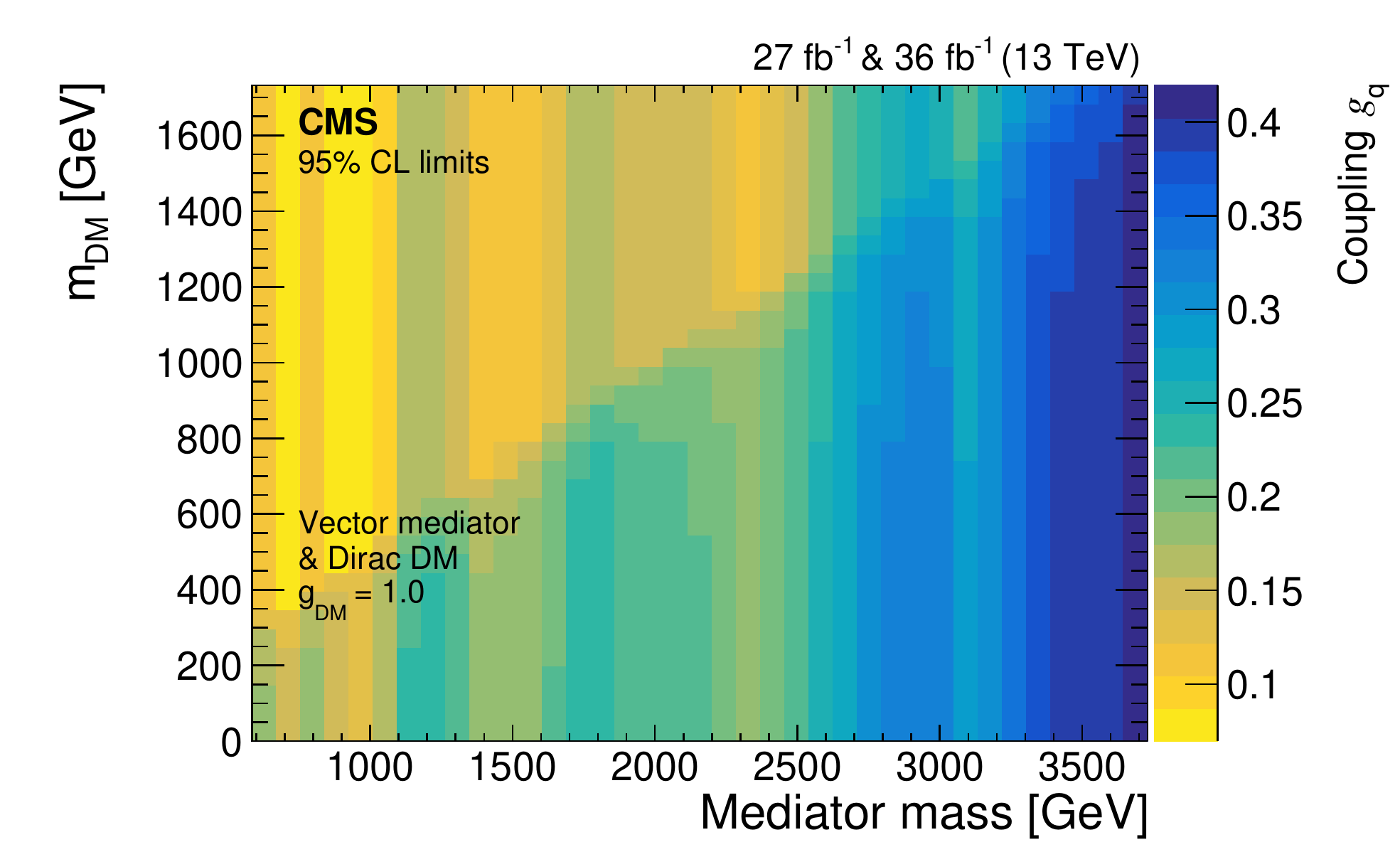}
  \caption{The 95\% \CL observed upper limits on a universal quark coupling \gq (color scale at right) in the plane of the dark matter
  particle mass versus mediator mass for an axial-vector mediator (upper) and a vector mediator (lower). }
         \label{fig:DMCouplingExclusion}
  \end{center}
\end{figure}

In Fig.~\ref{fig:DMCouplingExclusion} limits are presented on the coupling \gq as a function of \mDM and \mMed. The limits on
\gq decrease with increasing \mDM, again because the branching fraction to $\PQq\PAQq$ increases with \mDM. The minimum value of excluded \gq at
a fixed value of \mMed is obtained for \mDM greater than \mMed/2.

In Figs.~\ref{figCoupling} and \ref{fig:DMCouplingExclusion} we show exclusions from the narrow resonance search as a function of resonance mass and
quark coupling up to a maximum coupling value of approximately 0.4, corresponding to a maximum resonance mass of 3.7\TeV. At larger values of coupling
the natural width of the resonance influences significantly the observed width and our narrow resonance limits become
noticeably less accurate.  In the next
section we quantify more precisely the accuracy of our narrow-resonance limits, extend them to larger widths, and extend the limits on a dark matter mediator to
higher masses and couplings.

\clearpage

\section{Limits on broad resonances}
\label{sec:Wide}

The search for narrow resonances described in the previous sections assumes
the intrinsic resonance width $\Gamma$ is negligible compared to the experimental
dijet mass resolution.  Here we extend the search to cover broader resonances,
with the width up to 30\% of the resonance mass $M$. This
allows us to be sensitive to more models and larger couplings, and also
quantifies the level of approximation within the narrow-resonance search
by giving limits as an explicit function of $\Gamma/M$. We use the
same dijet mass data and background parameterization as in the high-mass
narrow resonance search.  The shapes of broad resonances are then used to
derive limits on such states decaying to $\Pq\Pq$ and $\Pg\Pg$.

\subsection{Breit--Wigner distributions}
The shape of a broad resonance depends on the relationship between the width and the resonance mass,
which in turn depends on the resonance spin and the decay channel.
The sub-process cross section for a resonance with mass $M$ as a function of di-parton
 mass $m$ is described by a relativistic Breit--Wigner (e.g. Eq.~(7.47) in Ref.~\cite{Sjostrand:2006za}):
\begin{equation}
\hat{\sigma} \propto \frac{\pi}{m^2} \,
\frac{[\Gamma^{(i)}M] \, [\Gamma^{(f)}M]}
{(m^2 - M^2)^2 + [\Gamma M]^2}\ ,
\label{eq:Breit-Wigner}
\end{equation}
where $\Gamma$ is the total width and $\Gamma^{(i,f)}$ are the partial widths for the initial state $i$ and final state $f$.
To obtain the correct expression when the di-parton mass is far from the
resonance mass, important for broad resonances,
generators like \PYTHIA~8 replace in Eq.~(\ref{eq:Breit-Wigner}) all $\Gamma M$ terms with $\Gamma(m) m$ terms, where $\Gamma(m)$ is the width
the resonance would have if its mass were $m$.  This general prescription for modifying the Breit--Wigner distribution is defined at Eq.~(47.58) in
Ref.~\cite{Patrignani:2016xqp}.
The replacement is done for the partial width terms in the numerator,
as well as the full width term in the denominator, and the resulting di-parton mass dependence within the numerator
significantly reduces the cross section at low values of $m$ far from the resonance pole.

We consider explicitly the shapes of spin-1 resonances in the quark-quark channel and the shape of spin-2 resonances in the
quark-quark and gluon-gluon channels.
For a spin-1 $\PZpr$ resonance in the quark-quark channel,
both for the CP-even vector and the CP-odd axial-vector cases, the partial width is proportional to the resonance mass ($\Gamma\propto M$)~\cite{Kim:2015vba} and generators make the
well known replacement
\begin{equation}
\Gamma M \to \left(\frac{m^2}{M^2}\right) \Gamma M
\label{eq:vectorReplacement}
\end{equation}
for the terms $[\Gamma^{(i)} M]$, $[\Gamma^{(f)} M]$ and $[\Gamma M]$ in Eq.~(\ref{eq:Breit-Wigner}). The factor $(m^2/M^2)$ in
Eq.~(\ref{eq:vectorReplacement})
converts the terms evaluated at the resonance mass to those evaluated at the di-parton mass for
the case of widths proportional to mass, as discussed at Eq.~(7.43) in Ref.~\cite{Sjostrand:2006za}.
For a spin-2 resonance,
a CP-even tensor such as a graviton, the partial widths in both the gluon-gluon channel~\cite{Kim:2015vba,Bijnens:2001gh} and the
quark-quark channel~\cite{Bijnens:2001gh} are proportional to the resonance mass cubed ($\Gamma\propto M^3$) and \PYTHIA~8 makes the following
replacement for an RS graviton:
\begin{equation}
\Gamma M \to \left(\frac{m^4}{M^4}\right) \Gamma M
\label{eq:tensorReplacement}
\end{equation}
for the above mentioned terms. The factor $(m^4/M^4)$ in
Eq.~(\ref{eq:tensorReplacement})
converts the terms evaluated at the resonance mass to those evaluated at the di-parton mass for
the case of widths proportional to mass cubed.

Applying the replacements in Eq.~(\ref{eq:vectorReplacement}) and (\ref{eq:tensorReplacement}) to the $[\Gamma^{(i)} M][\Gamma^{(f)} M]$ in the numerator of the Breit--Wigner
distribution results in an extra factor of $(m^2/M^2)(m^2/M^2)=m^4/M^4$ for a spin-2 resonance compared to a spin-1 resonance decaying to dijets.
At low di-parton mass, $m\ll M$, the replacement in the denominator does not matter, and the replacement in the numerator
will suppress the tail at low $m$ for spin-2 resonances compared to spin-1 resonances, as can be seen in the figures in the next section.
At high diparton mass, $m\gg M$, the replacement in the denominator will tend to cancel the replacement in the numerator and the high
mass tail is not significantly affected by the replacement.
This is true for the dijet decays of all spin-2 resonances calculated within
effective field theory~\cite{Kim:2015vba,Han:1998sg}. We note that spin-2 resonances decaying to dijets are required to be
CP-even, because the dijet decays of any spin-2 CP-odd resonances are suppressed~\cite{Kim:2015vba}.

Spin-0 resonances coupling directly to pairs of gluons
(e.g. color-octet scalars) or to pairs of gluons through fermion loops (e.g. Higgs-like bosons) will have a partial
width proportional to the resonance mass cubed~\cite{Chivukula:2014pma,Kim:2015vba,Ellis:1975ap}
and should have a similar shape as a spin-2 resonance in the gluon-gluon channel.
Spin-0 resonances coupling to quark-quark (e.g. Higgs-like bosons or scalar diquarks) will have
a partial width proportional to the resonance mass~\cite{Ellis:1975ap,Cakir:2005iw} and should have a similar shape as a spin-1 resonance in the
quark-quark channel. Therefore, the three shapes we consider in Section~\ref{sec:wideShape}, for spin-2 resonances coupling to quark-quark and gluon-gluon and for spin-1 resonances coupling to
quark-quark, are sufficient to determine the shapes of all broad resonances decaying to quark-quark or gluon-gluon. We do not consider
broad resonances with non-integer spin decaying to quark-gluon in this paper.
Further discussion of the model dependence of the
shape of broad resonances can be found in the Appendix of Ref.~\cite{Khachatryan:2015sja}.

\subsection{Resonance signal shapes and limits}
\label{sec:wideShape}
\begin{figure}[htb]
  \begin{center}

  \includegraphics[width=0.48\textwidth]{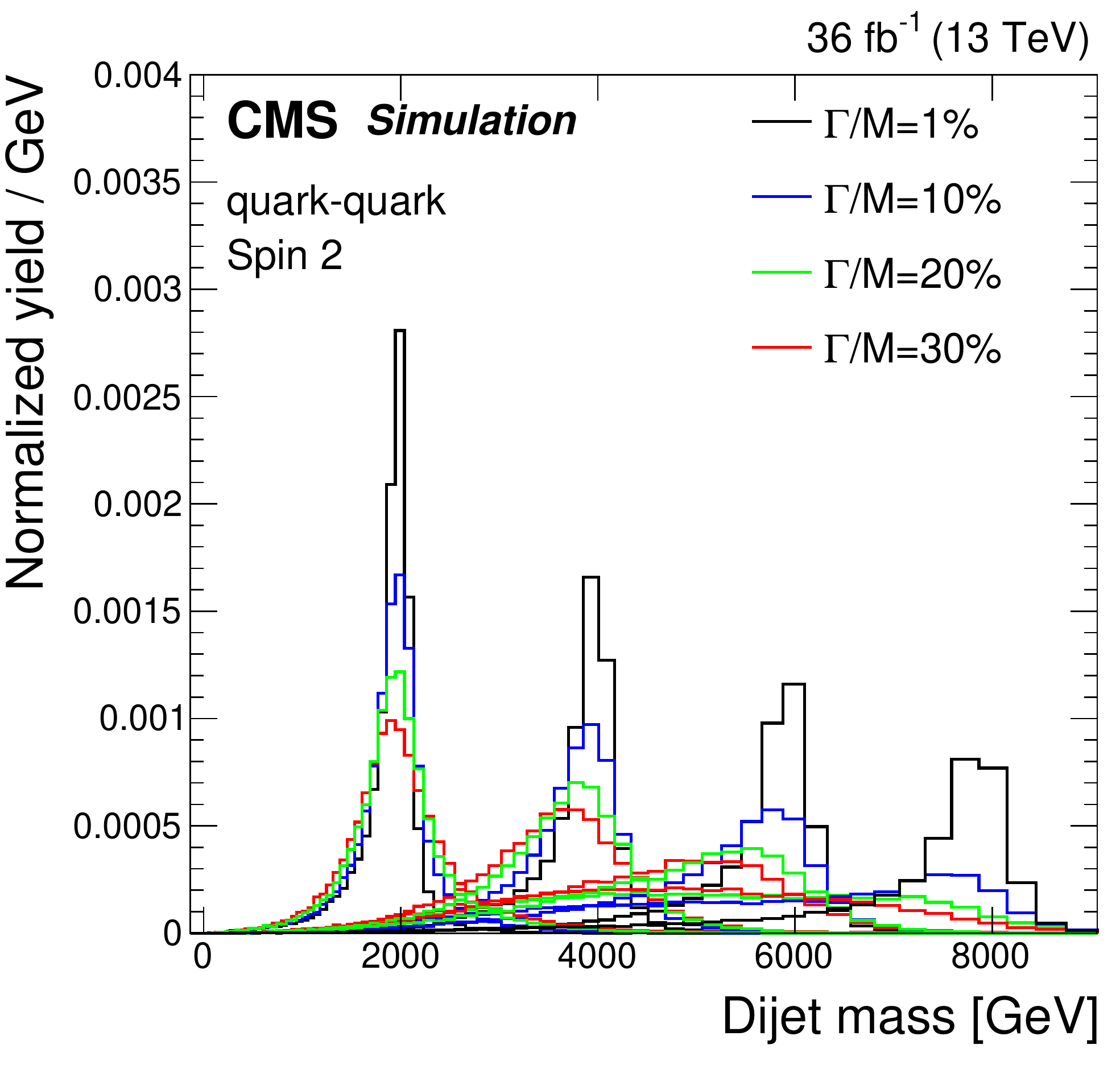}
  \includegraphics[width=0.48\textwidth]{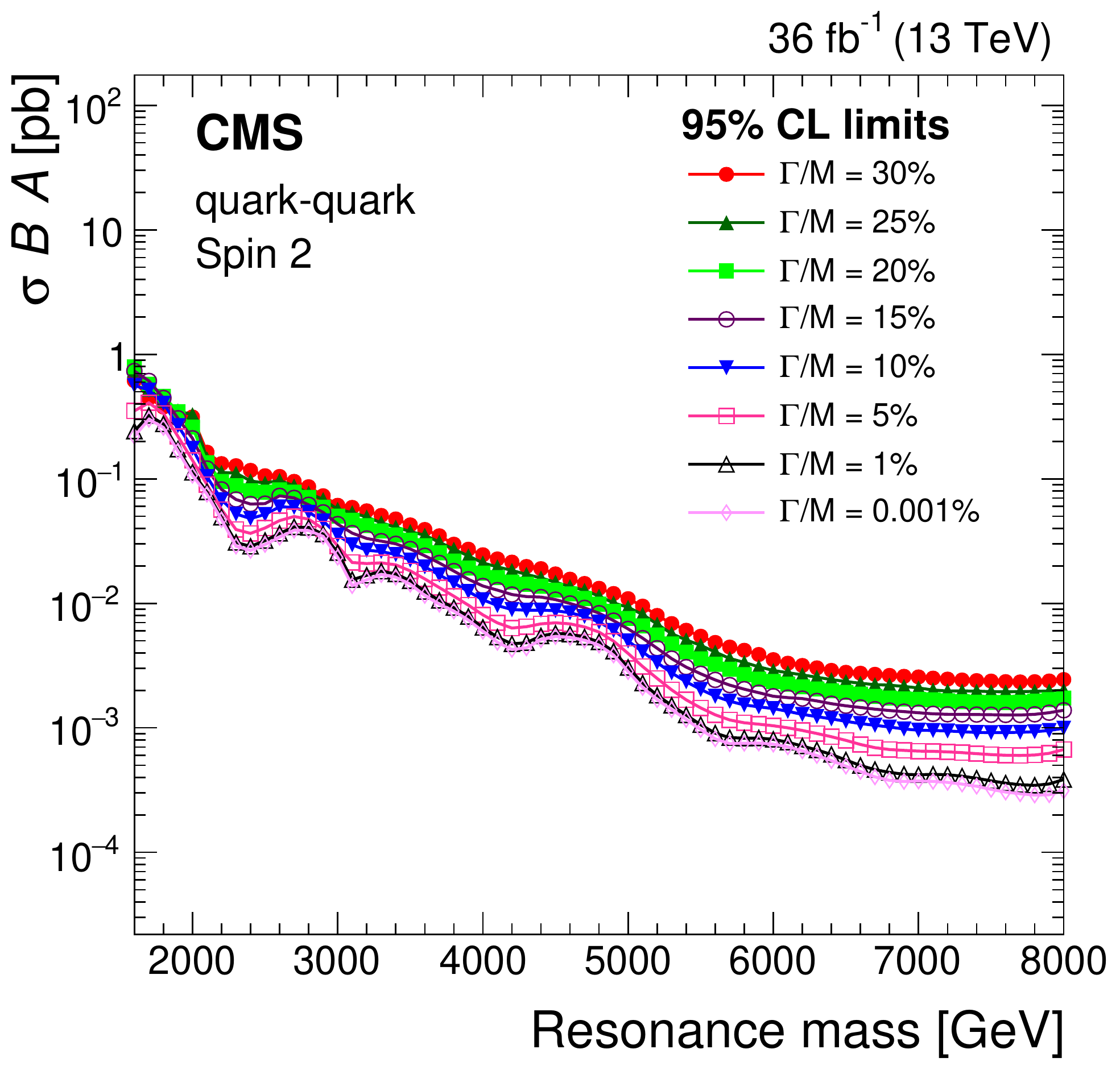}
  \caption{
The resonance signal shapes (\cmsLeft) and observed 95\% \CL upper limits on the product of the cross section, branching fraction, and acceptance (\cmsRight)
for spin-2 resonances produced and decaying in the quark-quark channel are shown for various values of intrinsic width and resonance mass.
The reconstructed dijet mass spectrum for these resonances is estimated from the {\PYTHIA8} MC event generator, followed by the simulation of the CMS detector response.
}
         \label{fig:wide_qq}
  \end{center}
\end{figure}

\begin{figure}[htb]
  \begin{center}

  \includegraphics[width=0.48\textwidth]{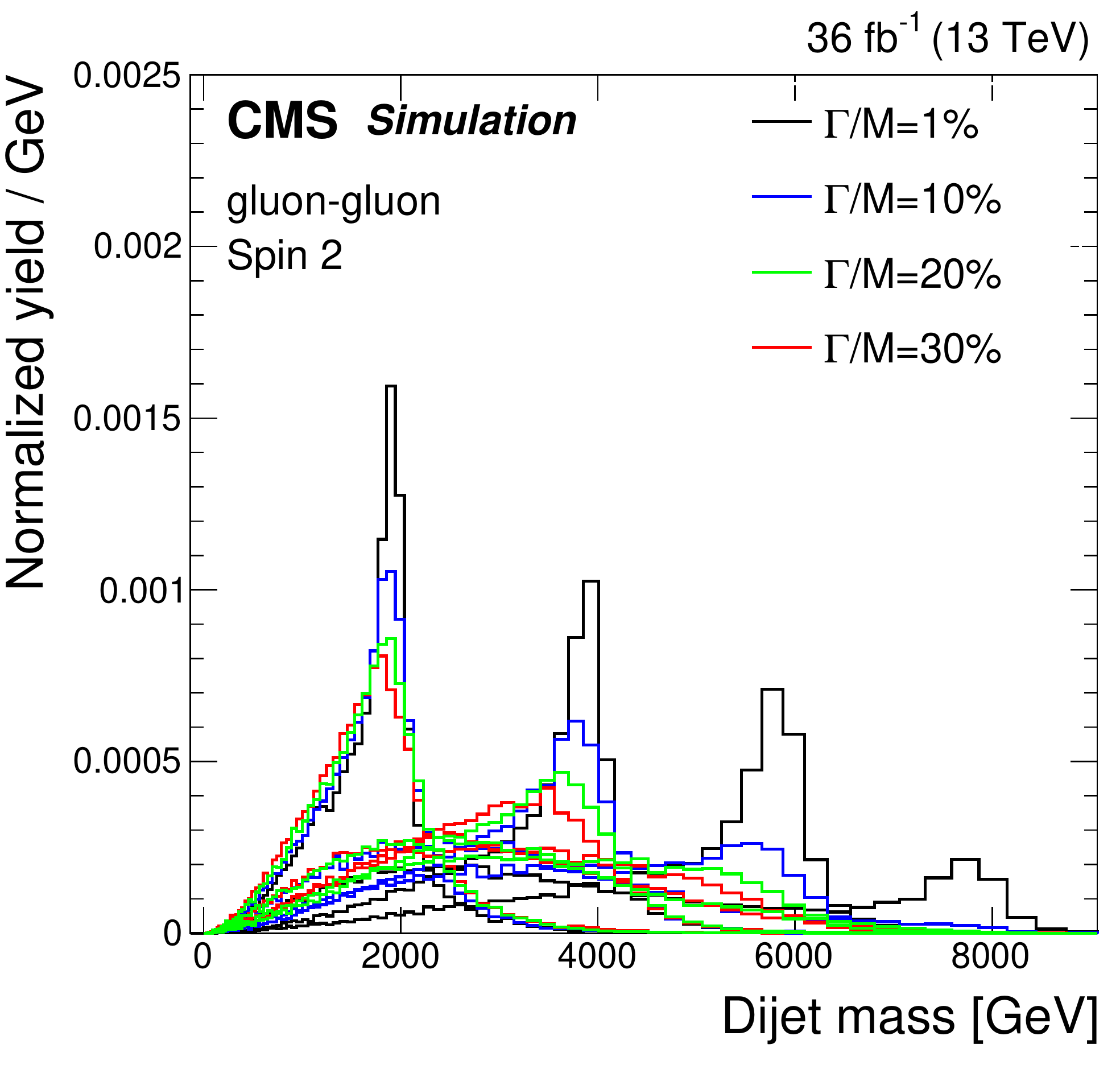}
  \includegraphics[width=0.48\textwidth]{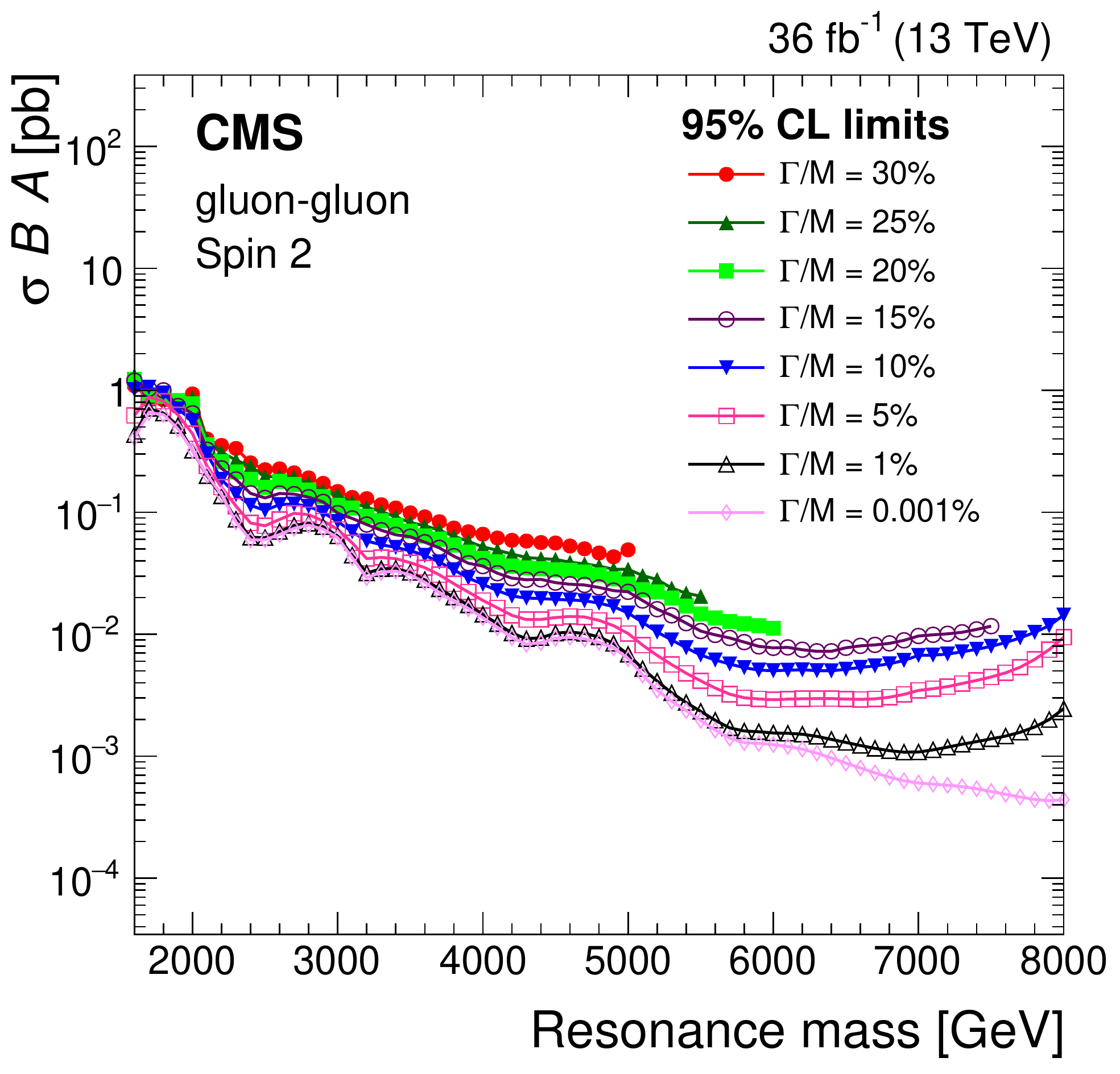}
  \caption{
The resonance signal shapes (\cmsLeft) and observed 95\% \CL upper limits on the product of the cross section, branching fraction, and acceptance (\cmsRight)
for spin-2 resonances produced and decaying in the gluon-gluon channel are shown for various values of intrinsic width and resonance mass.
The reconstructed dijet mass spectrum for these resonances is estimated from the {\PYTHIA8} MC event generator, followed by the simulation of the CMS detector response.
}
         \label{fig:wide_gg}
  \end{center}
\end{figure}

\begin{figure}[htb]
  \begin{center}

  \includegraphics[width=0.48\textwidth]{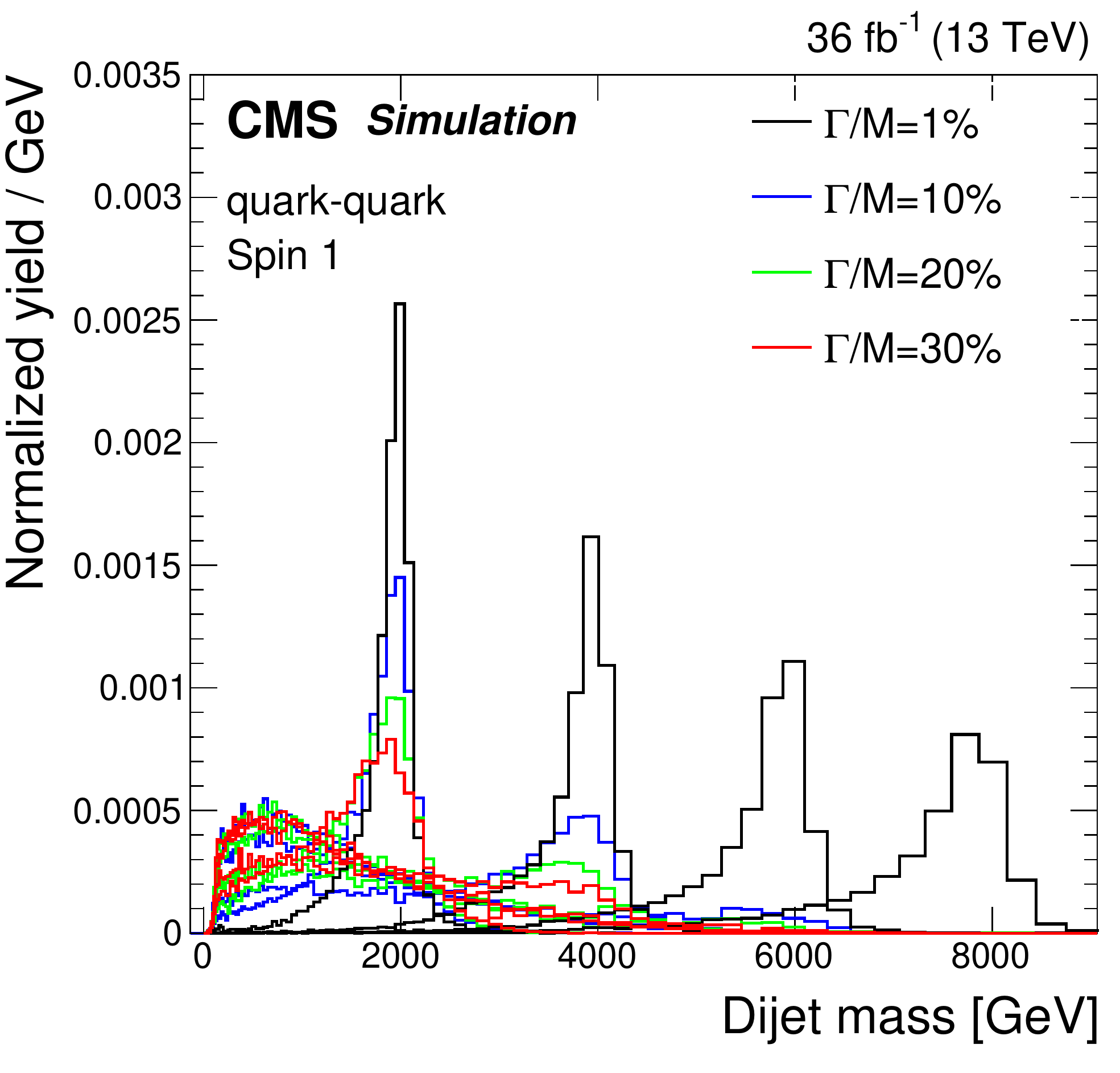}
  \includegraphics[width=0.48\textwidth]{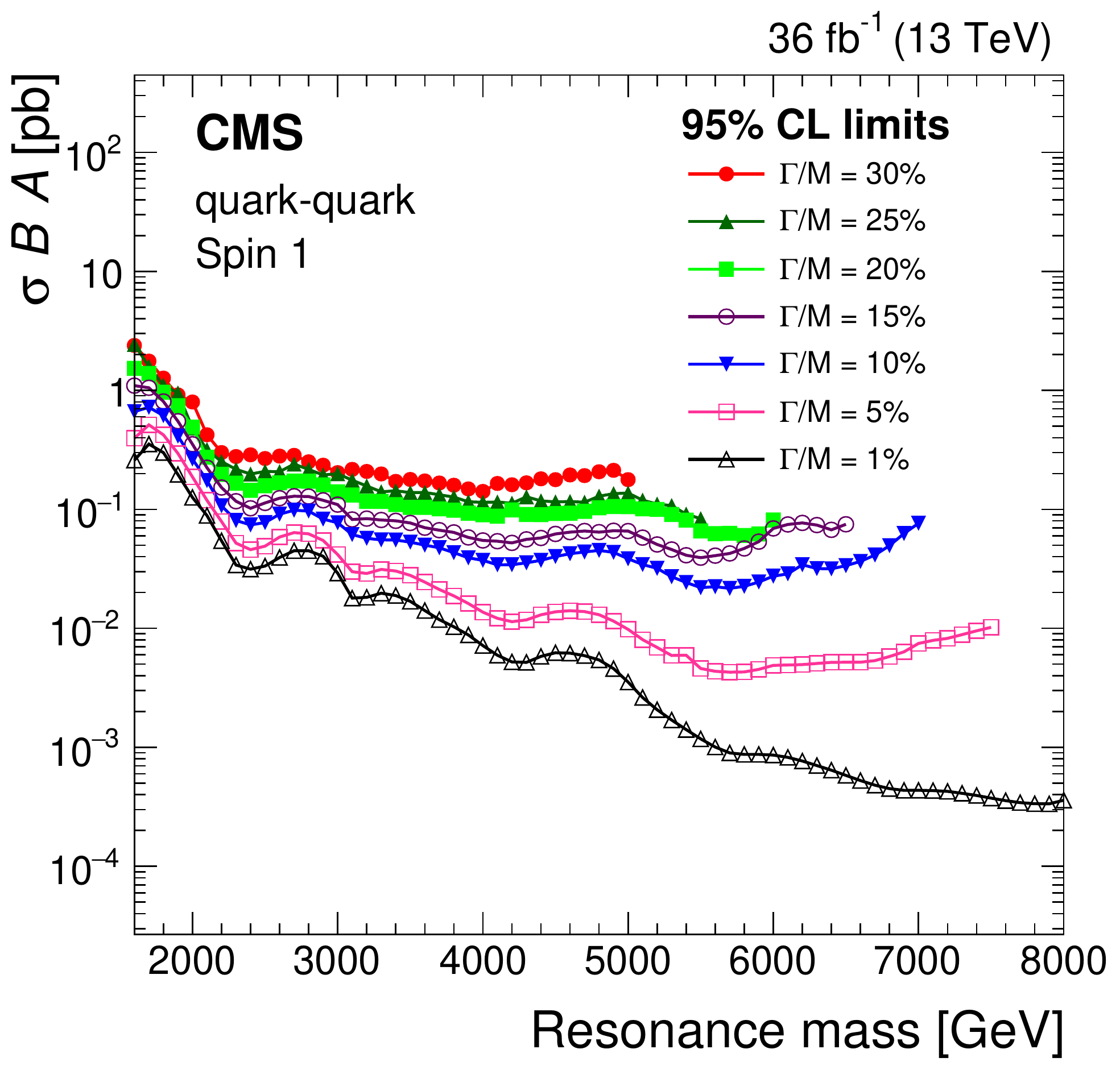}
  \caption{
The resonance signal shapes (\cmsLeft) and observed 95\% \CL upper limits on the product of the cross section, branching fraction, and acceptance (\cmsRight)
for spin-1 resonances produced and decaying in the quark-quark channel are shown for various values of intrinsic width and resonance mass.
The reconstructed dijet mass spectrum for these resonances is estimated from the {\PYTHIA8} MC event generator, followed by the simulation of the CMS detector response.
}
         \label{fig:wide_vector}
  \end{center}
\end{figure}

In Figs.~\ref{fig:wide_qq} and ~\ref{fig:wide_gg} we show resonance signal shapes and observed CMS
limits for various widths of spin-2 resonances modeled by an RS graviton signal in the quark-quark and
gluon-gluon channels, respectively.
The limits become less stringent as the resonance intrinsic
width increases. While the extra factor of $m^4/M^4$ in the Breit--Wigner distribution discussed in the previous section suppresses the tail at low dijet mass for $\Pq\Pq$ resonances, increased QCD radiation and a longer tail due to parton distributions partially compensates this
effect for $\Pg\Pg$ resonances. As a consequence and similar to narrow
resonances, the broad resonances decaying to $\Pg\Pg$ have a more
pronounced tail at low mass, and hence the limits for these resonances are weaker than those for resonances decaying to $\Pq\Pq$.
In Fig.~\ref{fig:wide_vector} we show the signal shapes and limits for spin-1 resonances in the quark-quark channel.
The spin-1 resonances in Fig.~\ref{fig:wide_vector} do not contain the extra factor of $m^4/M^4$ in the Breit--Wigner distribution and
are therefore significantly broader than the spin-2 $\Pq\Pq$ resonances in Fig.~\ref{fig:wide_qq}.
For the same reason, the limits in Fig.~\ref{fig:wide_vector} are weaker than those in Fig.~\ref{fig:wide_qq}. The difference in
the angular distribution of spin-1 and spin-2 resonances has a negligible effect on the resonance shapes and the cross section upper limits.
In Fig.~\ref{fig:wide_vector} we use a model of a vector DM mediator, and find the signal shapes and limits
indistinguishable from an axial-vector model.

\subsection{Validity tests of the limits}
The limits are calculated up to a resonance mass of 8\TeV but are only quoted up to the maximum resonance mass
for which the presence of the low-mass tails in the signal shape does not significantly affect the limit value.
For these quoted values, the limits on the resonance cross section are well understood, increasing monotonically
as a function of resonance width at each value of resonance mass.  To obtain this behavior in the limit, we find it is sufficient
to require that the expected limit derived for a truncated shape
agrees with that derived for the full shape within 15\%. The truncated shape is cut off at a dijet mass equal to 70\% of the nominal
resonance mass, while the full shape is cut off at a dijet mass of \RECOminMjjCut.  For both the truncated and the full limits, the
cross section limit of the resonance signal is corrected for the acceptance of this requirement on the dijet mass in order to obtain
limits on the total signal cross section.  The difference between the expected limits using the full shape and the truncated shape is
negligible for most resonance masses and widths, because the signal tail at low mass is insignificant compared to the steeply falling
background.  For some resonance masses beyond our maximum, the low dijet mass tail causes the limit to behave in an unphysical manner
as a function of increasing width. This condition does not affect the maximum resonance mass presented
for a spin-2 $\Pq\Pq$ resonance in Fig.\ref{fig:wide_qq}, but it does restrict the maximum masses presented for a
spin-2 $\Pg\Pg$ resonance in Fig.\ref{fig:wide_gg} and a vector resonance in Fig.\ref{fig:wide_vector}.
For example, for a vector resonance, we find that the highest resonance
mass that satisfies this condition is 5\TeV for a resonance with 30\% width, 6\TeV for
20\% width, 7\TeV for 10\% width, and 8\TeV for a narrow resonance.  It is useful to define the signal pseudo-significance
distribution $\text{S}/\sqrt{\text{B}}$ where S is the resonance signal and B is the QCD background.
The signal pseudo-significance indicates sensitivity to the signal in the presence of background as a function of dijet mass,
and has been used as an alternative method of evaluating the sensitivity of the search to the low mass tail.
The maximum resonance mass values we present correspond to a 70\% acceptance for the signal pseudo-significance,
when the signal shape is truncated at 70\% of the nominal resonance mass. This demonstrates that, for resonance masses
and widths which satisfy our resonance mass condition, the signals are being constrained mainly by data in the dijet mass region near the resonance pole.
Signal injection tests analogous to those already described for the narrow resonance search were
repeated for the broad resonance search, and the bias in the extracted signal was again found to be negligible.
As discussed in the previous CMS search for broad
dijet resonances~\cite{Khachatryan:2015sja}, our signal shapes consider only the $s$-channel process,
which dominates the signal, and our results do not include the possible effects of the $t$-channel exchange
of a new particle or the interference between the background and signal processes.

\subsection{Limits on the coupling to quarks of a broad DM mediator}
The cross section limits in Fig.~\ref{fig:wide_vector} have been used to derive constraints on a DM mediator.
The cross section for mediator production for $\mDM=1$ GeV and $\gDM=1$ is calculated at leading order using \MADGRAPH{5}\_a\MCATNLO version 2.3.2~\cite{Alwall:2014hca} for mediator masses within the range $1.6 < \mMed < 4.1$\TeV in
0.1\TeV steps and for quark couplings within the range $0.1<\gq<1.0$ in 0.1 steps. For these choices the
relationship between the width and $\gq$ given in Ref.~\cite{Boveia:2016mrp,Abdallah:2015ter} simplifies to
\begin{equation}
\Gamma_{\text{Med}} \approx \frac{(18\gq^2 + 1)\mMed}{12\pi},
\label{eqWidth}
\end{equation}
for both vector and axial-vector mediators.

\begin{figure}[hbt]
  \centering
    \includegraphics[width=0.48\textwidth]{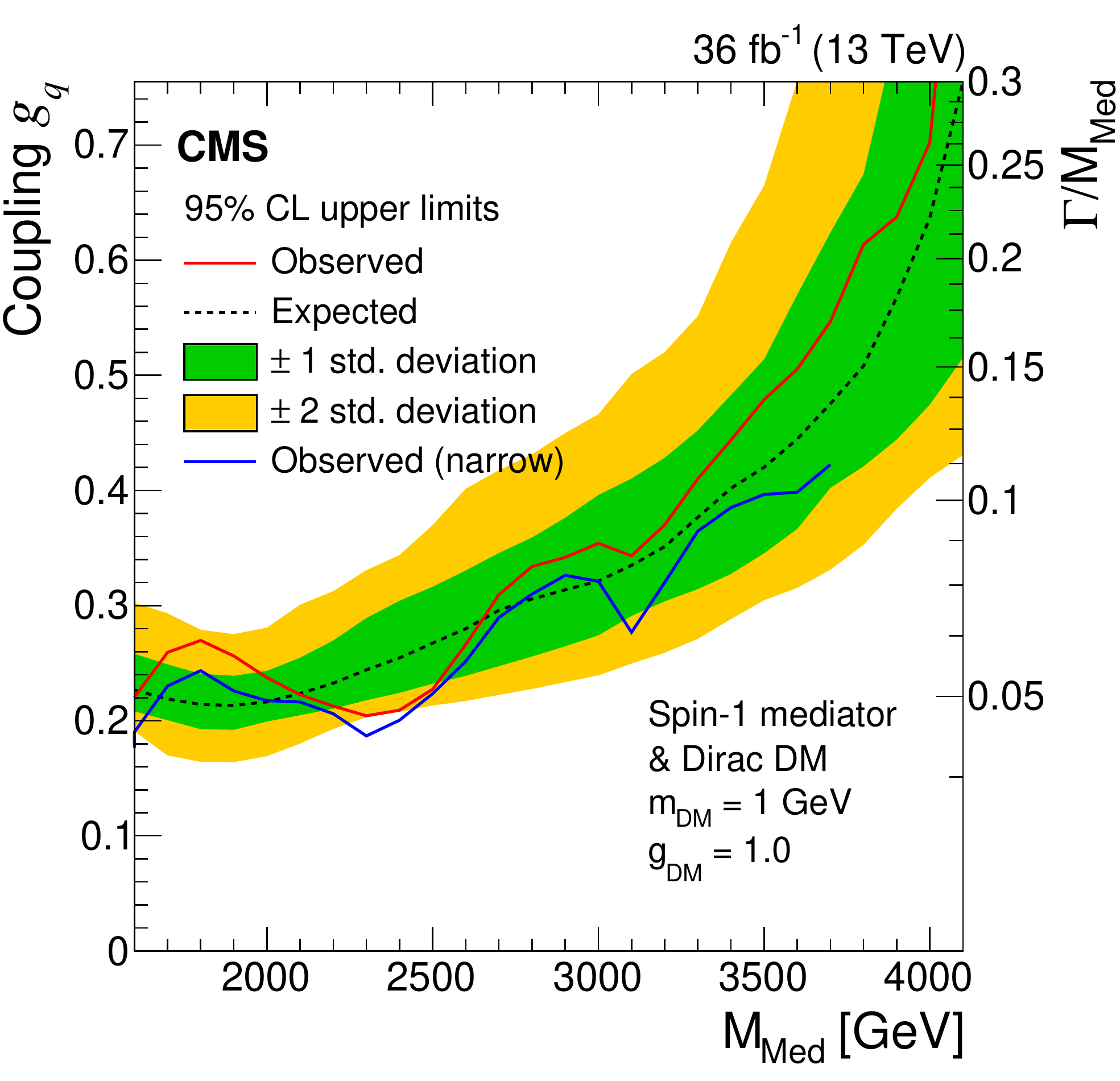}
\caption{  The 95\% \CL upper limits on the universal
quark coupling $\gq$ as a function of resonance mass for a vector mediator of interactions between quarks and DM particles.
The right vertical axis shows the natural width of the mediator divided by its mass.
The observed limits taking into account the natural width of the resonance are in red(upper
solid curve), expected
limits (dashed) and their variation at the 1 and 2 standard deviation levels (shaded bands) are shown.
The observed limits from the narrow resonance search are in blue (lower solid curve),
but are only valid for the width values up to approximately 10\% of the resonance mass.
The exclusions are computed for a spin-1 mediator and, Dirac DM particle with a mass $\mDM=1$\GeV and a coupling $\gDM=1.0$.}
    \label{figCouplingWide}
\end{figure}

For each mediator mass value, the predictions for the cross section for mediator production as
a function of $\gq$ are converted to a function of the width, using Eq.~(\ref{eqWidth}), and are then compared to our cross section limits
from Fig.~\ref{fig:wide_vector} to find
the excluded values of $\gq$ as a function of mass for a spin-1 resonance shown in Fig.~\ref{figCouplingWide}.
Also shown in Fig.~\ref{figCouplingWide} is the limit on $\gq$ from the quark-quark narrow
resonance shape we used in the previous sections to set narrow-resonance limits. These are equal to the limits on $\gq$ in Fig.~\ref{fig:DMCouplingExclusion}
and are derived from the limits on $\gq^{\prime}$ in Fig.~\ref{figCoupling} using the formula
\begin{equation}
\gq = \gq^{\prime} \sqrt{\frac{1}{2} + \sqrt{\frac{1}{4} + \frac{1}{18(\gq^{\prime})^2}}}.
\label{eqDMcoupling}
\end{equation}
Equation~(\ref{eqDMcoupling}) is applicable for a narrow mediator with $\gDM=1$ and mass much larger than the quark and DM particle masses.
The quark-quark narrow-resonance limits are derived from a
narrow spin-2 resonance shape, which is approximately the same as a spin-1 resonance shape for small values of $\gq$, and therefore in Fig.~\ref{figCouplingWide} at
small values of $\gq$ the narrow-resonance limits are roughly the same as the limits which take into account the width of the resonance.  For resonance masses
smaller than about 2.5\TeV, the acceptance of the dijet mass requirement ${\mjj>1.25}$\TeV is reduced by taking into account the
resonance natural width, resulting in a small increase in the limits compared to the narrow-resonance limits,
which can be seen in Fig.~\ref{figCouplingWide}. At 3.7\TeV,
the largest value of the resonance mass considered approximately valid for the narrow-resonance limits on \gq, the narrow-resonance limit is
${\gq>0.42}$, while the more accurate limit taking into account the width for the spin-1 resonance is $\gq>0.53$.  The limits taking into account the
natural width can be calculated up to a resonance mass of 4.1\TeV for a width up to 30\% of the resonance mass.
The limits from the narrow resonance search are approximately valid up to coupling values of about 0.4, corresponding to a width of 10\%, while the
limits taking into account the natural width of the resonance probe up to a coupling value of 0.76, corresponding to a natural width of 30\%.
We conclude that these limits on a vector DM mediator, taking into account the natural width of the resonance, improve on the accuracy of
the narrow-width limits and extend them to larger values of the resonance mass and coupling to quarks.

\section{Summary}

Searches have been presented for resonances decaying into pairs of jets using proton-proton
collision data collected at $\sqrt{s} = 13$\TeV corresponding to an integrated luminosity of up to \RunLumi.
A low-mass search, for resonances with masses between 0.6 and 1.6\TeV, is performed based on events with dijets reconstructed at
the trigger level from calorimeter information. A high-mass search, for resonances with masses above 1.6\TeV, is performed using
dijets reconstructed offline with a particle-flow algorithm.
The dijet mass spectra are observed to be smoothly falling
distributions. In the analyzed data samples, there is no evidence for resonant particle production.
Generic upper limits are presented on the product of the cross section, the branching fraction to dijets, and the acceptance for narrow quark-quark, quark-gluon, and gluon-gluon
resonances that are applicable to any model of narrow dijet resonance production.
String resonances with masses below 7.7\TeV are excluded at 95\% confidence level, as are scalar diquarks below 7.2\TeV,
axigluons and colorons below 6.1\TeV, excited quarks below 6.0\TeV, color-octet
scalars below 3.4\TeV, $\PWpr$ bosons with the SM-like couplings below 3.3\TeV, $\PZpr$ bosons with the SM-like couplings below 2.7\TeV,
Randall--Sundrum gravitons below 1.8\TeV and in the range 1.9 to 2.5\TeV, and dark matter mediators below 2.6\TeV.
The limits on both vector and axial-vector mediators, in a simplified model of interactions between quarks and dark matter particles,
are presented as functions of dark matter particle mass.
Searches are also presented for broad resonances, including for the first time spin-1 resonances with intrinsic widths as large
as 30\% of the resonance mass. The broad resonance search improves and extends the exclusions of a dark matter mediator to larger values of its mass and coupling to quarks.
The narrow and broad resonance searches extend limits previously reported by CMS in the dijet channel, resulting in the most stringent constraints on many of the
models considered.

\begin{acknowledgments}
\hyphenation{Bundes-ministerium Forschungs-gemeinschaft Forschungs-zentren Rachada-pisek} We congratulate our colleagues in the CERN accelerator departments for the excellent performance of the LHC and thank the technical and administrative staffs at CERN and at other CMS institutes for their contributions to the success of the CMS effort. In addition, we gratefully acknowledge the computing centers and personnel of the Worldwide LHC Computing Grid for delivering so effectively the computing infrastructure essential to our analyses. Finally, we acknowledge the enduring support for the construction and operation of the LHC and the CMS detector provided by the following funding agencies: the Austrian Federal Ministry of Science, Research and Economy and the Austrian Science Fund; the Belgian Fonds de la Recherche Scientifique, and Fonds voor Wetenschappelijk Onderzoek; the Brazilian Funding Agencies (CNPq, CAPES, FAPERJ, and FAPESP); the Bulgarian Ministry of Education and Science; CERN; the Chinese Academy of Sciences, Ministry of Science and Technology, and National Natural Science Foundation of China; the Colombian Funding Agency (COLCIENCIAS); the Croatian Ministry of Science, Education and Sport, and the Croatian Science Foundation; the Research Promotion Foundation, Cyprus; the Secretariat for Higher Education, Science, Technology and Innovation, Ecuador; the Ministry of Education and Research, Estonian Research Council via IUT23-4 and IUT23-6 and European Regional Development Fund, Estonia; the Academy of Finland, Finnish Ministry of Education and Culture, and Helsinki Institute of Physics; the Institut National de Physique Nucl\'eaire et de Physique des Particules~/~CNRS, and Commissariat \`a l'\'Energie Atomique et aux \'Energies Alternatives~/~CEA, France; the Bundesministerium f\"ur Bildung und Forschung, Deutsche Forschungsgemeinschaft, and Helmholtz-Gemeinschaft Deutscher Forschungszentren, Germany; the General Secretariat for Research and Technology, Greece; the National Research, Development and Innovation Fund, Hungary; the Department of Atomic Energy and the Department of Science and Technology, India; the Institute for Studies in Theoretical Physics and Mathematics, Iran; the Science Foundation, Ireland; the Istituto Nazionale di Fisica Nucleare, Italy; the Ministry of Science, ICT and Future Planning, and National Research Foundation (NRF), Republic of Korea; the Lithuanian Academy of Sciences; the Ministry of Education, and University of Malaya (Malaysia); the Mexican Funding Agencies (BUAP, CINVESTAV, CONACYT, LNS, SEP, and UASLP-FAI); the Ministry of Business, Innovation and Employment, New Zealand; the Pakistan Atomic Energy Commission; the Ministry of Science and Higher Education and the National Science center, Poland; the Funda\c{c}\~ao para a Ci\^encia e a Tecnologia, Portugal; JINR, Dubna; the Ministry of Education and Science of the Russian Federation, the Federal Agency of Atomic Energy of the Russian Federation, Russian Academy of Sciences and the Russian Foundation for Basic Research; the Ministry of Education, Science and Technological Development of Serbia; the Secretar\'{\i}a de Estado de Investigaci\'on, Desarrollo e Innovaci\'on, Programa Consolider-Ingenio 2010, Plan Estatal de Investigaci\'on Cient\'{\i}fica y T\'ecnica y de Innovaci\'on 2013-2016, Plan de Ciencia, Tecnolog\'{i}a e Innovaci\'on 2013-2017 del Principado de Asturias and Fondo Europeo de Desarrollo Regional, Spain; the Swiss Funding Agencies (ETH Board, ETH Zurich, PSI, SNF, UniZH, Canton Zurich, and SER); the Ministry of Science and Technology, Taipei; the Thailand Center of Excellence in Physics, the Institute for the Promotion of Teaching Science and Technology of Thailand, Special Task Force for Activating Research and the National Science and Technology Development Agency of Thailand; the Scientific and Technical Research Council of Turkey, and Turkish Atomic Energy Authority; the National Academy of Sciences of Ukraine, and State Fund for Fundamental Researches, Ukraine; the Science and Technology Facilities Council, UK; the US Department of Energy, and the US National Science Foundation.

Individuals have received support from the Marie-Curie program and the European Research Council and Horizon 2020 Grant, contract No. 675440 (European Union); the Leventis Foundation; the A. P. Sloan Foundation; the Alexander von Humboldt Foundation; the Belgian Federal Science Policy Office; the Fonds pour la Formation \`a la Recherche dans l'Industrie et dans l'Agriculture (FRIA-Belgium); the Agentschap voor Innovatie door Wetenschap en Technologie (IWT-Belgium); the F.R.S.-FNRS and FWO (Belgium) under the ``Excellence of Science - EOS" - be.h project n. 30820817; the Ministry of Education, Youth and Sports (MEYS) of the Czech Republic; the Lend\"ulet (``Momentum") program and the J\'anos Bolyai Research Scholarship of the Hungarian Academy of Sciences, the New National Excellence Program \'UNKP, the NKFIA research grants 123842, 123959, 124845, 124850 and 125105 (Hungary); the Council of Scientific and Industrial Research, India; the HOMING PLUS program of the Foundation for Polish Science, cofinanced from European Union, Regional Development Fund, the Mobility Plus program of the Ministry of Science and Higher Education, the National Science Center (Poland), contracts Harmonia 2014/14/M/ST2/00428, Opus 2014/13/B/ST2/02543, 2014/15/B/ST2/03998, and 2015/19/B/ST2/02861, Sonata-bis 2012/07/E/ST2/01406; the National Priorities Research Program by Qatar National Research Fund; the Programa de Excelencia Mar\'{i}a de Maeztu and the Programa Severo Ochoa del Principado de Asturias; the Thalis and Aristeia programs cofinanced by EU-ESF and the Greek NSRF; the Rachadapisek Sompot Fund for Postdoctoral Fellowship, Chulalongkorn University and the Chulalongkorn Academic into Its 2nd Century Project Advancement Project (Thailand); the Welch Foundation, contract C-1845; and the Weston Havens Foundation (USA).
\end{acknowledgments}

\bibliography{auto_generated}
\cleardoublepage \appendix\section{The CMS Collaboration \label{app:collab}}\begin{sloppypar}\hyphenpenalty=5000\widowpenalty=500\clubpenalty=5000\vskip\cmsinstskip
\textbf{Yerevan Physics Institute, Yerevan, Armenia}\\*[0pt]
A.M.~Sirunyan, A.~Tumasyan
\vskip\cmsinstskip
\textbf{Institut f\"{u}r Hochenergiephysik, Wien, Austria}\\*[0pt]
W.~Adam, F.~Ambrogi, E.~Asilar, T.~Bergauer, J.~Brandstetter, E.~Brondolin, M.~Dragicevic, J.~Er\"{o}, A.~Escalante~Del~Valle, M.~Flechl, M.~Friedl, R.~Fr\"{u}hwirth\cmsAuthorMark{1}, V.M.~Ghete, J.~Grossmann, J.~Hrubec, M.~Jeitler\cmsAuthorMark{1}, A.~K\"{o}nig, N.~Krammer, I.~Kr\"{a}tschmer, D.~Liko, T.~Madlener, I.~Mikulec, E.~Pree, N.~Rad, H.~Rohringer, J.~Schieck\cmsAuthorMark{1}, R.~Sch\"{o}fbeck, M.~Spanring, D.~Spitzbart, A.~Taurok, W.~Waltenberger, J.~Wittmann, C.-E.~Wulz\cmsAuthorMark{1}, M.~Zarucki
% ...
\vskip\cmsinstskip
% ...
1:  Also at Vienna University of Technology, Vienna, Austria\\
% ...
\end{sloppypar}
\end{document}